\documentclass[conference]{IEEEtran}

\usepackage[T1]{fontenc}
\usepackage[utf8]{inputenc}

\usepackage{amsmath}
\usepackage{amssymb}

\usepackage{graphicx}
\usepackage{subcaption}   
\usepackage{float}        

\usepackage{booktabs}     
\usepackage{multirow}     
\usepackage{array}        

\usepackage{enumitem}     

\usepackage[hidelinks,breaklinks=true]{hyperref}

\usepackage{booktabs}
\usepackage{url} 
\usepackage{amsmath} 
\usepackage{amsfonts} 
\usepackage{multicol} 
\usepackage{graphics} 
\usepackage[sort,nocompress]{cite} 

\usepackage{amsmath,amsfonts}
\usepackage{algorithmic}
\usepackage{algorithm}
\usepackage{array}
\usepackage{textcomp}
\usepackage{stfloats}
\usepackage{verbatim}
\usepackage{balance}
\usepackage{colortbl}
\usepackage{arydshln}
\usepackage{titlecaps}

\usepackage[dvipsnames]{xcolor}

\usepackage{tikz}
\usepackage{todonotes}
\usepackage{graphicx}
\usepackage{caption}
\usepackage[normalem]{ulem}
\usepackage{multirow}
\usepackage{xspace}
\usepackage{tabularx}
\usepackage{hyperref}
\usepackage{orcidlink}

\useunder{\uline}{\ul}{}
\usepackage{subcaption}

\newcommand{\espresso}{\textsc{ESPRESSO}\xspace}

\newcommand{\paragraphX}[1]{\vskip 4pt \noindent \textbf{#1} \hskip .05in}

\begin{document}

\title{Tracing the Chain: Deep Learning for\\Stepping-Stone Intrusion Detection}

\author{
    \IEEEauthorblockN{
        Nate Mathews\textsuperscript{\orcidlink{0000-0001-6186-7001}},\;
        Matthew Wright\textsuperscript{\orcidlink{0000-0002-8489-6347}}
    }
    \IEEEauthorblockA{
        \textit{Department of Cybersecurity}\\
        \textit{Rochester Institute of Technology}\\
        Rochester, NY, USA\\
        \{njm3308, matthew.wright\}@rit.edu
    }
    \and
    \IEEEauthorblockN{
        Nicholas Hopper\textsuperscript{\orcidlink{0000-0003-2536-9587}}
    }
    \IEEEauthorblockA{
        \textit{Department of Computer Science \& Engineering}\\
        \textit{University of Minnesota}\\
        Minneapolis, MN, USA\\
        hoppernj@umn.edu
    }
}

\maketitle

\begin{abstract}
Stepping-stone intrusions (SSIs) are a prevalent network evasion
technique in which attackers route sessions through chains of
compromised intermediate hosts to obscure their origin. Effective
SSI detection requires correlating the incoming and outgoing flows
at each relay host at extremely low false positive rates---a
stringent requirement that renders classical statistical methods
inadequate in operational settings. We apply ESPRESSO, a deep
learning flow correlation model combining a transformer-based
feature extraction network, time-aligned multi-channel interval
features, and online triplet metric learning, to the problem of
stepping-stone intrusion detection. To support training and
evaluation, we develop a synthetic data collection tool that
generates realistic stepping-stone traffic across five tunneling
protocols: SSH, SOCAT, ICMP, DNS, and mixed multi-protocol chains.
Across all five protocols and in both host-mode and network-mode
detection scenarios, ESPRESSO substantially outperforms the
state-of-the-art DeepCoFFEA baseline, achieving a true positive
rate exceeding 0.99 at a false positive rate of $10^{-3}$ for
standard bursty protocols in network-mode. We further demonstrate
chain length prediction as a tool for distinguishing malicious
from benign pivoting, and conduct a systematic robustness analysis
revealing that timing-based perturbations are the primary
vulnerability of correlation-based stepping-stone detectors.
\end{abstract}

\begin{IEEEkeywords}
stepping-stone intrusion detection,
network traffic correlation,
flow correlation,
deep learning,
triplet metric learning,
transformer networks,
network security
\end{IEEEkeywords}



\section{Introduction}
\label{sec:intro}

Stepping-stone intrusions (SSIs), also known as pivoting, are a fundamental
evasion tactic in the modern attacker's playbook. Rather than connecting
directly to a target, an adversary routes their session through a chain of
compromised intermediate hosts---stepping stones---each of which relays the
traffic to the next, obscuring the true origin of the attack. This technique
has been widely exploited in high-profile campaigns, malware operations, and
botnets~\cite{international2016panamaExample,lee2016tlpExample,ayala2016activeExample,mcafee2011nightdragonExample,tankard2011operationAuroraExample},
and the European Union Agency for Cybersecurity (ENISA) has identified SSIs
as one of the top ten threats to IoT
security~\cite{EUISA2017IoTReport}. The core defensive challenge is
therefore \emph{flow correlation}: given the traffic observed at a
host or network boundary, determine whether an incoming flow and an
outgoing flow belong to the same end-to-end connection.

Because attackers routinely encrypt their traffic, content-based inspection
is largely ineffective against modern SSIs~\cite{staniford1995_ContentSSID}.
Passive detection methods instead exploit subtle statistical similarities
in the timing, size, and rate of network
flows~\cite{zhang2000_IntervalSSID,yoda2000_deviationSSID,stepping-stones,donoho2002_multiscaleSSID,blum2004_PoissonRWSSID,Yang2011_ContextDependedSSID,KUMAR2016_NeuralNetSSID}.
Active methods go further, embedding timing-based watermarks into flows that
survive the transformations introduced by intermediate
hosts~\cite{wang2001sleepy,Wang2003_Watermark,houmansadr2009rainbow,Mo2021_QuaternaryWatermark}.
Despite two decades of research in both directions, a critical and often
overlooked challenge remains unsolved: the \emph{extremely low base rate}
of stepping-stone activity in real networks.

In a typical large-scale deployment, malicious relayed traffic constitutes
only a tiny fraction of total connection volume.
Consider a network gateway that observes one million connection pairs per
hour, of which only ten are part of a true stepping-stone chain.
A detector with a seemingly excellent 1\% false positive rate (FPR)
would still generate approximately 10{,}000 false alarms per hour
while correctly flagging all ten true intrusions---an entirely
intractable workload for any security analyst.
For stepping-stone detection to be operationally useful, its FPR must be
driven far below 1\%, ensuring that genuine alerts are not buried under a
flood of false positives.
This stringent low-FPR requirement is a regime in which most existing
passive and active SSID methods fall short.

The challenge of correlating encrypted, transformed flows under tight FPR
constraints mirrors the problem of traffic correlation attacks against
anonymity networks like Tor, where adversaries attempt to link ingress
and egress flows to deanonymize users. Recent deep learning approaches
to this problem---most notably DeepCoFFEA
(DCF)~\cite{oh2022deepcoffea}---have demonstrated that triplet metric
learning can yield significant improvements in the low-FPR regime
compared to classical statistical methods. The key insight is to train
a neural network to \emph{embed} flows into a space where
correlated pairs cluster together and uncorrelated pairs are pushed
apart, rather than performing direct pairwise classification. This
embedding approach generalizes well to unseen traffic conditions and
scales efficiently to large numbers of candidate pairs.

In this paper, we investigate the application of \espresso~\cite{Chawla2024ESPRESSO_poster}, a deep
learning flow correlation model that extends DCF with a
transformer-based feature extraction network, time-aligned
multi-channel interval features, and online triplet mining, to the
domain of stepping-stone intrusion detection. The primary 
contributions of this paper are the adaptation, 
evaluation, and analysis of this correlation approach in the SSID
setting, which has received comparatively little attention from the
deep learning community:

\begin{itemize}

    \item \textbf{Synthetic stepping-stone dataset.}
    A primary obstacle to progress in this field is the absence of
    large-scale, labeled datasets suitable for training and evaluating
    modern correlation methods. We develop a flexible synthetic data
    collection tool that simulates stepping-stone attack chains over
    five tunneling protocols (SSH, SOCAT/TCP, ICMP, DNS, and mixed
    multi-protocol chains) using realistic traffic distributions derived
    from real-world SSH captures.

    \item \textbf{Strong baseline results.}
    We benchmark \espresso\ against DCF across all five datasets in both a
    \emph{host-mode} setting (correlating flows at a single intermediate
    host) and the more demanding \emph{network-mode} setting (correlating
    traffic at the network boundary). \espresso\ consistently and
    substantially outperforms DCF across all protocols and both detection
    modes.

    \item \textbf{Loss function enhancements.}
    We investigate two auxiliary training objectives---temporal alignment
    and feature decorrelation---designed to improve the quality of the
    learned embeddings. Temporal alignment yields measurable gains,
    particularly for challenging mixed-protocol traffic, while feature
    decorrelation provides inconsistent benefits.

    \item \textbf{Chain length prediction.}
    Since long connection chains are a strong indicator of malicious
    intent, we explore two complementary approaches to estimating chain
    length: a standalone CNN-based regression model and a multi-task
    extension of \espresso. We find that the standalone model offers the
    best accuracy, while chain reconstruction via full-path correlation
    provides high-confidence results where correlation signals are strong.

    \item \textbf{Robustness evaluation.}
    We systematically evaluate \espresso's resilience against packet
    padding and timing perturbation obfuscation strategies. While
    \espresso\ is highly robust to padding-only defenses, we find that
    even modest artificial timing jitter causes significant performance
    degradation, identifying the primary vulnerability of timing-based
    correlation methods.

    \item \textbf{Source code.} We release the traffic model code\footnote{\href{https://github.com/notem/ESPRESSO-Flow-Correlation}{https://github.com/notem/ESPRESSO-Flow-Correlation}}\footnote{\href{https://github.com/notem/SSID-ESPRESSO}{https://github.com/notem/SSID-ESPRESSO}}, data generation tools\footnote{\href{https://github.com/notem/SSI-Simulator}{https://github.com/notem/SSI-Simulator}}, and datasets\footnote{\href{https://drive.google.com/drive/folders/1tPEkS6BXcSXXtwoBGICEmA-UoiH0Lkiv}{https://drive.google.com/drive/folders/1tPEkS6BXcSXXtwoBGICEmA-UoiH0Lkiv}} for community usage.

\end{itemize}

The remainder of this paper is organized as follows.
Section~\ref{sec:background} reviews the stepping-stone intrusion
threat model and prior detection approaches, and introduces the
transformer architecture and triplet metric learning framework that
underpin our model.
Section~\ref{sec:espresso} describes the \espresso\ architecture in
detail, including its time-interval feature representation,
transformer backbone, and online mining strategy.
Section~\ref{sec:ssid} presents our stepping-stone detection
experiments: dataset generation, baseline comparisons, loss
enhancements, chain length prediction, and robustness against
obfuscation.
Section~\ref{sec:discussion} discusses the broader implications of our
findings and directions for future work, and
Section~\ref{sec:conclusion} concludes.
Appendix~\ref{app:tor} reports the original evaluation of \espresso\
on Tor traffic correlation, which motivated its application to the
SSID domain.
Appendix~\ref{app:dcf_ablation} provides supplemental ablation
experiments on the DCF architecture.

\section{Background}
\label{sec:background}

This section provides the background necessary to understand the
design and evaluation of \espresso.
We begin by surveying the stepping-stone intrusion threat and the
landscape of prior detection approaches
(Section~\ref{sec:bg_ssid}).
We then describe Tor traffic correlation---the problem domain in which
\espresso\ was originally developed---and explain why the two problems
are deeply related (Section~\ref{sec:bg_tor}).
Finally, we review the transformer architecture
(Section~\ref{sec:bg_transformer}) and the triplet metric learning
framework (Section~\ref{sec:bg_triplet}) that form the technical
foundations of \espresso.

\subsection{Stepping-Stone Intrusions and Detection}
\label{sec:bg_ssid}


Stepping-stone attacks, also known as pivoting, involve adversaries
using multiple compromised intermediary hosts to relay traffic, making
it difficult to trace the origin of the attack.
In a \emph{stepping stone intrusion (SSI)}, an attacker establishes a
sequence of connections through multiple compromised intermediary hosts,
known as stepping stones.
These connections form a \emph{connection chain}, where each pair of
hosts communicates using \emph{bidirectional flows} (e.g., a single SSH
connection), which can be further divided into unidirectional
\emph{send} and \emph{echo} (or reply) flows.

The primary motivations for using SSIs include evading detection,
concealing the attacker's identity, and facilitating lateral movement
within a target network. SSIs are particularly effective and have been
widely utilized in high-profile attacks, malware campaigns, and
botnets~\cite{international2016panamaExample,lee2016tlpExample,ayala2016activeExample,mcafee2011nightdragonExample,tankard2011operationAuroraExample}.
Recognizing their potency, the European Union Agency for Cybersecurity
(ENISA) has identified SSIs as one of the top ten threats to IoT
security~\cite{EUISA2017IoTReport}.

\begin{figure}
   \centering
   \includegraphics[width=0.9\linewidth, trim={0.2cm 0.9cm 0.2cm
   0.6cm}, clip]{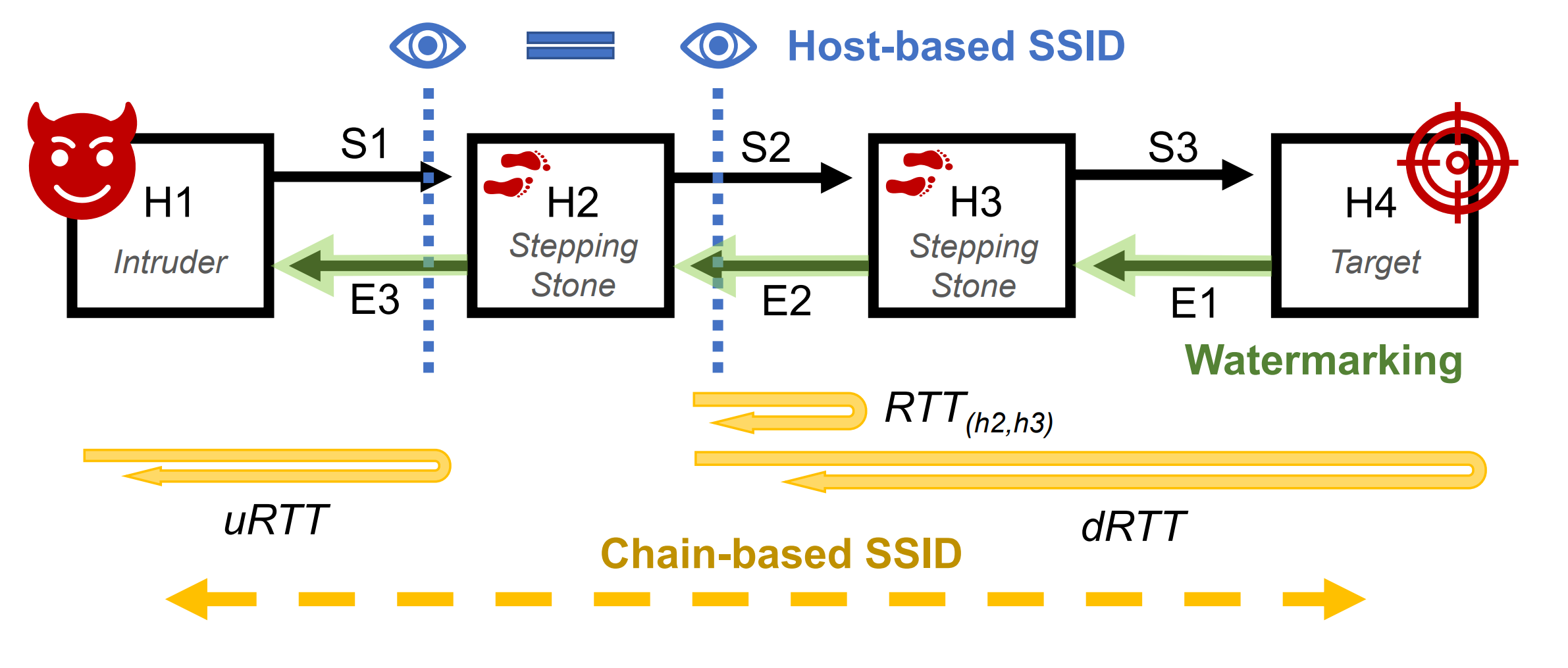}
   \caption{The standard stepping-stone intrusion detection model with
   a 3-hop chain containing two stepping stones (H2 and H3).}
   \label{fig:StandardSSIDModel}
\end{figure}

In response to this threat, various \emph{stepping stone intrusion
detection (SSID)} methods have been proposed over the last two decades.
We present a standard model for SSID in
Figure~\ref{fig:StandardSSIDModel}.
These methods can be divided into two main categories: \emph{passive
SSID} and \emph{active SSID}.

\subsubsection{Passive SSID}

Passive SSID techniques detect stepping-stone intrusions by analyzing
network traffic without interfering. These methods are further
subdivided into \emph{host-based}, \emph{network-based}, and
\emph{chain-based} detection
approaches~\cite{Wang2018_SSIDSurvey}.

\paragraph{Host \& Network-Based SSID}
Host-based methods focus on identifying whether a specific host within
the network is acting as a stepping stone. For example, in a three-host
chain (H1, H2, H3), host-based SSID would analyze H2 to determine if
it is relaying traffic between H1 and H3.

Early host-based methods relied on content-based fingerprinting, which
proved ineffective against encrypted communications prevalent in modern
networks~\cite{staniford1995_ContentSSID}. Consequently, researchers
shifted towards analyzing packet timing information to correlate
flows~\cite{zhang2000_IntervalSSID,yoda2000_deviationSSID,stepping-stones,donoho2002_multiscaleSSID}.
Techniques included:

\begin{itemize}
    \item \textbf{Traffic Activity Patterns:} Zhang et
    al.~\cite{zhang2000_IntervalSSID} utilized periods of traffic
    activity (ON/OFF) to correlate flows between hosts.
    \item \textbf{Time Lag Analysis:} Yoda et
    al.~\cite{yoda2000_deviationSSID} employed average time lag
    measurements between connections to establish correlations.
    \item \textbf{Inter-Packet Delay Similarity:} Wang et
    al.~\cite{stepping-stones} compared flows based on the similarity
    of their inter-packet delays.
    \item \textbf{Wavelet Transformation:} Donoho et
    al.~\cite{donoho2002_multiscaleSSID} applied wavelet
    transformations to inter-arrival times, separating short-term and
    long-term traffic behaviors.
\end{itemize}

To enhance resistance against jitter and \emph{chaff} (fake or padding
packets), methods were developed to model traffic as a Poisson
process~\cite{blum2004_PoissonRWSSID,he2007_PoissonSSID,Yang2008_PoissonSSID}.
Additionally, packet length features have been identified as valuable
for flow
correlation~\cite{Yang2021_PacketLengthSSID}. More recent approaches
incorporate Pearson's correlation
coefficient~\cite{Yang2011_ContextDependedSSID} and neural networks to
improve detection performance~\cite{KUMAR2016_NeuralNetSSID}.

Despite these advancements, host-based SSID methods have limitations,
particularly in distinguishing between benign and hostile intrusions.
Moreover, recent research has shown that many traditional techniques
achieve low true positive rates with minimal false positive rates when
faced with advanced evasion strategies like
chaffing~\cite{YANG2018_Chaff,Clausen2020_Chaff}.

\paragraph{Chain-Based SSID}
Chain-based SSID techniques assess a connection chain's overall length
and structure to identify potential intrusions. The rationale is that
benign stepping stones are unlikely to form long chains, whereas
malicious intrusions often involve extended chains to obscure the
attacker's origin.

Key methods in chain-based SSID include:

\begin{itemize}
    \item \textbf{Round-Trip Time (RTT) Estimation:} Early methods
    estimate RTT for each hop and compare it against expected values to
    infer chain
    lengths~\cite{yung2002_detectingChainSSID,yang2004_chainSSID}.
    \item \textbf{Clustering and Data Mining:} More recent approaches
    use clustering algorithms to handle outliers and improve RTT
    estimation
    accuracy~\cite{wang2020_kmeansSSID,wang2021_kmeans,YANG2007_ChainSSID}.
\end{itemize}

Chain-based SSID is most effective when sensors are positioned early in
the network, allowing for more accurate RTT measurements. However,
accurately estimating both upstream and downstream RTT remains
challenging, and current methods have not fully addressed robustness
against sophisticated evasion techniques.

\subsubsection{Active SSID (Watermarking)}

A separate line of work pursues active flow correlation techniques that
insert ``watermarks'' into network flows---by delaying or dropping
packets---that can survive the transformations introduced by various
network
conditions~\cite{Wang2003_Watermark,Rezaei2021_FINNWatermarking,Mo2021_QuaternaryWatermark,Yu2021_DynamicIntervalWatermarking,wang2001sleepy,houmansadr2009rainbow,Gong2013_WatermarkErrors,Rezaei2017_TagIt,HOUMANSADR2013_BotMosaic,Iacovazzi2018_DropWat,houmansadr2013non,wang2007network}.
These techniques have been applied successfully against botnet
detection~\cite{HOUMANSADR2013_BotMosaic}, anonymity system flow
correlation~\cite{wang2007network}, and SSID~\cite{wang2001sleepy}.
In comparison to passive SSID, watermarking incurs a network overhead
that must be considered for real-world deployment.
These methods can be categorized into \emph{blind} and
\emph{non-blind} watermarking techniques.

Blind watermarking techniques embed watermarks without requiring any
prior knowledge of the traffic flow. The primary challenge is ensuring
the watermark remains detectable despite potential network-induced
perturbations.
Non-blind techniques combine watermark embedding with traffic feature
analysis. For example, RAINBOW~\cite{houmansadr2009rainbow} integrates
watermarking with statistical traffic features to enhance detection
robustness.

Key challenges in watermarking-based SSID include ensuring that
watermarks survive network perturbations like jitter and packet loss;
making watermarks imperceptible to avoid adversarial evasion; and
efficiently applying watermarking across large networks without
significant overhead.
Recent advancements have focused on improving watermark robustness and
efficiency~\cite{Mo2021_QuaternaryWatermark}. However, adversaries have
also devised methods to detect and neutralize
watermarks~\cite{Jia2013_BlindWatermarkDetection,Lin2012_TimingWatermarkAttacks,Peng2006_OnTimingWatermarkSecrecy,Luo2011_BACKLIT,kiyavash2008multi},
and few techniques address deliberate noise addition by attackers.

\subsection{Tor Traffic Correlation}
\label{sec:bg_tor}

The challenge posed by stepping-stone connection chains mirrors that of
flow correlation attacks against anonymity networks such as Tor, where
an adversary positioned at two points in the network attempts to link
an ingress flow (entering the anonymity system) with its corresponding
egress flow (exiting toward the destination) to deanonymize a user.
This structural parallel is important: in both settings, the two flows
being correlated have traversed one or more intermediate relays that
introduce timing jitter, packet re-sizing, and other
transformations---yet they must retain enough statistical similarity to
be identified as belonging to the same end-to-end connection.

Early Tor correlation attacks used classical statistical
methods~\cite{Nasr18DeepCorr}, but the introduction of deep learning
transformed the state of the art.
DeepCorr~\cite{Nasr18DeepCorr} was among the first to demonstrate that
a convolutional neural network trained end-to-end on raw packet
sequences could substantially outperform hand-crafted features.
DeepCoFFEA (DCF)~\cite{oh2022deepcoffea} advanced the field further by
replacing direct classification with \emph{triplet metric learning}
(described in Section~\ref{sec:bg_triplet}), which embeds flows into a
low-dimensional space where correlated pairs cluster together. This
approach generalized better to unseen network conditions and achieved
significantly lower false positive rates than prior methods.

\espresso\ was designed to further improve upon DCF for Tor correlation,
and its performance on that problem is documented in
Appendix~\ref{app:tor}.
Crucially, the low-FPR requirements of Tor correlation---where even
1-in-a-million false positives can expose an innocent user---closely
align with the operational constraints of stepping-stone detection
discussed in Section~\ref{sec:intro}. This makes Tor correlation
research a natural source of techniques for SSID, and motivates the
application of \espresso\ to the stepping-stone domain.

\subsection{Transformer Networks}
\label{sec:bg_transformer}

The transformer~\cite{vaswani2017attention} is a deep learning
architecture that excels at capturing contextual relationships in
sequential data. By using an attention mechanism to process all tokens
in an input sequence simultaneously, the model can attend to all
positions in the input, making it effective for capturing relationships
throughout the sequence.

The transformer's self-attention mechanism computes a weighted sum of
all tokens in the input for each token, with weights determined by how
relevant each token is to the others. Specifically, the mechanism uses
three representations: \emph{query}, \emph{key}, and \emph{value}. The
relevance between a pair of tokens is computed by taking the dot
product of their \emph{query} and \emph{key} representations and
applying a softmax function. The outcome is used to weight the value
representations, such that the output is a contextually enriched
embedding for each token.
Layers in the transformer consist of a multi-head self-attention (MHSA)
mechanism and a position-wise feed-forward network. The multi-head
attention allows the model to focus on different parts of the input for
different tasks or reasons, amplifying its capacity to discern complex
patterns.

Despite its success across many domains, self-attention has a critical
limitation: computational cost scales quadratically with sequence
length, since every token attends to every other token. This is
problematic for applications that operate on long sequences, such as
the traffic traces used in flow correlation.

Since the self-attention mechanism has no inherent notion of order,
\emph{positional encodings} are added to the input embeddings to
preserve the positional context of each token.
In \espresso, this role is fulfilled naturally by the time-interval
feature representation described in Section~\ref{sec:time_interval_features},
which embeds positional information directly into the input.

\subsubsection{The Convolutional Vision Transformer (CvT)}

The Convolutional vision Transformer
(CvT)~\cite{Wu2021} improves vision transformers in terms of
performance and efficiency by strategically adding convolutional layers,
combining the shift and scale invariance of convolutions with the
transformer's ability to model long-range dependencies. This allows the
architecture to consider both local and global structures efficiently.

The CvT arranges transformers hierarchically across multiple stages.
At the onset of each stage, a convolutional layer increases the feature
dimension while performing spatial downsampling via strided convolutions.
As one progresses through the stages, the feature dimension grows while
the sequence length is reduced, decreasing the cost of attention.

\begin{figure}[t]
\centering
\includegraphics[width=0.7\linewidth]{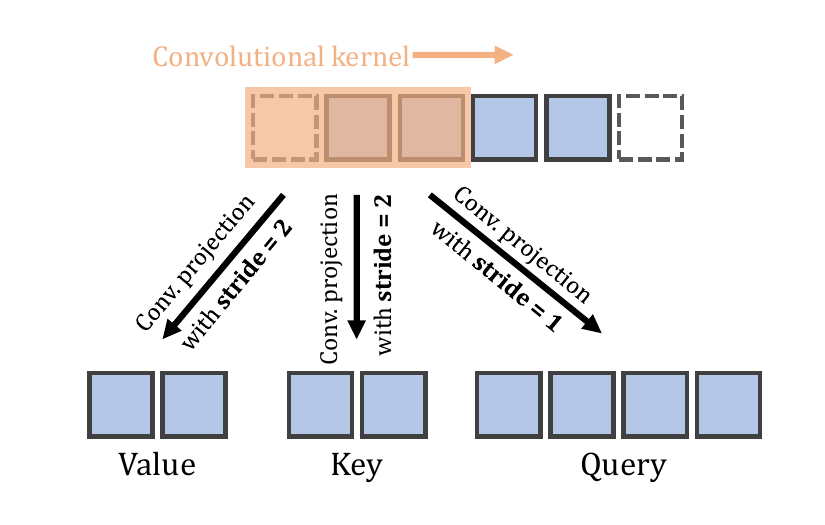}
\vskip -0.3cm
\caption{Convolutional projection during transformer self-attention. Strided convolutional projection of the \emph{value} and \emph{key} vectors reduces the sequence length and thus significantly reduces the impact of the quadratic computational complexity of attention.}
\label{fig:conv_attention}
\end{figure}

Furthermore, CvT substitutes the conventional linear projection used to
produce the \emph{key}, \emph{query}, and \emph{value} representations
with \emph{convolutional} projection (see
Figure~\ref{fig:conv_attention}).
The convolutional kernel captures local relationships to embed into the
attention mechanism, and strided convolutions on the \emph{key} and
\emph{value} sequences reduce their length, providing additional
computational efficiency.
The \espresso\ architecture builds on these concepts to construct an
efficient transformer backbone for network traffic sequences, as
described in Section~\ref{sec:transformer_backbone}.

\subsection{Triplet Metric Learning for Flow Correlation}
\label{sec:bg_triplet}

Framing flow correlation as a standard classification task is a poor
fit for the problem: the goal is not to assign flows to discrete
classes, but to map \emph{related} flows from different vantage points
onto one another in a shared embedding space. Triplet metric learning
provides a powerful framework for this. DeepCoFFEA
(DCF)~\cite{oh2022deepcoffea}, inspired by Triplet
Fingerprinting~\cite{TripletFingerprinting2019}, was the first to apply
this approach to flow correlation and demonstrated substantial gains
over classification-based methods like
DeepCorr~\cite{Nasr18DeepCorr}.

\subsubsection{The Triplet Loss Function}

At the heart of triplet metric learning is the \emph{triplet loss
function}, designed to ensure that, in the learned embedding space,
correlated traffic flows are closer together than uncorrelated flows.
The loss is computed using three inputs: an anchor sample $x_a$, a
positive sample $x_p$ (correlated with the anchor), and a negative
sample $x_n$ (uncorrelated). The network learns an embedding function
$f(x)$ such that the anchor--positive distance is smaller than the
anchor--negative distance by a margin $\alpha$:

\[
d(f(x_a), f(x_p)) + \alpha < d(f(x_a), f(x_n))
\]

where $d(\cdot, \cdot)$ denotes the distance metric\footnote{Measures
of similarity, such as cosine similarity used by DCF and \espresso, can
be used in place of distance by reversing the inequality.} and $\alpha$
is a margin that prevents the model from collapsing all distances to
zero. The corresponding loss function $L$ is:

\[
L = \max \left( 0,\; d(f(x_a), f(x_p)) - d(f(x_a), f(x_n)) + \alpha
\right)
\]

\begin{figure}[h!]
  \centering
  \includegraphics[width=0.45\textwidth,trim={4cm 6cm 2.5cm
  0},clip]{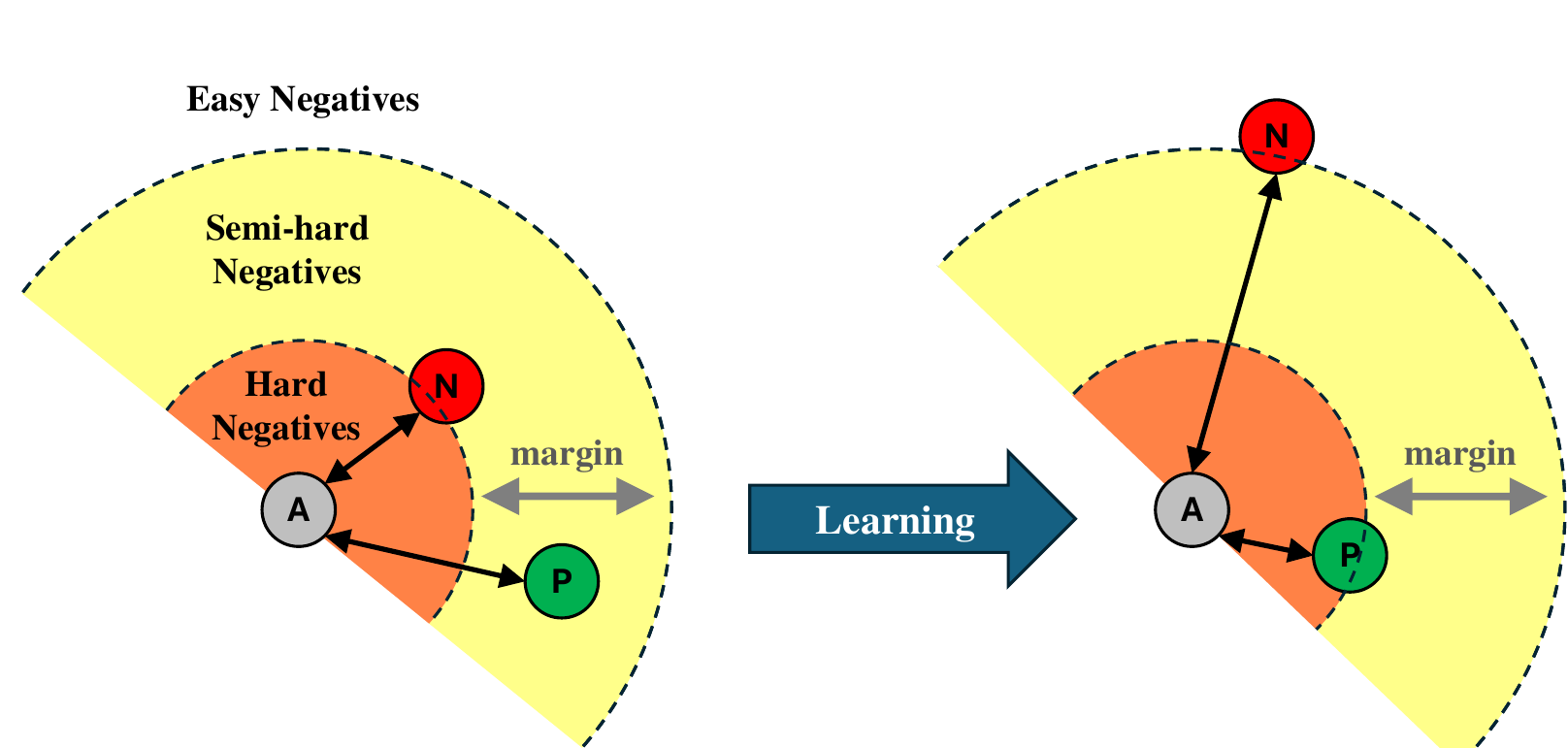}
  \caption{Illustration of triplet metric learning. The anchor and
  positive flows are pushed closer together, while the negative flow is
  pushed away in the learned embedding space.}
  \label{fig:triplet_learning}
\end{figure}

This loss minimizes the distance between correlated flows (anchor and
positive) while maximizing the distance between uncorrelated flows
(anchor and negative), as illustrated in
Figure~\ref{fig:triplet_learning}.

\subsubsection{Triplet Mining Strategies}

In practice, training with triplet loss requires carefully selecting
triplets from the dataset. Naive random sampling is inefficient, as
many triplets may already satisfy the constraint (i.e., the anchor is
already closer to the positive than the negative) and contribute no
gradient signal.
Two key strategies address this:

\paragraph{Hard Negative Mining.}
Hard negative mining selects triplets where the negative sample is very
close to the anchor in the embedding space, forcing the network to
sharpen its decision boundaries. A hard negative satisfies:

\[
d(f(x_a), f(x_n)) < d(f(x_a), f(x_p)) + \alpha
\]

\paragraph{Semi-Hard Negative Mining}
Semi-hard mining selects negatives that are farther from the anchor
than the positive but still within the margin $\alpha$, balancing
difficulty with training stability:

\[
d(f(x_a), f(x_p)) < d(f(x_a), f(x_n)) < d(f(x_a), f(x_p)) + \alpha
\]

A variant that combines both hard and semi-hard negatives (excluding
only easy triplets) exposes the model to a diverse range of difficult
distinctions, improving generalization without destabilizing
convergence. This combined strategy is used by DCF.

\subsubsection{Key Contributions of DeepCoFFEA}
\label{sec:bg_dcf}

DCF introduced several innovations beyond standard triplet
learning that make it particularly effective for flow correlation.

\paragraph{Separate Feature Embedding Networks (FENs)}
DCF uses separate feature embedding networks for Tor ingress and egress
flows, recognizing that traffic characteristics differ significantly
between the two: ingress traffic is encapsulated in Tor cells of fixed
size, while egress traffic follows the protocol of the underlying
application. Two distinct networks allow the model to capture the
unique nature of traffic at each observation point, leading to more
accurate embeddings.

\paragraph{Amplification with Windowing}
Rather than comparing entire flows at once, DCF divides flows into
short overlapping windows, computes embeddings for each window, and
aggregates results through a threshold-based voting mechanism. This
windowing strategy amplifies the correlation signal while reducing
false positives and allows the model to handle traffic variability
within different time frames.

\noindent
\espresso\ retains both of these innovations while introducing
architectural improvements that substantially boost performance,
particularly in the low-FPR regime. These improvements are described
in detail in the following section.

\section{Deep Learning for Flow Correlation}
\label{sec:espresso}

This section describes the deep learning flow correlation model at the
core of our stepping-stone detection system.
The model, \espresso, was originally developed for Tor traffic
correlation and is described and evaluated in that context in
Appendix~\ref{app:tor}; here we present the components necessary to
understand its application to SSID.
The design rests on three pillars: a multi-channel time-interval
feature representation (Section~\ref{sec:features}), a
transformer-based feature extraction network
(Section~\ref{sec:transformer_backbone}), and a triplet learning
training framework with online mining
(Section~\ref{sec:dcf_framework}).

\subsection{Multi-Channel Traffic Feature Representation}
\label{sec:features}

The choice of input representation has an outsized effect on the
performance of deep learning models for traffic analysis.
Early deep learning attacks on network
traffic~\cite{Sirinam2018DeepFingerprinting,Rimmer2018AWF} used a
single-channel direction-only representation---a sequence of $+1$ for
each outgoing packet and $-1$ for each incoming packet.
Subsequent work showed that incorporating timing
information~\cite{rahman2020tiktok,bhat2019varcnn} and exploring
richer feature spaces~\cite{Shen2023RobustFingerprintingRF} can
substantially improve model accuracy, since different feature forms
leak variable amounts of information depending on the traffic
conditions~\cite{Mathews2023sok,Li2018infoleak}.

Motivated by these findings, we use a \emph{multi-channel} input
representation that provides the model with a variety of complementary
views of the same traffic sequence, allowing it to learn more robust
internal representations.
Our features fall into two categories.

\paragraph{Packet-Level Features}
Packet-level representations operate directly on the ordered sequence
of individual packets.
The two fundamental packet-level feature vectors are \texttt{times}
--- raw packet timestamps --- and \texttt{dirs} --- a sequence of $+1$
for each outgoing packet and $-1$ for each incoming packet, padded to
the maximum sequence length with zeros.
Additional packet-level features are derived from these base vectors by
applying transformations; for example, \texttt{burst\_edges} is a
sparse vector computed as the difference between adjacent \texttt{dirs}
values, taking on values of $+2$, $-2$, or (most commonly) $0$, and
thus compactly encodes the boundaries between traffic bursts.

\paragraph{Interval Features}
\label{sec:time_interval_features}
Packet-level representations have an important limitation for flow
correlation: two correlated flows, having passed through intermediate
relays, will generally contain different numbers of packets due to
re-segmentation.
Comparing them on a packet-index basis therefore introduces a
systematic misalignment.
Interval features address this by aggregating packets into fixed-width
time bins, producing representations whose indices correspond to
identical periods of real time in both flows, regardless of how many
packets each interval contains.

Concretely, for a time interval of width $\Delta t$, the traffic is
divided into consecutive windows and a set of statistics is computed
per window.
For SSID, \espresso\ uses nine interval features, combining direction
and size information:

\begin{itemize}[nosep]
    \item \texttt{interval\_dirs\_up} / \texttt{interval\_dirs\_down}:
    count of upstream / downstream packets per interval.
    \item \texttt{interval\_dirs\_sum} / \texttt{interval\_dirs\_sub}:
    sum and difference of the upstream and downstream packet counts.
    \item \texttt{interval\_size\_up} / \texttt{interval\_size\_down}:
    total bytes sent upstream / downstream per interval.
    \item \texttt{interval\_size\_sum} / \texttt{interval\_size\_sub}:
    sum and difference of upstream and downstream byte counts.
    \item \texttt{interval\_cumul\_norm}: cumulative normalized
    directional byte count, providing a global view of the traffic's
    directional trend.
\end{itemize}

The \texttt{interval\_size\_*} features are included because, unlike
Tor traffic (where all cells have a fixed size), the raw egress traffic
in the SSID setting is not encapsulated in uniform-length cells, making
packet size a meaningful discriminating signal.

Each feature vector is computed over the same set of time bins and
stacked along a channel dimension, forming a multi-channel tensor of
shape $[C \times T]$ where $C = 9$ is the number of channels and $T$
is the number of time intervals.
The interval width used in all experiments is $\Delta t = 30$\,ms, and
the input is truncated or padded to $T = 1{,}200$ bins (corresponding
to 36 seconds of traffic).

\subsection{Transformer-Based Feature Extraction}
\label{sec:transformer_backbone}

The feature extraction network (FEN) maps an input traffic sequence to
a sequence of embedding vectors that are subsequently used for
correlation.
\espresso\ uses a transformer-based FEN, building on concepts from the
Convolutional vision Transformer (CvT)~\cite{Wu2021} described in
Section~\ref{sec:bg_transformer}.

The FEN begins with a convolutional embedding layer that projects the
multi-channel input into a sequence of $d$-dimensional tokens (kernel
size 3, stride 3).
A stack of $L$ transformer blocks is then applied.
Each block uses a multi-head self-attention (MHSA) layer with
CvT-style convolutional projection for the \emph{key} and \emph{value}
sequences (kernel size 3, stride 2), which reduces the effective
sequence length during attention and captures local structure within
the attention computation.
The self-attention layer is followed by a position-wise feed-forward
(MLP) sublayer.
By processing the \emph{full} traffic sequence before any windowing
operation, the FEN can capture global temporal context that would be
lost if windows were processed in isolation, as in DCF.

\begin{figure}
  \centering
  \includegraphics[trim={2cm 4cm 2.5cm 0},clip,width=0.95\columnwidth]{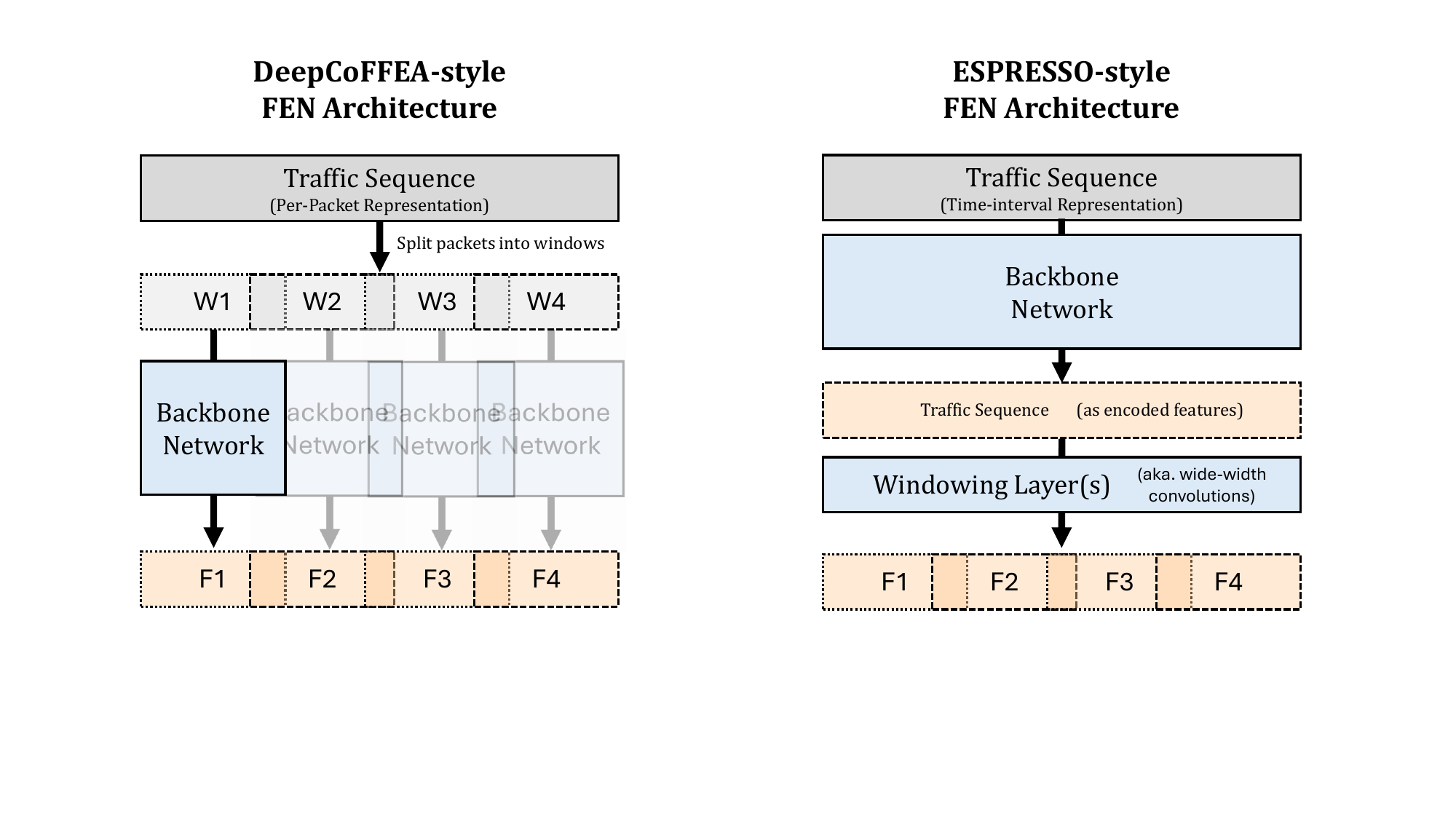}
  \caption{Architecture comparison between DCF and \espresso.
  \espresso\ generates a global feature sequence first and then applies
  windowing, whereas DCF performs windowing before feature extraction.}
  \label{fig:espresso_architecture}
\end{figure}

After the transformer blocks, a large-kernel convolutional layer
(kernel size 50, stride 3) maps the resulting feature stream into a
sequence of window embeddings.
This \emph{post-FEN windowing} strategy (illustrated in
Figure~\ref{fig:espresso_architecture}) has two advantages over the
DCF approach of windowing before feature extraction: (i) the entire
sequence is encoded in a single forward pass rather than one pass per
window, and (ii) the windows are guaranteed to be temporally aligned
because the interval feature representation ensures that each index
corresponds to the same absolute time in both flows.

The hyperparameters of the FEN are summarized in
Table~\ref{tab:espresso_hyperparams}.
The effective temporal span of each output window is
$0.03\,\text{s} \times 3 \times 50 = 4.5$\,seconds, which is
comparable to the default 5-second window used by DCF.
Two architectural ablations --- \textbf{GreenTea} (MHSA replaced with
a local convolutional mixer, kernel width 7) and \textbf{HotWater}
(MHSA replaced with an identity layer) --- are evaluated in
Appendix~\ref{app:tor} to isolate the contribution of global
self-attention.

\begin{table}[]
\centering
\begin{tabular}{l|l|p{0.35\linewidth}}
\toprule
\toprule
\textbf{Hyperparameter}    & \textbf{Value} & \textbf{Description} \\
\hline
\textbf{Input Size}        & 1200  & Length of the input sequence. \\ \hline
\textbf{Feature Dim.}      & 64    & Output feature size. \\ \hline
\textbf{Hidden Dim.}       & 96    & Token feature size in the Transformer. \\ \hline
\textbf{Depth}             & 9     & Number of Transformer blocks. \\ \hline
\textbf{Embedding Conv.}   &
  \begin{tabular}[c]{@{}l@{}}\textit{kernel:} 3\\ \textit{stride:} 3\end{tabular} &
  Initial feature extraction layer. \\ \hline
\textbf{Windowing Conv.}   &
  \begin{tabular}[c]{@{}l@{}}\textit{kernel:} 50\\ \textit{stride:} 3\end{tabular} &
  Final windowing layer. \\ \hline
\textbf{Mixer Type}        &
  \begin{tabular}[c]{@{}l@{}}\textit{Type:} MHSA\\ \textit{head\_dim:} 16\\ \textit{conv\_proj:} True\end{tabular} &
  Token mixer used in each block. \\ \hline
\textbf{Mixer Conv.}       &
  \begin{tabular}[c]{@{}l@{}}\textit{kernel:} 3\\ \textit{stride:} 2\end{tabular} &
  Parameters for CvT-style MHSA. \\ \hline
\textbf{Feedforward}       &
  \begin{tabular}[c]{@{}l@{}}\textit{Style:} MLP\\ \textit{ratio:} 4\\ \textit{dropout:} 0.0\end{tabular} &
  Feedforward sublayer parameters. \\ \hline
\textbf{Block Dropout}     & 0.1   & Dropout after each Transformer block. \\ \hline
\textbf{Interval Width}    & 30\,ms & Time bin width for interval features. \\
\bottomrule
\bottomrule
\end{tabular}
\caption{Hyperparameters of the \espresso\ feature extraction network.}
\label{tab:espresso_hyperparams}
\end{table}

\subsection{Triplet Learning Framework}
\label{sec:dcf_framework}

Building on the triplet metric learning background described in
Section~\ref{sec:bg_triplet}, \espresso\ extends DCF's framework with
two improvements: online mining and a learned classification head.

\paragraph{Online Triplet Mining.}
DCF relies on \emph{offline} semi-hard mining, pre-computing embeddings
for the entire training set at the start of each epoch to identify
triplets. This is computationally expensive and scales poorly to large
datasets.
\espresso\ instead uses \emph{online} mining strategies that select
triplets dynamically within each training batch, eliminating the
pre-computation step:

\begin{itemize}
    \item \textbf{Batch-All:} All valid (hard and semi-hard) triplets
    within a batch are used; easy triplets with zero loss are excluded.
    \item \textbf{Batch-Hard:} For each anchor, only the hardest
    positive and hardest negative within the batch are selected,
    maximizing training efficiency.
\end{itemize}

Online batch-hard mining significantly accelerates training while also
improving correlation accuracy, since the model is continuously
challenged by the most difficult triplets available.

\paragraph{Classification-Based Correlation Decision}
DCF uses a threshold-based voting scheme: each window's cosine
similarity is compared against a fixed threshold, and the majority vote
determines whether the overall flow pair is correlated.
Tuning this threshold requires a separate optimization step over the
full set of candidate pairs.
\espresso\ replaces this with a small multi-layer perceptron (MLP)
that takes the vector of per-window cosine similarity scores as input
and is trained to predict whether the pair is correlated.
This learned decision boundary requires no manual threshold tuning,
naturally exploits the large number of overlapping windows that
\espresso's efficient single-pass architecture produces, and can
accommodate non-monotonic relationships between window-level scores and
overall correlation.

\subsection{Adaptation for Stepping-Stone Detection}
\label{sec:ssid_adaptation}

The \espresso\ model requires only minimal adaptation for the
stepping-stone detection task.

In the Tor setting, separate FENs are trained for ingress and egress
flows because the two differ structurally: ingress traffic is
encapsulated in fixed-size Tor cells, while egress traffic follows the
underlying application protocol.
In the stepping-stone setting there is no such asymmetry---both flows
being correlated are raw network traffic observed at different points
along the same connection chain.
A single, shared FEN is therefore sufficient to capture the critical
features needed for correlating traffic across all intermediate hosts.

Under the triplet learning setup, the traffic captured at the
\emph{attacker's host} (the connection entering the first stepping
stone) serves as the anchor, and the traffic captured at the
\emph{target host} (the connection exiting the final stepping stone)
serves as the positive example.
Negative examples are drawn from traffic flows belonging to different,
unrelated stepping-stone chains.
All other aspects of the training procedure---online batch-hard mining,
the MLP classification head, and the triplet loss with cosine
similarity---carry over unchanged.

\section{Synthetic Stepping-Stone Dataset}
\label{sec:dataset}

One key challenge in developing effective stepping-stone intrusion
detection methods is the scarcity of suitable datasets for training and
evaluation. Specifically, no publicly available datasets provide
network traffic samples with the packet-level information required to
train \espresso. Existing datasets are often incomplete, outdated, or
lack the necessary metadata to simulate stepping stone attacks
accurately. This limitation motivated the development of a synthetic
network environment capable of generating realistic stepping-stone
intrusion traffic.

To address this gap, we have developed a synthetic data collection tool
to simulate stepping stone chains and capture the associated network
traffic. This tool automates the data generation process, providing a
flexible and scalable method for creating high-quality datasets
tailored to the unique challenges of stepping-stone detection. By
incorporating various tunneling protocols and customizable network
conditions, this approach enables the generation of diverse traffic
patterns that reflect real-world attack scenarios.

\subsection{Synthetic Stepping-Stone Dataset Generation}

\begin{figure}[]
    \centering
    \includegraphics[width=0.9\linewidth]{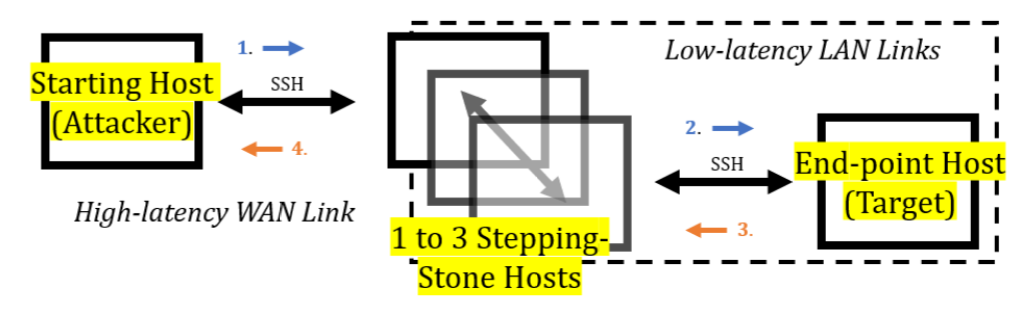}
    \caption{Diagram of synthetic stepping stone data collection
    scheme. A multi-hop tunnel is established between the attacker and
    the target through intermediate hosts (stepping stones) over a WAN
    link.}
    \label{fig:stepping-stone-collection}
\end{figure}

Our data collection tool simulates stepping-stone attacks across a
multi-hop network using tunneling protocols such as SSH, SOCAT (TCP),
DNS, and ICMP. These tunneling techniques mimic real-world scenarios
where attackers attempt to relay their traffic across intermediate
hosts (stepping stones) to obscure their presence.
Figure~\ref{fig:stepping-stone-collection} illustrates the design of
the stepping-stone simulator.
We developed this tool using Docker containers to manage and simulate
the network hosts through which the stepping-stone intrusion chain is
constructed.

Our collection tool supports the use of multiple protocols and
tunneling methods within a single stepping stone chain, offering
flexibility in simulating various forms of traffic obfuscation and
evasion techniques. This ensures that the dataset covers a range of
different intrusion techniques. A unique aspect of our approach is the
ability to randomize the tunneling tool used at each hop in the chain,
simulating more complex multi-protocol tunneling chains that could be
used by sophisticated attackers and pose a large challenge for flow
correlation methods.

\subsection{Simulating Stepping Stone Chains}

The simulation begins by establishing an artificial stepping-stone
chain of multiple hosts. In this chain, the attacker initiates a
series of connections from a starting host (the attacker) to an
end-point host (the target) through one or more intermediate hosts
(stepping stones). These hosts are represented as Docker containers,
allowing for easy scalability and deployment in experimental settings.
A variable number of stepping-stone hosts can be used in each chain to
create different levels of complexity; in our experiments, we use a
range from 1 to 3 intermediate hosts.

A network delay is simulated between hosts using the
\texttt{netem}~\cite{netem} tool, which introduces artificial network
latencies between the first and second hosts to mimic the
high-latency wide-area network (WAN) links typically seen between
remote nodes. The target and intermediate hosts, on the other hand,
are simulated as being on a local-area network (LAN) with lower
latency. The delay emulation ensures that the simulated network
conditions resemble realistic stepping-stone intrusion scenarios.

\subsection{Emulating Realistic Traffic Properties}

The traffic collection tool we developed uses a generic transmission
mechanism designed around \textit{bursts} of data. The tool sends
bursts of data into the tunnel consisting of two phases: (i) the
attacker sends \textit{N} bytes of data upstream (i.e., toward the
target), and (ii) the target responds by sending \textit{M} bytes
downstream. This is accomplished using an \textsc{echo} Linux command.
After completing a burst, the attacker waits for \textit{Z} seconds
before initiating the next burst, repeating this process for
\textit{J} bursts in total. In an effort to generate network traffic
that reflects real-world attacker behavior, we have constructed
distributions based on real-world traffic data to parameterize our
traffic-sending mechanism.

\paragraph{Burst Definition and Threshold Selection}
To create the distributions needed for our collector, we first parse
real-world traffic samples into bursts that approximate the
functionality of our tool. For this purpose, we define a burst as a
contiguous sequence of packets transmitted over a connection, separated
from adjacent sequences by an extended period of inactivity. To
identify suitable burst boundaries, we analyzed a publicly available
DEFCON 25 capture-the-flag dataset~\cite{defcon25}, focusing on the
timing characteristics of SSH traffic. Based on the global
inter-packet delay distribution observed within these flows, we
selected the 97th percentile of the inter-packet delays (approximately
6.5\,ms) as the threshold for distinguishing distinct bursts of
activity. This choice captures the majority of meaningful bursts while
minimizing the classification of sporadic or isolated packets as
separate bursts, and is reasonably modellable by our transmission
mechanism.

With this definition, we parse the DEFCON SSH flows into bursts and
extract four key statistics to build global distributions of traffic
properties:
\begin{enumerate}
    \item The distribution of the total number of bursts in a given
    traffic sample;
    \item The distribution of inter-burst time intervals;
    \item The distribution of upstream bytes per burst; and
    \item The distribution of downstream bytes per burst.
\end{enumerate}
These empirical distributions then govern the timing and volume of
data transmissions, allowing the simulation to emulate bursty traffic
patterns characteristic of interactive sessions. Specifically, the
simulation samples the total number of bursts (\textit{J}), the amount
of data sent upstream and downstream for each burst (\textit{N} and
\textit{M}), and the inter-burst interval (\textit{Z}) from these
distributions.

To generate the stepping stone traffic, the collection tool samples
bursts of traffic from the aforementioned distributions and relays them
across the stepping stone chain using a variety of tunneling protocols.
SSH tunnels are created using the \texttt{ssh} command, raw TCP tunnels
using \texttt{socat}~\cite{socat}, and covert tunnels using
\texttt{ptunnel-ng}~\cite{ptunnel-ng} (for ICMP) or
\texttt{dnscat2}~\cite{dnscat2} (for DNS). The variety of tunneling
protocols allows the dataset to capture a wide range of attack
techniques, from more overt forms of tunneling (SSH, TCP) to covert
methods designed to evade detection (ICMP, DNS). These are also
standard and openly available tools that real-world adversaries are
likely to use.

\subsection{Dataset Collection and Structure}

Using this data collection tool, we generated five distinct datasets,
each containing at least 10,000 artificial stepping stone chains.
The protocols used for each dataset are as follows:

\begin{enumerate}
    \item \textbf{SSH-only:} All stepping stone traffic is relayed via
    SSH tunnels.
    \item \textbf{SOCAT-only:} Raw TCP sockets are used for tunneling.
    \item \textbf{DNS covert tunnels:} All traffic is relayed through
    covert DNS tunnels, simulating techniques designed to evade
    traditional intrusion detection systems.
    \item \textbf{ICMP covert tunnels:} ICMP is used for covert
    traffic relaying, further increasing the difficulty of detection.
    \item \textbf{Mixed protocol:} Each hop in the stepping stone chain
    uses a different randomly selected tunneling protocol from the
    above options, simulating a more complex attack chain.
\end{enumerate}

Each dataset contains network traffic captures for stepping stone
chains with different numbers of intermediate hosts and varying
latencies between hosts. This variety ensures that the dataset can be
used to evaluate the effectiveness of stepping-stone detection methods
under different conditions. The inclusion of covert protocols (DNS and
ICMP) further challenges detection methods by introducing tunnels that
are traditionally harder to detect.

\section{Stepping-Stone Detection Experiments}
\label{sec:ssid}

\subsection{Experimental Setup}

\begin{figure}[h]
    \centering
    \includegraphics[width=0.75\linewidth]{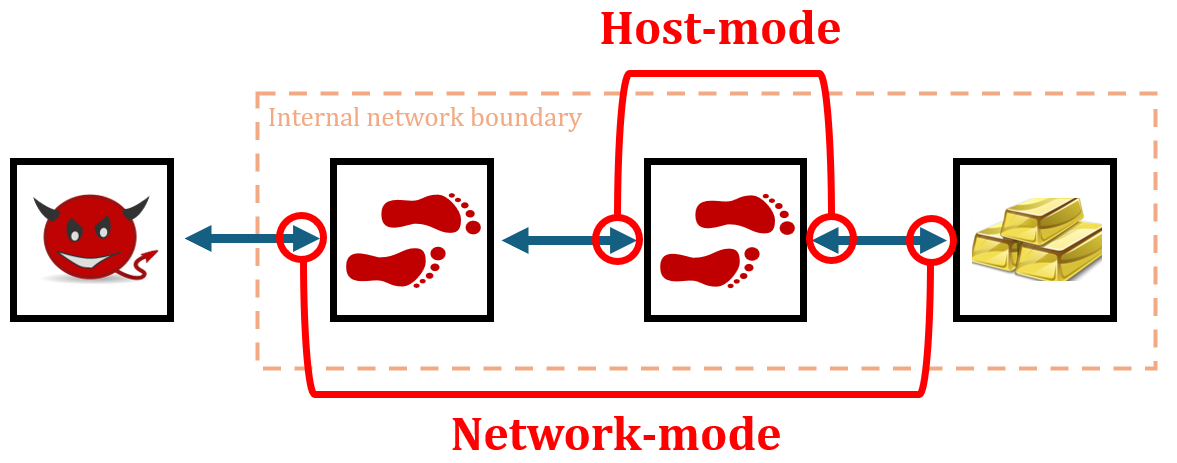}
    \caption{Illustration of the network traffic correlation scenarios
    considered in our evaluations.}
    \label{fig:stepping-stone-diagram}
\end{figure}

Our experimental evaluation uses five distinct datasets, each
containing stepping-stone intrusion traffic generated through tunneling
protocols: SSH-only, SOCAT-only, ICMP-only, and DNS-only. The datasets
include 9,000 stepping-stone chains for training and 1,000 chains for
testing. Each chain represents a sequence of relayed traffic through
one or more intermediary hosts (stepping stones) between an attacker
and a target, utilizing one of the selected tunneling protocols.
Correlated pairs are traffic flows that are present on the same chain.
Uncorrelated pairs are traffic flows from different chains.

We evaluate the detection performance of our models in two modes,
depicted in Figure~\ref{fig:stepping-stone-diagram}:
\begin{itemize}
    \item \textbf{Host-mode detection:} The model is trained and tested
    to correlate traffic flows on the same host. In this mode,
    correlated flow pairs must exist within the same machine. In real
    environments, this mode of detection can work alongside traditional
    host and system logging tools, but it is also uniquely capable of
    defending devices without robust on-device logging capabilities
    (such as IoT devices) where external traffic sensors must be relied
    on.
    \item \textbf{Network-mode detection:} The model is trained and
    tested to correlate traffic that leaves a host with incoming
    traffic into the network, enabling the detection of stepping-stone
    chains that span multiple hosts in a network. In real environments,
    this strategy will be most valuable when applied to detect when
    traffic from specific sensitive devices or regions of the network
    are exiting the organization's network at the boundary.
    Network-mode detection is considerably more difficult but has
    considerable utility in practice, so most of our evaluation focuses
    on this setting.
\end{itemize}

The evaluation metrics used in our experiments are consistent with
those applied in the Tor correlation evaluation in
Appendix~\ref{app:tor}. For flow correlation performance, we measure
the maximum True Positive Rate (TPR), partial Area Under the Receiver
Operating Characteristic curve (pAUC), and present full ROC curves for
comparison. We compare against the performance of the
DeepCoFFEA~\cite{oh2022deepcoffea} (DCF) correlation method as a
baseline. Additionally, we performed a limited exploration of window
size and model structure; details are reported in
Appendix~\ref{app:dcf_ablation},
Section~\ref{sec:appendix_dcf_windows}. From this exploration, we
include a tertiary baseline in
Table~\ref{tab:net-mode-max-tpr-pauc-datasets-corr} that we call
``Modified DCF,'' which represents a variant of the \espresso\
framework that uses a DCF network architecture adapted to ingest
complete traffic sequences (in \espresso's time-interval
representation) and return features with a window dimension.

\subsection{Baseline Stepping Stone Correlation Results}

\begin{figure}[h!]
    \centering
    \begin{subfigure}[b]{0.48\textwidth}
        \centering
        \includegraphics[width=\linewidth]{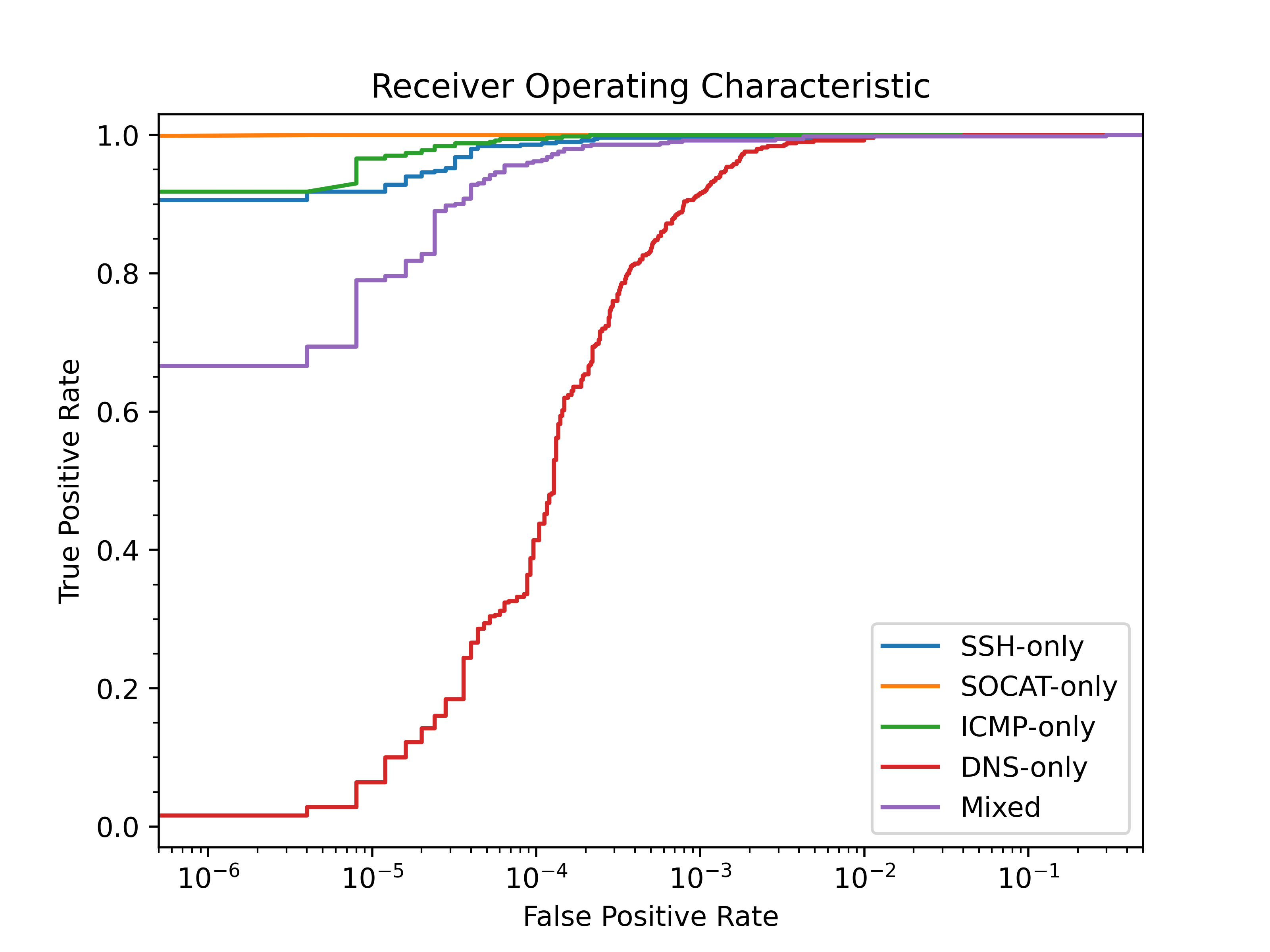}
        \caption{\espresso\ \emph{host-mode} ROC}
        \label{fig:ssid-host-roc-espresso}
    \end{subfigure}
    \hfill
    \begin{subfigure}[b]{0.48\textwidth}
        \centering
        \includegraphics[width=\linewidth]{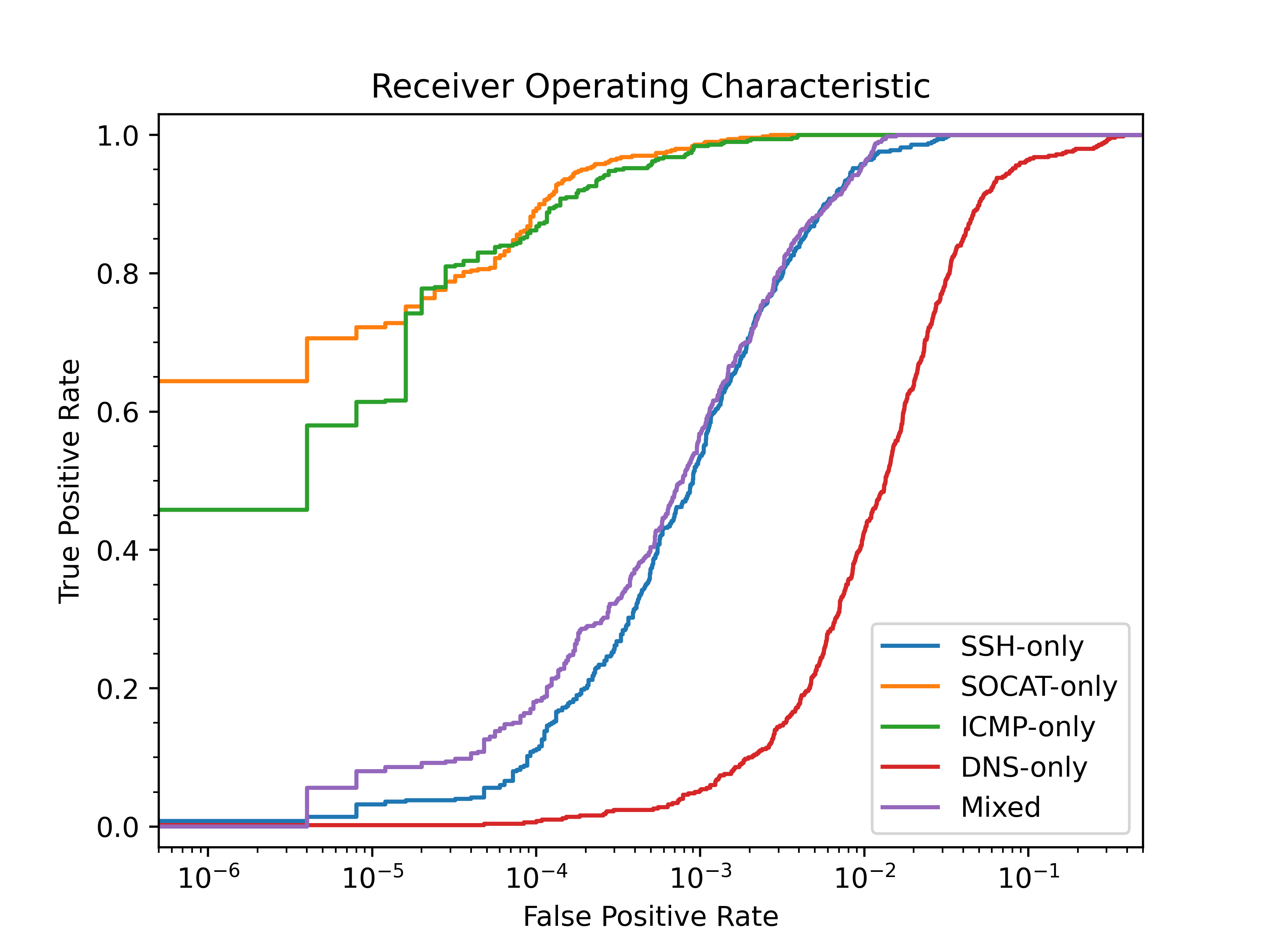}
        \caption{DCF \emph{host-mode} ROC}
        \label{fig:ssid-host-roc-dcf}
    \end{subfigure}
    \caption{ROC curves presenting correlation efficacy of (a)
    \espresso\ and (b) DCF on datasets in the \textit{host-mode}
    detection scenario.}
    \label{fig:ssid-host-roc-combined}
\end{figure}

\begin{figure}[h!]
    \centering
    \begin{subfigure}[b]{0.48\textwidth}
        \centering
        \includegraphics[width=\linewidth]{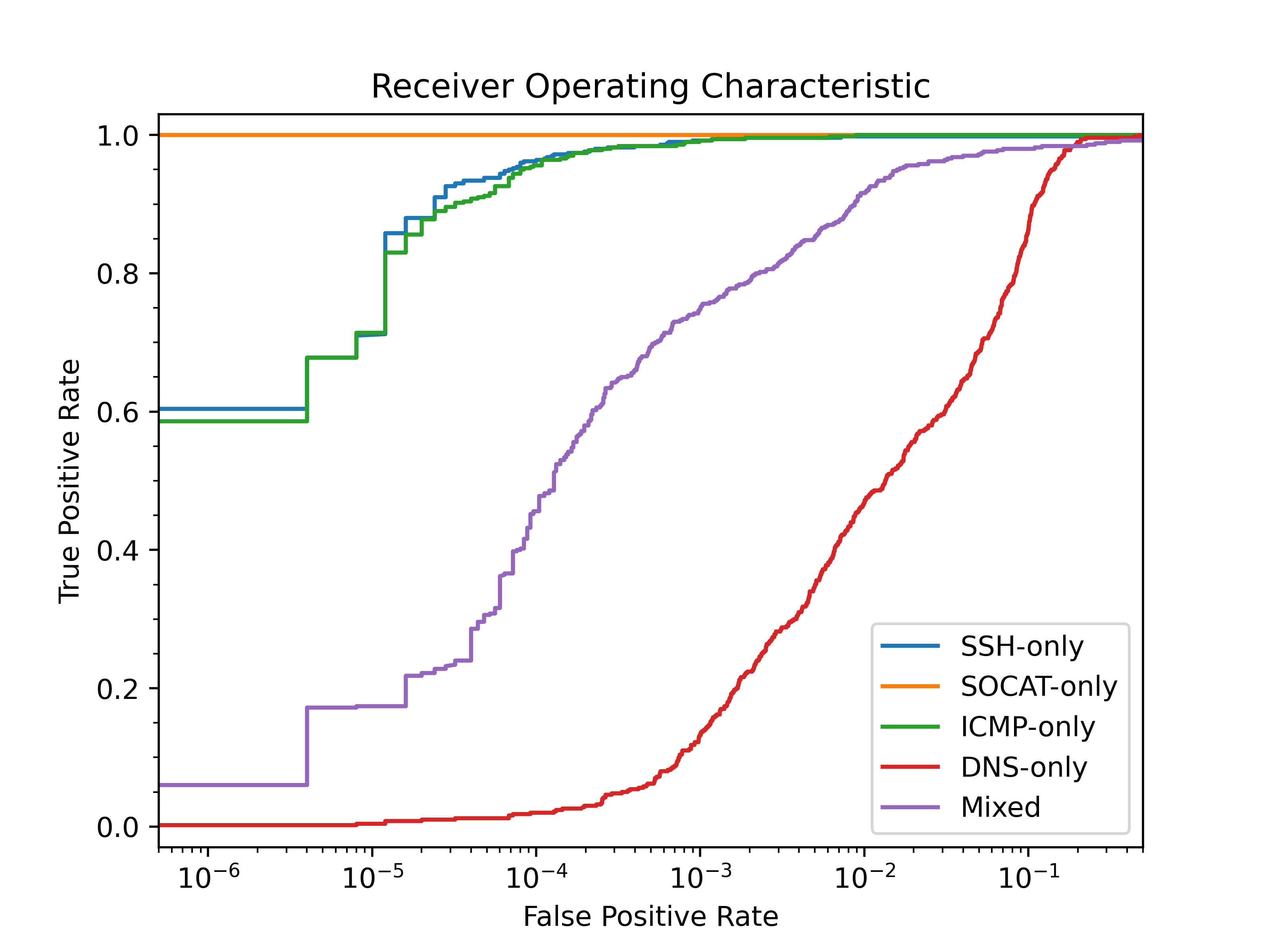}
        \caption{\espresso\ \emph{network-mode} ROC}
        \label{fig:ssid-net-roc-espresso}
    \end{subfigure}
    \hfill
    \begin{subfigure}[b]{0.48\textwidth}
        \centering
        \includegraphics[width=\linewidth]{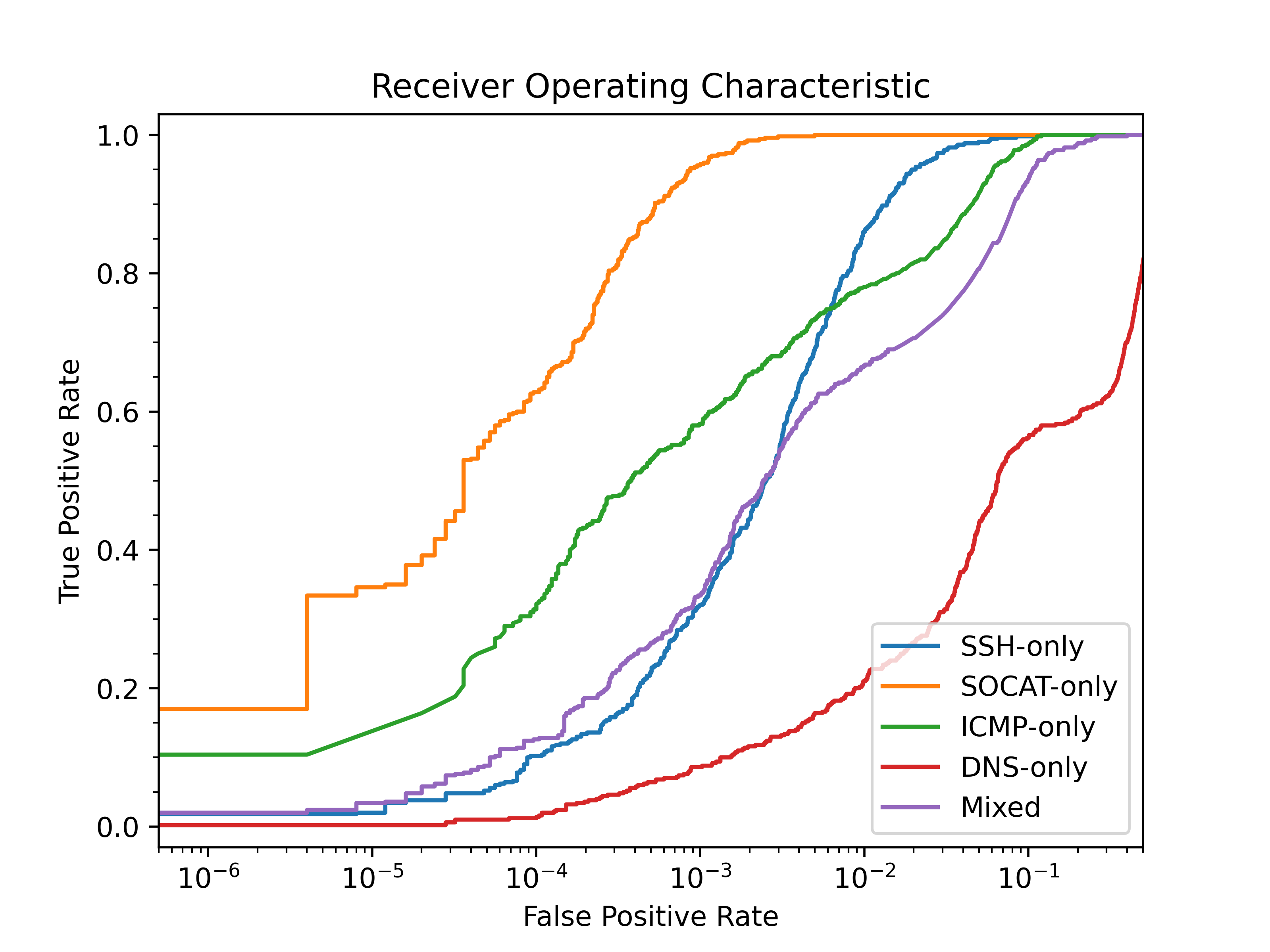}
        \caption{DCF \emph{network-mode} ROC}
        \label{fig:ssid-net-roc-dcf}
    \end{subfigure}
    \caption{ROC curves presenting correlation efficacy of (a)
    \espresso\ and (b) DCF on datasets in the \textit{network-mode}
    detection scenario.}
    \label{fig:ssid-net-roc-combined}
\end{figure}

The results from our completed experiments benchmarking \espresso\
against the DCF baseline are presented in this section. The overall
correlation performance for both models is visualized as ROC curves in
Figure~\ref{fig:ssid-host-roc-combined} for the \textit{host-mode}
scenario and Figure~\ref{fig:ssid-net-roc-combined} for the
\textit{network-mode} scenario. A detailed breakdown of performance
metrics, including maximum TPR and pAUC at three low FPR thresholds,
is provided in Table~\ref{tab:max-tpr-pauc-datasets-corr} for
host-mode and Table~\ref{tab:net-mode-max-tpr-pauc-datasets-corr} for
network-mode.

\begin{table*}[]
    \centering
    \begin{tabular}{c|c|cc|cc|cc}
    \toprule
    \toprule
    \multirow{2}{*}{\textbf{Dataset}} &
    \multirow{2}{*}{\textbf{Model}} &
    \multicolumn{2}{c|}{\textbf{FPR} $\leq 10^{-3}$} &
    \multicolumn{2}{c|}{\textbf{FPR} $\leq 10^{-4}$} &
    \multicolumn{2}{c}{\textbf{FPR} $\leq 10^{-5}$} \\
    & & \textbf{TPR} & \textbf{pAUC} & \textbf{TPR} & \textbf{pAUC} &
    \textbf{TPR} & \textbf{pAUC} \\
    \midrule
    \multirow{2}{*}{SSH-only} & DCF & 0.536 & 0.339 & 0.110 & 0.057 & 0.032 & 0.015 \\
    & ESPRESSO & \textbf{0.996} & \textbf{0.990} & \textbf{0.986} & \textbf{0.986} & \textbf{0.918} & \textbf{0.790} \\
    \midrule
    \multirow{2}{*}{SOCAT-only} & DCF & 0.986 & 0.952 & 0.890 & 0.803 & 0.722 & 0.685 \\
    & ESPRESSO & \textbf{1.000} & \textbf{0.995} & \textbf{1.000} & \textbf{0.959} & \textbf{1.000} & \textbf{0.599} \\
    \midrule
    \multirow{2}{*}{ICMP-only} & DCF & 0.984 & 0.938 & 0.862 & 0.776 & 0.614 & 0.446 \\
    & ESPRESSO & \textbf{1.00} & \textbf{0.988} & \textbf{0.998} & \textbf{0.890} & \textbf{0.966} & \textbf{0.909} \\
    \midrule
    \multirow{2}{*}{DNS-only} & DCF & 0.052 & 0.027 & 0.006 & 0.003 & 0.002 & 0.002 \\
    & ESPRESSO & \textbf{0.916} & \textbf{0.757} & \textbf{0.414} & \textbf{0.245} & \textbf{0.064} & \textbf{0.034} \\
    \midrule
    \multirow{2}{*}{Mixed Protocols} & DCF & 0.568 & 0.384 & 0.180 & 0.118 & 0.080 & 0.038 \\
    & ESPRESSO & \textbf{0.992} & \textbf{0.977} & \textbf{0.962} & \textbf{0.883} & \textbf{0.790} & \textbf{0.576} \\
    \bottomrule
    \bottomrule
    \end{tabular}
    \caption{\textbf{Host-mode:} Maximum TPR and pAUC for FPR
    thresholds of $10^{-3}$, $10^{-4}$, and $10^{-5}$ across five
    stepping stone datasets.}
    \label{tab:max-tpr-pauc-datasets-corr}
\end{table*}

\begin{table*}[]
    \centering
    \begin{tabular}{c|c|cc|cc|cc}
    \toprule
    \toprule
    \multirow{2}{*}{\textbf{Dataset}} &
    \multirow{2}{*}{\textbf{Model}} &
    \multicolumn{2}{c|}{\textbf{FPR} $\leq 10^{-3}$} &
    \multicolumn{2}{c|}{\textbf{FPR} $\leq 10^{-4}$} &
    \multicolumn{2}{c}{\textbf{FPR} $\leq 10^{-5}$} \\
    & & \textbf{TPR} & \textbf{pAUC} & \textbf{TPR} & \textbf{pAUC} &
    \textbf{TPR} & \textbf{pAUC} \\
    \midrule
    \multirow{3}{*}{SSH-only}
    & DCF & 0.320 & 0.210 & 0.102 & 0.056 & 0.02 & 0.012 \\
    & ESPRESSO & \textbf{0.992} & \textbf{0.975} & \textbf{0.962} & \textbf{0.8905} & \textbf{0.710} & \textbf{0.535} \\
    & Modified DCF & 0.922 & 0.855 & 0.848 & 0.699 & 0.484 & 0.392 \\
    \midrule
    \multirow{3}{*}{SOCAT-only}
    & DCF & 0.956 & 0.823 & 0.628 & 0.507 & 0.346 & 0.285 \\
    & ESPRESSO & \textbf{1.00} & \textbf{0.989} & \textbf{1.00} & \textbf{0.989} & \textbf{1.00} & \textbf{0.989} \\
    & Modified DCF & \textbf{1.00} & \textbf{0.997} & \textbf{0.996} & \textbf{0.978} & 0.976 & 0.880 \\
    \midrule
    \multirow{3}{*}{ICMP-only}
    & DCF & 0.582 & 0.487 & 0.314 & 0.232 & 0.104 & 0.090 \\
    & ESPRESSO & \textbf{0.992} & \textbf{0.972} & \textbf{0.956} & \textbf{0.877} & \textbf{0.714} & \textbf{0.559} \\
    & Modified DCF & \textbf{1.00} & \textbf{0.985} & \textbf{0.984} & \textbf{0.874} & 0.60 & 0.445 \\
    \midrule
    \multirow{3}{*}{DNS-only}
    & DCF & 0.086 & 0.056 & 0.012 & 0.008 & 0.002 & 0.002 \\
    & ESPRESSO & \textbf{0.132} & \textbf{0.067} & 0.020 & 0.012 & 0.004 & 0.002 \\
    & Modified DCF & 0.110 & \textbf{0.066} & 0.026 & 0.013 & 0.002 & 0.000 \\
    \midrule
    \multirow{3}{*}{Mixed Protocols}
    & DCF & 0.334 & 0.244 & 0.126 & 0.086 & 0.034 & 0.024 \\
    & ESPRESSO & \textbf{0.748} & \textbf{0.640} & 0.456 & 0.302 & \textbf{0.174} & \textbf{0.133} \\
    & Modified DCF & 0.404 & 0.310 & \textbf{0.492} & \textbf{0.431} & \textbf{0.172} & \textbf{0.130} \\
    \bottomrule
    \bottomrule
    \end{tabular}
    \caption{\textbf{Network-mode:} Maximum TPR and pAUC for FPR
    thresholds of $10^{-3}$, $10^{-4}$, and $10^{-5}$ across five
    stepping stone datasets.}
    \label{tab:net-mode-max-tpr-pauc-datasets-corr}
\end{table*}

\paragraphX{Analysis.}
The primary finding from these experiments is that \espresso\
substantially outperforms the DCF baseline across all evaluated
datasets and in both detection modes. The performance gap is evident
in the ROC curves (Figures~\ref{fig:ssid-host-roc-combined}
and~\ref{fig:ssid-net-roc-combined}) and is quantified in the metrics
tables. For instance, in the host-mode scenario against the challenging
mixed-protocol dataset (Table~\ref{tab:max-tpr-pauc-datasets-corr}),
\espresso\ achieves a TPR of 0.992 at an FPR of $10^{-3}$, whereas DCF
reaches only 0.568. This margin of improvement underscores the superior
feature representation learned by \espresso.

As expected, correlation performance is generally higher in the
host-mode setting compared to the network-mode setting for both models.
However, \espresso\ demonstrates greater resilience to the challenges of
network-mode detection. For example, while performance on the
mixed-protocol dataset drops in network-mode, \espresso\ still maintains
a TPR of 0.748 (at FPR $\leq 10^{-3}$), more than double DCF's TPR of
0.334 (Table~\ref{tab:net-mode-max-tpr-pauc-datasets-corr}).
Interestingly, for SOCAT-tunneled traffic, \espresso\ shows no
performance degradation between modes, achieving a perfect TPR of 1.00
in both scenarios. This suggests that the traffic patterns within
SOCAT connections are exceptionally stable between hops, making them
equally correlatable from either a host or network vantage point.

When examining the impact of the tunneling protocol, DNS-based tunnels
prove to be the most difficult to correlate for both models,
particularly in network-mode. We attribute this to the polling-based
communication mechanism of the dnscat2 tool~\cite{dnscat2}, which
disrupts typical traffic timing patterns. Despite this challenge,
\espresso's performance in host-mode against the DNS dataset is
remarkably strong, achieving a TPR of 0.916 (FPR $\leq 10^{-3}$),
while DCF's performance is near zero. Among the other protocols,
performance is consistently high. In host-mode, \espresso\ achieves a
perfect TPR on both SOCAT and ICMP traffic, though the higher pAUC for
ICMP at an FPR of $10^{-5}$ (0.909 vs.\ 0.599) suggests that ICMP
correlations are slightly more robust at extremely low FPRs. In
network-mode, SOCAT connections are the most reliably correlated by
\espresso, further highlighting the stability of that tunneling
protocol.

To isolate the factors driving \espresso's performance gains, we also
evaluated a ``Modified DCF'' variant which adapts the CNN-based DCF
backbone to process full-traffic sequences. While the per-packet
feature variant suffered from misalignment issues, the
\emph{time-interval} feature variant achieved performance competitive
with \espresso, particularly at low FPRs, as highlighted in
Table~\ref{tab:net-mode-max-tpr-pauc-datasets-corr}. This result
suggests that the primary driver of correlation efficacy is the
combination of full-sequence context and time-aligned feature
representation, rather than the specific choice of the Transformer
architecture over a CNN. Notably, this modified architecture matched
\espresso's performance across most protocols, with the exception of
DNS-tunneled traffic, which appears to benefit uniquely from the
original DCF's windowed, packet-level representation (see
Appendix~\ref{app:dcf_ablation}).

In the remainder of this section, we explore several different
strategies to further improve the performance and functionality of
\espresso\ on the SSID correlation task. \emph{Given the increased
challenge presented by the network-mode detection, further methods will
be evaluated in this setting solely.}

\subsection{Exploring Loss Enhancements for Improved Feature Representation}

\paragraphX{Motivation and Rationale.}
The \espresso\ method has so far demonstrated robust performance for
stepping-stone intrusion detection. However, two key limitations of the
current approach hinder its ability to achieve optimal performance:

\begin{enumerate}
    \item The standard triplet loss function does not enforce temporal
    alignment of features, which may lead to suboptimal correlations
    across traffic windows, especially in noisy or variable network
    conditions.
    \item The amplification strategy relies on diverse information
    being present between different windows to enhance the overall
    correlation performance. However, under the current loss scheme, no
    mechanism encourages information represented within different time
    windows to be uncorrelated.
\end{enumerate}

Addressing these limitations aims to enhance \espresso's ability to
generalize across diverse network conditions, improve robustness to
false positives, and effectively capture both local and global traffic
correlations.

\subsubsection{Amplification with Temporal Alignment}

\paragraphX{Motivation.}
The \espresso\ framework independently computes window-based
similarities without explicitly aligning temporal patterns across
samples. This lack of alignment may fail to capture the sequential
nature of network traffic, reducing the robustness of similarity
calculations.

\begin{figure}[h]
    \centering
    \includegraphics[width=0.55\linewidth]{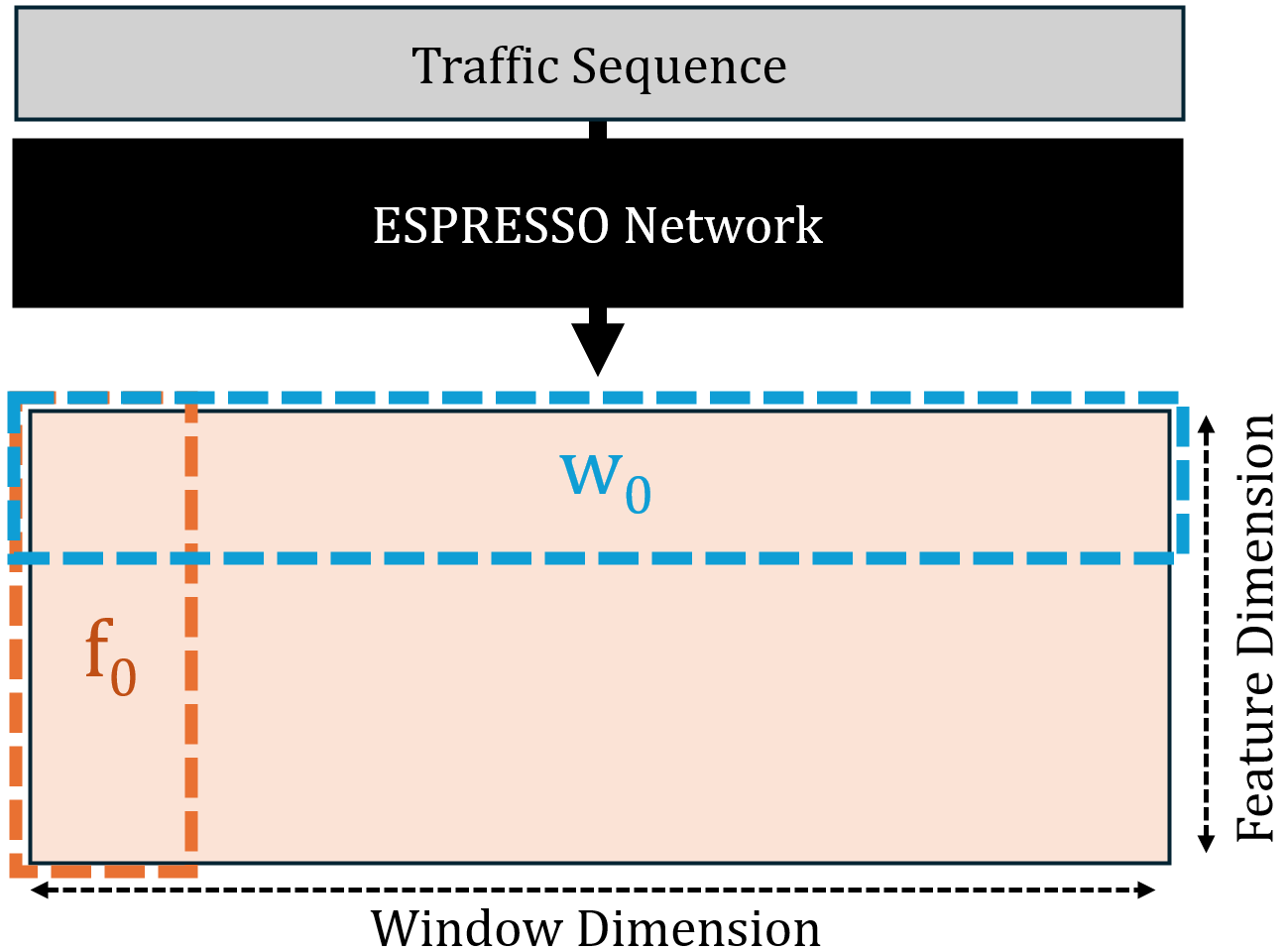}
    \caption{Diagram that visually represents the feature vectors used
    for temporal feature similarity.}
    \label{fig:temporal_amplification}
\end{figure}

\paragraphX{Our Approach.}
Traditional DCF computed cosine similarity between traffic samples
along the feature dimension, focusing on how individual features
extracted from packet windows compare. While this approach effectively
captured local feature similarities, it limited the model's ability to
generalize to traffic variations and overlooked possible temporal
patterns across windows.

Recent experimental results showed an unexpected performance
improvement when similarity was calculated along the \textbf{window}
dimension rather than the \textbf{feature} dimension. This approach,
which we refer to as \textit{temporal similarity}, captures temporal
structure in the evolution of traffic flows over time. Based on these
insights, we propose a \textbf{hybrid similarity strategy}, which
incorporates both feature-based and window-based similarity to better
capture correlations between traffic samples.

The core computation in DCF involves calculating the cosine similarity
for each window:
\[
\text{Sim}_{\text{window}}(i, j) = \frac{\mathbf{f}_i \cdot
\mathbf{f}_j}{\|\mathbf{f}_i\| \|\mathbf{f}_j\|}
\]
where $\mathbf{f}_i$ and $\mathbf{f}_j$ are the feature vectors for
traffic windows $i$ and $j$. This method measures feature similarity
for individual windows within a singular slice of time, with no
consideration of the temporal behavior of the features.

An alternative and complementary strategy is to change the dimension
across which feature similarity is calculated. We illustrate this
visually in Figure~\ref{fig:temporal_amplification}. By computing the
cosine similarity across the \textbf{window} dimension, we capture
correlations based on how the traffic evolves over time:
\[
\text{Sim}_{\text{temporal}}(i, j) = \frac{\mathbf{w}_i \cdot
\mathbf{w}_j}{\|\mathbf{w}_i\| \|\mathbf{w}_j\|}
\]
where $\mathbf{w}_i$ and $\mathbf{w}_j$ represent the traffic
sequences across windows for samples $i$ and $j$, respectively. This
method measures the similarity of a singular feature across all windows
of time between samples. This simple adjustment provides an additional
mechanism to encourage the model to produce features that are
explicitly aligned both within a window and temporally across all
windows during training.

To incorporate both dimensions into a hybrid learning strategy, we
propose a new loss function:
\[
\mathcal{L}_{\text{triplet}} = \alpha \cdot \mathcal{L}_{\text{window}}
+ (1 - \alpha) \cdot \mathcal{L}_{\text{temporal}},
\]
where $\alpha$ is a hyperparameter that controls the balance between
feature-based and window-based
similarity.\footnote{To simplify hyperparameter selection, we use
$\alpha = 0.5$, effectively taking the average of the two losses.}

This \textit{Amplification with Temporal Alignment} ensures that
correlations are identified within the intrinsic feature space of each
window (as done in DCF) and across the temporal progression of traffic
patterns. By combining both similarities, \espresso\ may produce richer
embeddings that reflect local and global traffic correlations, making
it more robust to noise and traffic variability.

\subsubsection{Feature Diversity and Decorrelation}

\paragraphX{Motivation.}
When features between windows are highly correlated, the model's
ability to amplify meaningful signals diminishes, as overlapping
information replaces novel perspectives. This lack of diversity weakens
discrimination between correlated and non-correlated samples, reducing
the robustness of the amplification process. In contrast, embedding
spaces with independent, decorrelated features enable the model to
leverage diverse and complementary information, enhancing the system's
ability to identify correlations across dimensions.

\paragraphX{Our Approach.}
Correlated feature dimensions within embeddings reduce the diversity of
learned representations, introducing redundancy and inefficiency in
similarity calculations. To mitigate this, we augment the triplet loss
function to encourage feature decorrelation:

\begin{itemize}
    \item \textbf{Orthogonality.} This regularization term imposes a
    direct geometric constraint on the embedding space, forcing the
    feature dimensions to be perpendicular and thus minimizing their
    correlation:
    \[
    \mathcal{L}_{\text{orth}} = \| \mathbf{F}^\top \mathbf{F} -
    \mathbf{I} \|_F^2,
    \]
    where $\mathbf{F} \in \mathbb{R}^{d \times n}$ is the embedding
    matrix, $\mathbf{I}$ is the identity matrix, and $\|\cdot\|_F^2$
    is the Frobenius norm. By penalizing the deviation of the Gram
    matrix from the identity, this loss encourages a basis of feature
    vectors that are structurally independent. This strategy is
    inspired by the Orthogonal Projection Loss (OPL)~\cite{ranasinghe2021orthogonal}.

    \item \textbf{Covariance.} This loss component promotes a broader
    and more expressive coverage of the embedding space by maximizing
    the determinant of the feature covariance matrix:
    \[
    \mathcal{L}_{\text{cov}} = -\log \det (\text{Cov}(\mathbf{F})),
    \]
    where $\text{Cov}(\mathbf{F})$ is the covariance matrix of the
    feature embeddings. Maximizing the determinant incentivizes high
    variance for each individual feature and low covariance between
    features, ensuring that the embeddings are spread out while
    reducing redundancy.
\end{itemize}

By incorporating these regularization terms, the goal is to guide the
model toward learning a more efficient and expressive feature
representation. When primary correlation signals are weak or
obfuscated, a diverse set of features offers more independent sources
of information, increasing the chance that a stable, meaningful signal
will be amplified.

\subsubsection{Experimental Results}

\paragraph{Temporal Alignment.}
The results of incorporating the temporal alignment loss are presented
in Table~\ref{tab:baseline_results}. Overall, this loss enhancement
provides a tangible, albeit modest, improvement to correlation
performance, particularly for the SSH and Mixed-protocol datasets. For
SSH traffic, the model with temporal loss achieved a max TPR of 0.946
at an FPR of $10^{-5}$, a significant increase from the baseline's
0.710. Similarly, for the mixed-protocol data, the max TPR improved
from 0.174 to 0.413. The improvement is less pronounced for SOCAT and
ICMP traffic, which already exhibited near-perfect performance with the
baseline model. Unfortunately, the temporal alignment loss offered no
benefit for DNS traffic, where correlation remains a significant
challenge. This indicates that while temporal alignment is beneficial
for protocols with more consistent traffic patterns, it is not
sufficient to overcome the fundamental timing disruptions inherent in
the polling mechanism of DNS-based tunnels.

\begin{table*}[t]
\centering
\begin{tabular}{llccccccccc}
\toprule
\toprule
\multicolumn{2}{c}{\textbf{Model}} & \textbf{AUC} & \multicolumn{2}{c}{\textbf{FPR $\leq 10^{-3}$}} & \multicolumn{2}{c}{\textbf{FPR $\leq 10^{-4}$}} & \multicolumn{2}{c}{\textbf{FPR $\leq 10^{-5}$}} \\
\cmidrule(lr){4-5} \cmidrule(lr){6-7} \cmidrule(lr){8-9}
& & & \textbf{pAUC} & \textbf{max TPR} & \textbf{pAUC} & \textbf{max TPR} & \textbf{pAUC} & \textbf{max TPR} \\
\midrule
\multicolumn{10}{l}{\textbf{SSH Traffic}} \\
& ESPRESSO & 0.998 & 0.975 & 0.992 & 0.891 & 0.962 & 0.535 & 0.710 \\
& + Temporal Loss & 0.998 & \textbf{0.984} & 0.992 & \textbf{0.955} & \textbf{0.972} & \textbf{0.861} & \textbf{0.946} \\
\midrule
\multicolumn{10}{l}{\textbf{SOCAT Traffic}} \\
& ESPRESSO & 1.000 & 0.989 & 1.000 & 0.989 & 1.000 & 0.989 & 1.000 \\
& + Temporal Loss & 1.000 & 0.992 & 1.000 & 0.992 & 1.000 & 0.992 & 1.000 \\
\midrule
\multicolumn{10}{l}{\textbf{ICMP Traffic}} \\
& ESPRESSO & 0.999 & 0.972 & 0.992 & 0.877 & 0.959 & 0.558 & 0.714 \\
& + Temporal Loss & 0.999 & 0.971 & 0.991 & 0.875 & 0.958 & \textbf{0.573} & \textbf{0.731} \\
\midrule
\multicolumn{10}{l}{\textbf{DNS Traffic}} \\
& ESPRESSO & 0.959 & 0.067 & \textbf{0.132} & 0.012 & 0.020 & 0.002 & 0.004 \\
& + Temporal Loss & 0.953 & 0.060 & 0.121 & 0.011 & 0.019 & 0.002 & 0.004 \\
\midrule
\multicolumn{10}{l}{\textbf{Mixed Traffic}} \\
& ESPRESSO & 0.988 & 0.640 & 0.748 & 0.302 & 0.456 & 0.133 & 0.174 \\
& + Temporal Loss & 0.991 & \textbf{0.781} & \textbf{0.852} & \textbf{0.553} & \textbf{0.689} & \textbf{0.291} & \textbf{0.413} \\
\bottomrule
\bottomrule
\end{tabular}%
\caption{Benchmark performance of baseline models across various traffic protocols. We compare our standard ESPRESSO model and its variant with a temporal alignment loss. Performance is evaluated using overall AUC, as well as partial AUC (pAUC) and max true positive rate (TPR) at low false positive rate (FPR) thresholds.}
\label{tab:baseline_results}
\end{table*}

\begin{figure}[h!]
    \centering
    \begin{subfigure}[b]{0.48\textwidth}
        \centering
        \includegraphics[width=\linewidth]{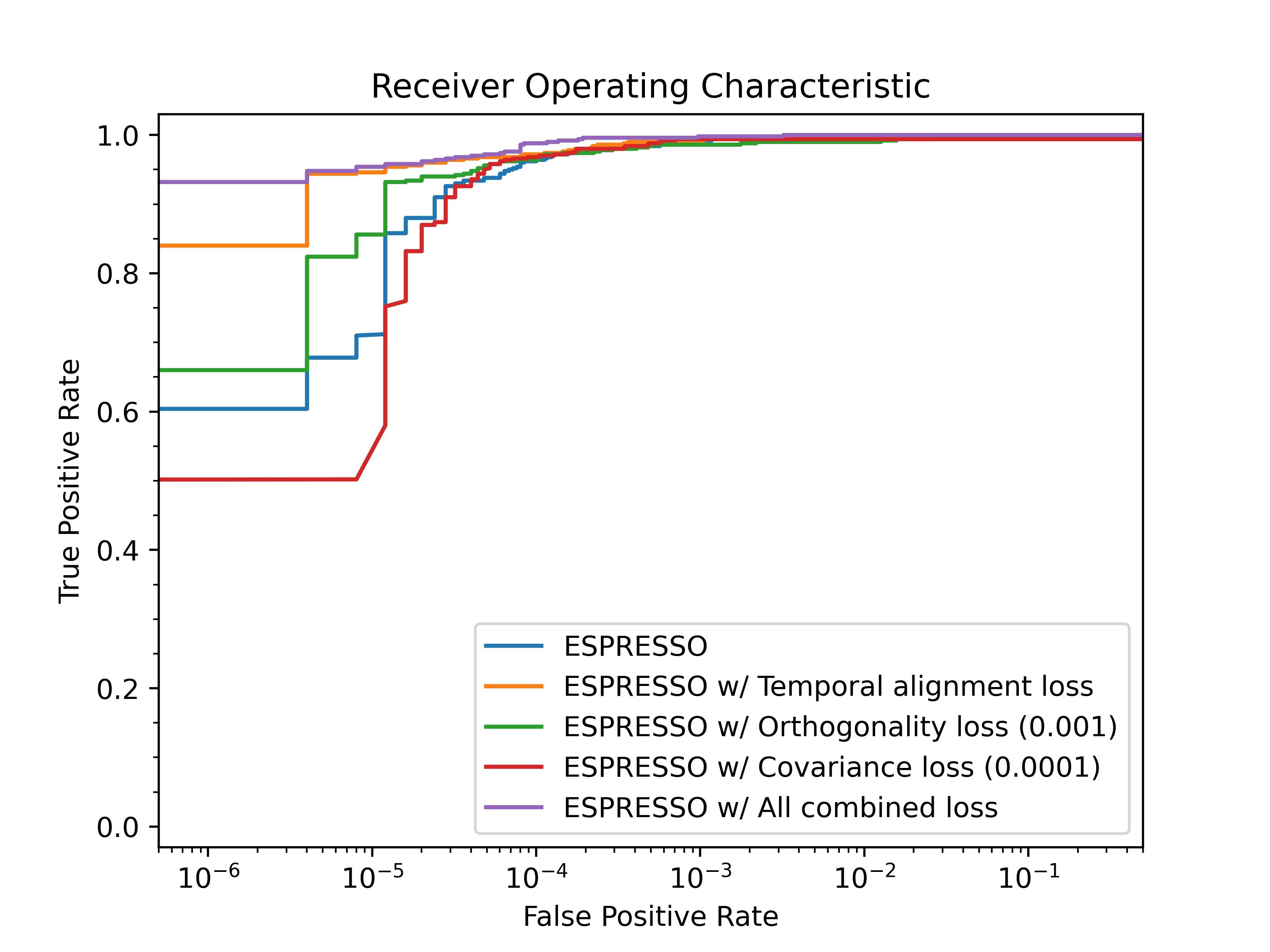}
        \caption{SSH-only ROC}
        \label{fig:roc-ssh}
    \end{subfigure}
    \hfill
    \begin{subfigure}[b]{0.48\textwidth}
        \centering
        \includegraphics[width=\linewidth]{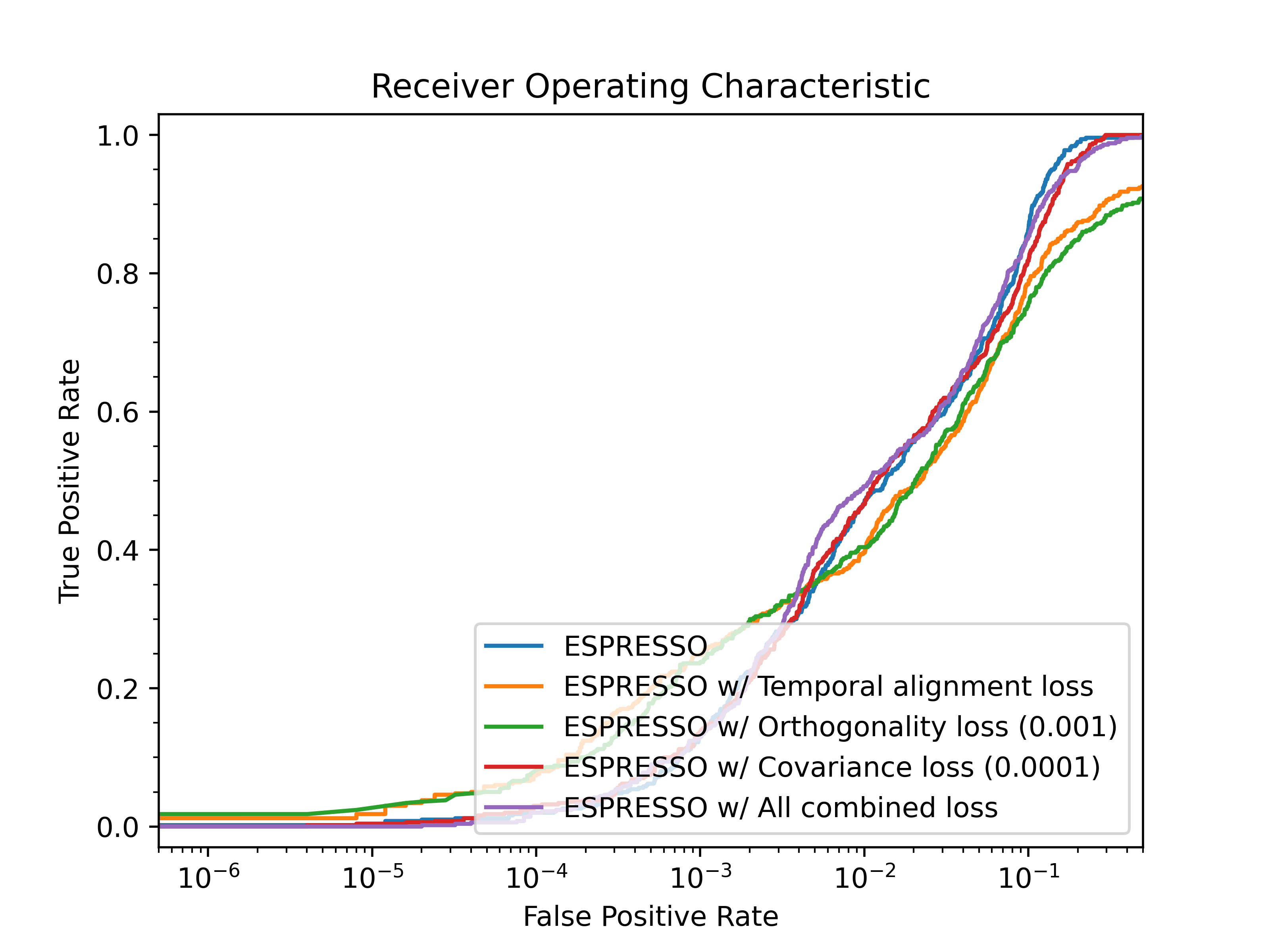}
        \caption{DNS-only ROC}
        \label{fig:roc-dns}
    \end{subfigure}
    \vspace{0.5cm}
    \begin{subfigure}[b]{0.48\textwidth}
        \centering
        \includegraphics[width=\linewidth]{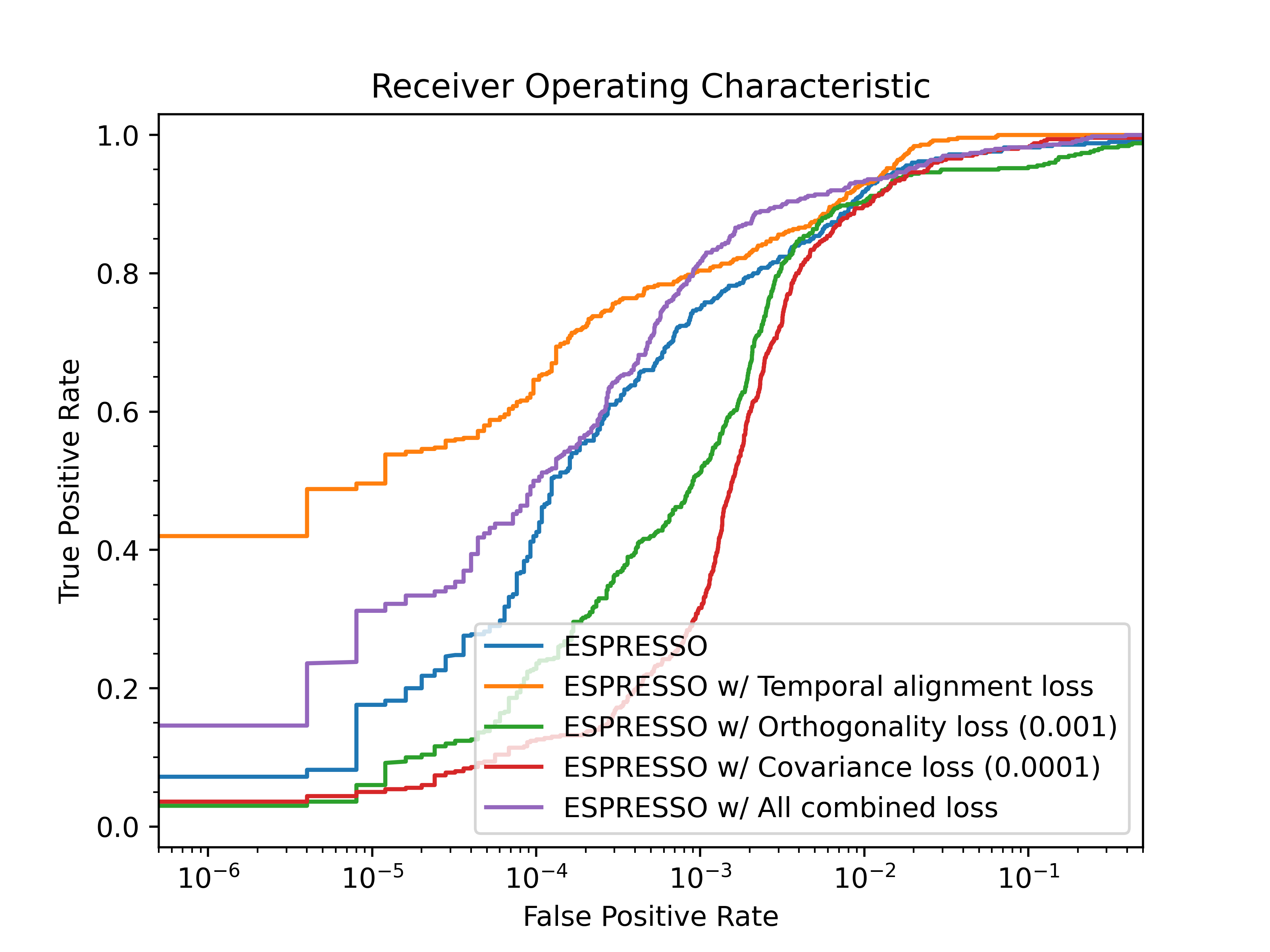}
        \caption{Mixed Protocol ROC}
        \label{fig:roc-mixed}
    \end{subfigure}
    \caption{ROC curves demonstrating correlation performance on (a)
    SSH-only, (b) DNS-only, and (c) Mixed protocol datasets.}
    \label{fig:roc-protocols-combined-espresso}
\end{figure}

\paragraph{Feature De-correlation}
The results of our benchmarks are presented in
Table~\ref{tab:loss_augmentation_results_all_protocols}, and full ROC
curves for selected experiments are presented in
Figure~\ref{fig:roc-protocols-combined-espresso}. Our investigation of
feature decorrelation losses found that these augmentations generally
do not yield consistent or significant improvements in correlation
performance. While the training process successfully optimized for
these auxiliary objectives---for example, the orthogonality loss on
SSH data converged from an extremely high initial value to as low as
35.1, and the covariance loss decreased from 708 to 550 without any
explicit weight---this did not translate into improved TPR or pAUC.
The covariance loss, in particular, appears not to correlate with
performance, as it decreased naturally during standard training.

There was one notable edge case: the model trained on SSH data with all
loss functions combined achieved the highest performance in that
specific setting, with a pAUC of 0.928 and a max TPR of 0.950 at an
FPR of $10^{-5}$. However, this result did not generalize. When the
same combined-loss configuration was applied to the Mixed-protocol
dataset, the model using only the temporal loss outperformed it. This
suggests that explicitly enforcing feature decorrelation is not a
primary factor for improving \espresso's performance and may even be
counterproductive for more diverse datasets.

\begin{figure*}[h]
    \centering
    \includegraphics[width=0.8\textwidth]{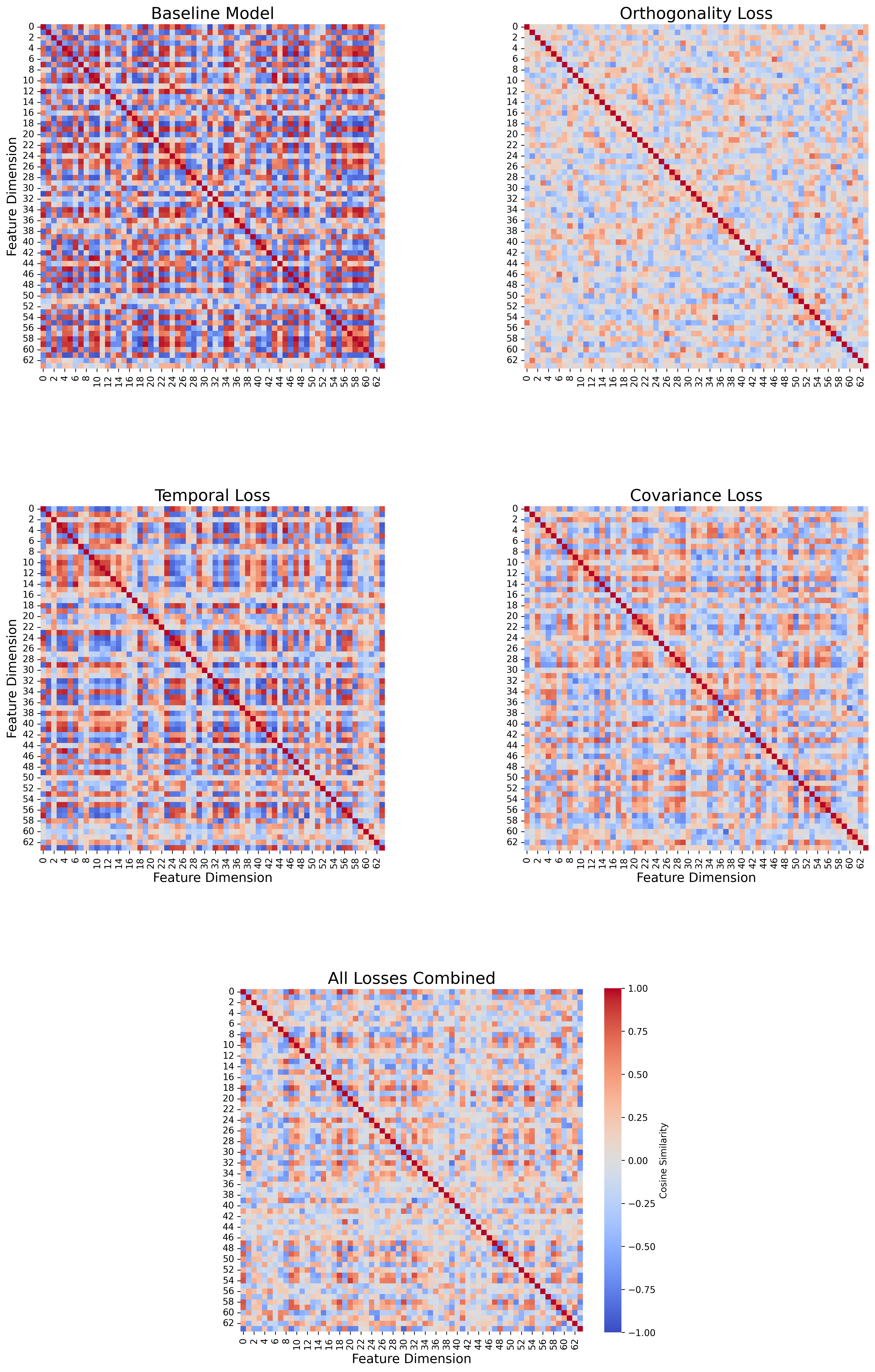}
    \caption{Comparison of window feature similarity of \espresso\
    trained with different loss strategies on one randomly chosen
    traffic sample.}
    \label{fig:decorrelation_sims}
\end{figure*}

Additionally, we plot the cosine similarity of window features within a
single sample in Figure~\ref{fig:decorrelation_sims}. We observe that
the model trained with orthogonality loss exhibits a clear decrease in
overall feature similarity between traffic windows. The covariance loss
exhibits a similar trend, albeit with a lesser impact. The temporal
loss shows higher similarity magnitudes in comparison and looks to have
more spatial clustering of window similarity (perhaps a natural result
of temporal alignment).

Supplemental experiments benchmarking these losses when applied to the
Modified DCF model are included in
Appendix~\ref{app:dcf_ablation},
Section~\ref{sec:appendix_dcf_loss}. While loss augmentation provided
distinct gains in the Mixed-protocol scenario---improving the baseline
TPR from 0.360 to over 0.620 at $10^{-4}$ FPR---results on
single-protocol data were largely stagnant or unstable, suggesting
that the benefits of these strategies are inconsistent and do not
strictly generalize to the CNN-based backbone.

\begin{table*}[t]
\centering
\begin{tabular}{llcccc}
\toprule
\toprule
\textbf{Protocol} & \textbf{Model Configuration} & \textbf{Loss Weight ($\lambda$)} & \textbf{AUC-ROC} & \textbf{pAUC-ROC} & \textbf{Max TPR} \\
\midrule
\multirow{15}{*}{\textbf{SSH}}
& Standard ESPRESSO & - & 0.998 & 0.535 & 0.710 \\
& + Temporal Loss & - & 0.998 & \textbf{0.861} & \textbf{0.946} \\
\cmidrule{2-6}
& + $L_{\text{orth}}$ & 0.1 & 0.994 & 0.043 & 0.074 \\
& & 0.01 & 0.996 & 0.677 & 0.680 \\
& & 0.001 & 0.998 & 0.656 & 0.856 \\
& & 0.0001 & 0.997 & \textbf{0.727} & \textbf{0.870} \\
\cmidrule{2-6}
& + $L_{\text{cov}}$ & 0.01 & 0.917 & 0.001 & 0.004 \\
& & 0.001 & 0.998 & 0.058 & 0.126 \\
& & 0.0001 & 0.996 & 0.421 & 0.502 \\
& & 0.00001 & 0.997 & \textbf{0.793} & \textbf{0.908} \\
\cmidrule{2-6}
& + $L_{\text{orth}} + L_{\text{cov}}$ & (0.01, 0.01) & 0.999 & 0.211 & 0.278 \\
& & (0.001, 0.001) & 0.999 & \textbf{0.599} & 0.644 \\
& & (0.0001, 0.0001) & 0.999 & 0.519 & \textbf{0.658} \\
\cmidrule{2-6}
& + All Losses Combined & (0.001, 0.0001) & 0.999 & \textbf{0.928} & \textbf{0.950} \\
\midrule
\midrule
\multirow{2}{*}{\textbf{DNS}}
& Standard ESPRESSO & - & 0.960 & 0.064 & 0.124 \\
& + Temporal Loss & - & 0.893 & \textbf{0.178} & \textbf{0.248} \\
\cmidrule{2-6}
& + $L_{\text{orth}}$ & 0.01 & 0.963 & 0.075 & 0.138 \\
& & 0.001 & 0.877 & \textbf{0.164} & \textbf{0.238} \\
& & 0.0001 & 0.840 & 0.140 & 0.192 \\
\cmidrule{2-6}
& + $L_{\text{cov}}$ & 0.001 & 0.954 & 0.081 & 0.152 \\
& & 0.0001 & 0.954 & 0.076 & 0.013 \\
& & 0.00001 & 0.884 & \textbf{0.119} & \textbf{0.162} \\
\cmidrule{2-6}
& + All Losses Combined & (0.001, 0.0001) & 0.955 & 0.074 & 0.13 \\
\midrule
\midrule
\multirow{2}{*}{\textbf{Mixed}}
& Standard ESPRESSO & - & 0.988 & 0.640 & 0.748 \\
& + Temporal Loss & - & 0.998 & \textbf{0.749} & \textbf{0.804} \\
\cmidrule{2-6}
& + $L_{\text{orth}}$ & 0.01 & 0.997 & 0.432 & 0.562 \\
& & 0.001 & 0.981 & 0.386 & 0.512 \\
& & 0.0001 & 0.965 & \textbf{0.468} & \textbf{0.646} \\
\cmidrule{2-6}
& + $L_{\text{cov}}$ & 0.001 & 0.989 & 0.113 & 0.202 \\
& & 0.0001 & 0.984 & 0.040 & 0.068 \\
& & 0.00001 & 0.992 & \textbf{0.209} & \textbf{0.316} \\
\cmidrule{2-6}
& + All Losses Combined & (0.001, 0.0001) & 0.994 & 0.673 & \textbf{0.816} \\
\bottomrule
\bottomrule
\end{tabular}%
\caption{Performance of \espresso\ with feature decorrelation loss
augmentations across various traffic protocols. The detailed breakdown
of loss augmentation experiments was performed on the SSH dataset. The
SSH-only model is compared on key detection metrics at an FPR threshold
of $10^{-5}$, while the DNS-only and Mixed-protocols models are
compared at an FPR threshold of $10^{-3}$.}
\label{tab:loss_augmentation_results_all_protocols}
\end{table*}

These loss augmentations have a very slight impact on computational
efficiency. With none of the strategies applied, it takes 253.3\,ms on
average to complete an optimization step on a single
batch.\footnote{Training is performed on an NVIDIA RTX 2080ti with a
batch size of 64.} With temporal loss the computation time is increased
by 1.3\% (to 256.6\,ms). When orthogonality loss is used time is
increased by 1.9\% (to 258.7\,ms). And finally, covariance loss has
the greatest impact of 2.8\% (to 260.5\,ms). With this low cost in
mind, the temporal loss augmentation seems to be a worthwhile strategy
to include.

\subsection{Investigating Performance Factors in Mixed-Protocol Environments}

The results presented in the prior sections demonstrate that
correlation performance in mixed-protocol environments lags behind
single-protocol baselines. To understand the drivers of this
degradation and explore potential remedies, we investigate two
hypotheses: (1) that a single shared feature extraction network lacks
the capacity to represent diverse protocols simultaneously, and (2)
that specific protocols with unique statistical properties (specifically
DNS) act as performance bottlenecks that skew the aggregate results.

\subsubsection{Multi-Model Architecture Strategy}
To improve efficacy in the multi-protocol setting, we also explore a
simple adjustment to the model structure. Rather than constrain
learning to a single set of network weights to learn all types of
protocol traffic, we initialize dedicated model weights for each
protocol that are optimized simultaneously during triplet learning.
Although this strategy notably increases training computational costs
and the memory footprint, the training process is otherwise identical
to the single-model strategy.

\begin{figure}[h]
    \centering
    \includegraphics[width=0.8\linewidth]{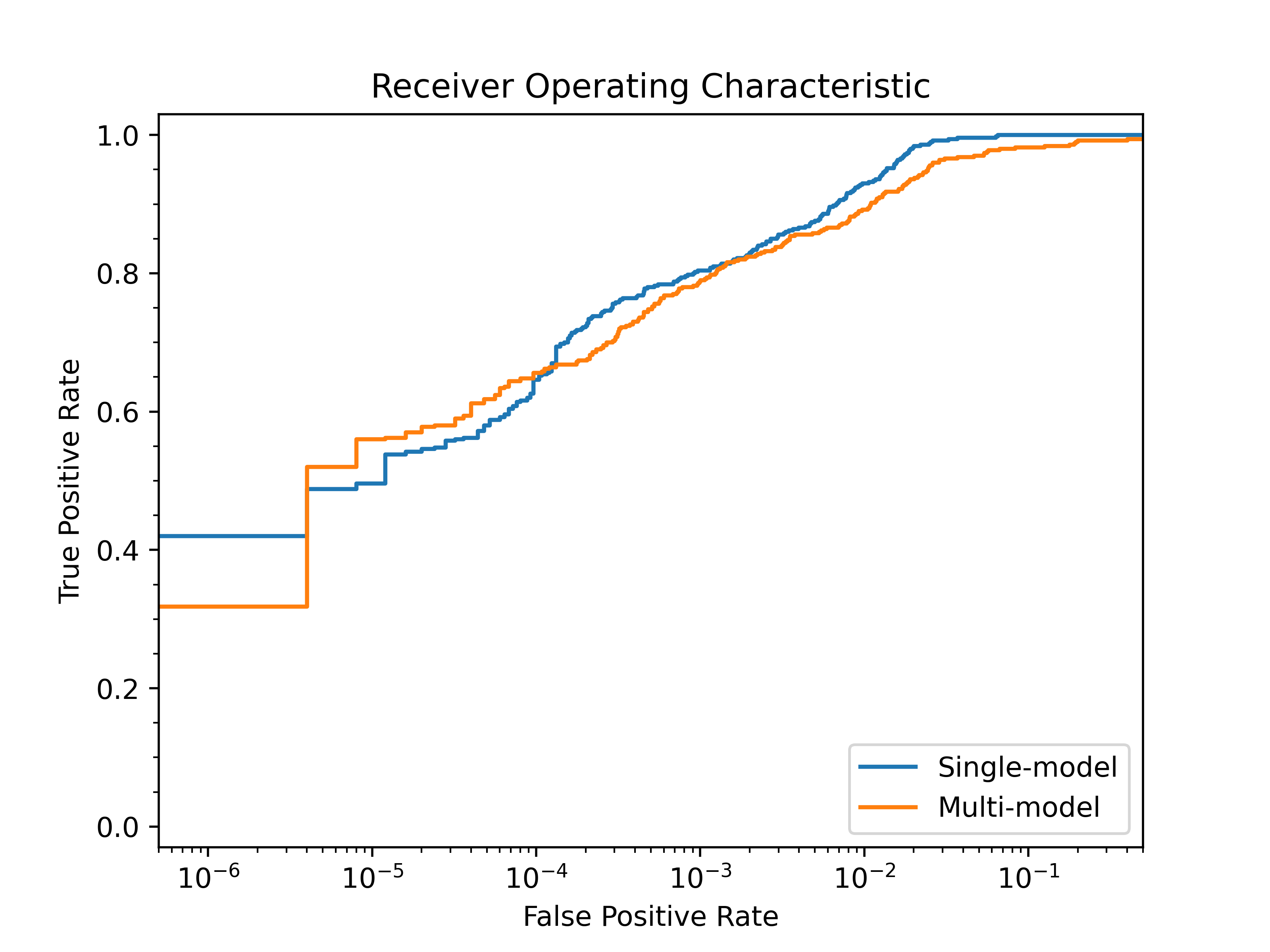}
    \caption{ROC comparison of \espresso\ trained using single-model
    and multi-model settings against the Mixed-protocol data in
    \textit{network-mode}.}
    \label{fig:mixed_multi_models}
\end{figure}

The performance of this strategy when applied with \espresso\ is
presented in Figure~\ref{fig:mixed_multi_models}. As the ROC curves
show, the performance of the two strategies is neck-and-neck. The
multi-model strategy outperforms the single-model strategy slightly in
maximum TPR at $10^{-5}$ FPR, but performance degrades somewhat
sharper than the single-model strategy soon after. Given the increased
computational cost presented by the multi-model strategy, the trade-off
would appear poor. We do not explore this strategy further.

\subsubsection{Impact of DNS Traffic on Detection Efficacy}

Our second investigation focuses on the composition of the dataset
itself. We previously observed that the specific tool used for DNS
tunneling in our dataset, \texttt{dnscat2}, employs a unique messaging
strategy to regulate data transmission rates and ensure reliability
over the connectionless DNS protocol. We hypothesized that the
resulting traffic patterns---driven by this distinct
application-layer behavior---present a unique correlation challenge
that differs fundamentally from the bursty dynamics of TCP-based
protocols like SSH and SOCAT, thereby disproportionately impacting the
aggregate metrics of the mixed-protocol dataset.

To assess this, we applied a filter to the multi-protocol dataset to
remove all samples from any chain that utilized the DNS protocol for
at least one hop. This curation yielded a dataset of approximately
4,900 samples, which we partitioned into 3,900 for training, 500 for
validation, and 500 for testing. Using this filtered data, we
benchmarked four model configurations: \espresso, \espresso\ with
Temporal Loss, the baseline DCF model, and the modified DCF model
utilizing time-interval features.

\begin{figure}[h]
    \centering
    \includegraphics[width=0.8\linewidth]{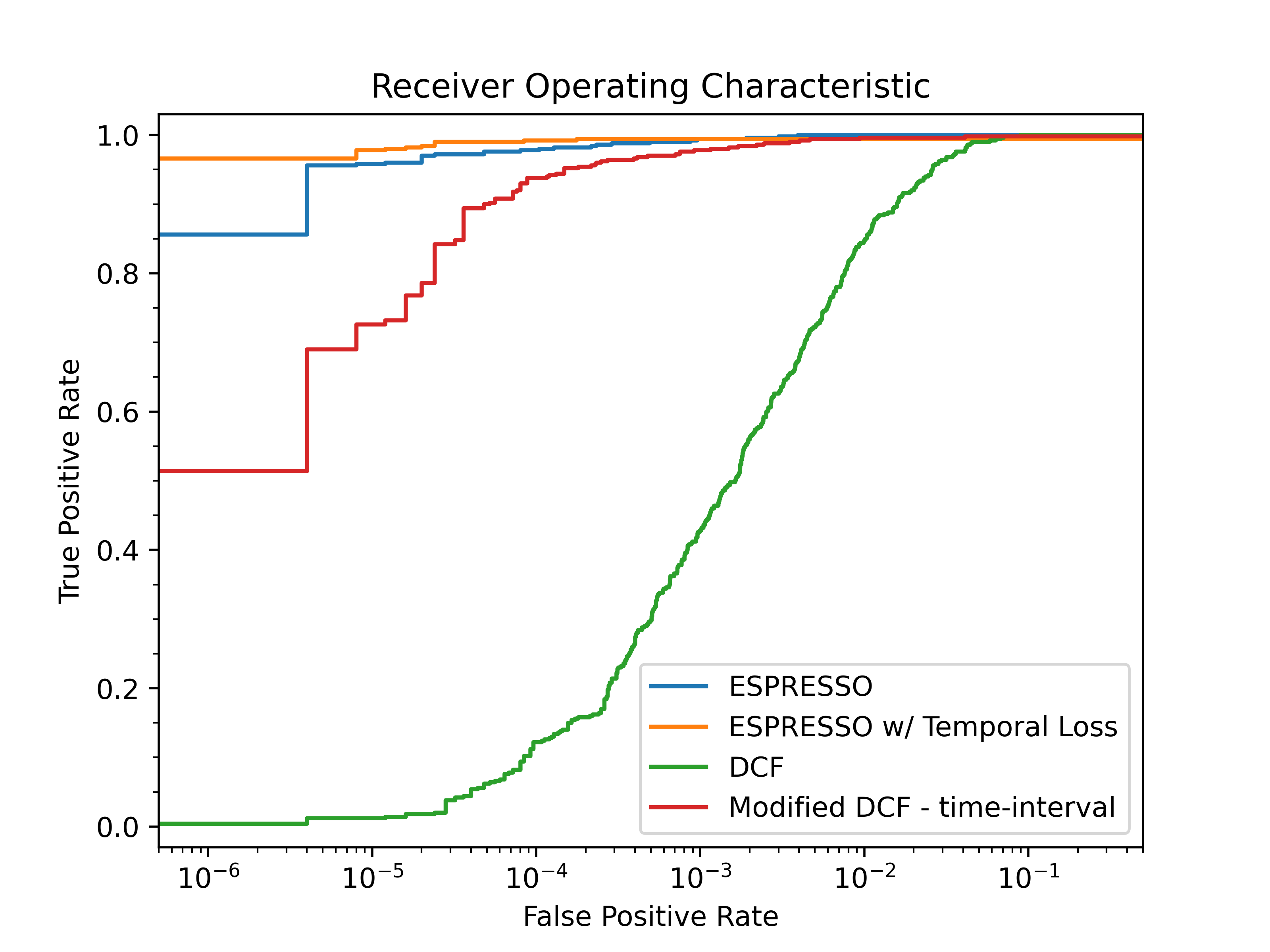}
    \caption{ROC results for additional benchmarks on the
    Multi-protocol dataset without chains that include DNS-tunneled
    traffic.}
    \label{fig:roc_nodns}
\end{figure}

Figure~\ref{fig:roc_nodns} presents the results from these benchmarks.
These results confirm that, as anticipated, DNS traffic is a primary
factor in performance degradation. With DNS excluded, detection
efficacy improved substantially across all evaluated models. Most
notably, \espresso\ trained with Temporal Loss achieved a True Positive
Rate (TPR) exceeding 95\% at a False Positive Rate (FPR) below
$10^{-5}$ (0.001\%). These findings validate that the core correlation
capabilities of \espresso\ remain robust across standard bursty
protocols (HTTP, HTTPS, SSH), even when mixed, and that future
improvements should focus on handling the unique ``polling'' behavior
of protocols such as DNS.

\subsection{Chain Length Prediction for Stepping Stone Detection}

While identifying the presence of a stepping-stone is a critical first
step, a network defender's ultimate goal is to assess intent: is this
connection relay benign or part of an attack? Not all instances of
pivoting are malicious. System administrators routinely use SSH jump
hosts to securely manage internal systems, and NAT gateways inherently
relay connections. These legitimate use-cases, however, are typically
characterized by operational simplicity and efficiency. As such, they
almost always involve very short connection chains, often just a single
hop.

In contrast, a malicious actor's primary motivation for using
stepping-stones is to obfuscate their origin and hinder forensic
investigation. Each additional hop in a connection chain, particularly
across different administrative domains, exponentially increases the
difficulty of tracing the attack back to its source. This provides a
strong incentive for attackers to construct longer, more convoluted
chains.

This discrepancy makes chain length a powerful heuristic for
distinguishing between benign and malicious stepping-stones. While a
single stepping-stone might be legitimate administrative traffic, a
five- or six-hop chain is highly indicative of a sophisticated attempt
at evasion and is far more likely to be malicious. Accurately
predicting the length of a stepping-stone chain is therefore not just
an academic exercise; it provides network defenders with a crucial tool
for risk assessment and alert triage.

Building on this premise, we explore two approaches to predict the
number of hosts in a stepping-stone chain. Both approaches treat the
problem as a regression task, predicting the number of hosts as a
floating-point value.

\subsubsection{Standalone Chain Length Prediction with a CNN Model}

The first approach involves training a deep learning model to predict
the number of hosts in upstream and downstream directions. We based
this model on a CNN architecture originally designed for website
fingerprinting tasks~\cite{Mathews2024Laserbeak}. The model takes per-packet
feature representations of the traffic as input, leveraging features
such as packet size, direction, and inter-packet timings to infer the
chain length.

The model outputs two floating-point numbers representing the predicted
number of upstream and downstream hosts, with the learning objective
being to minimize the mean squared error (MSE) between the predicted
value and the actual number of hosts. The value is rounded to a
discrete representation of the predicted host counts when measuring
accuracy.

\begin{equation}
    \text{MSE} = \frac{1}{N} \sum_{i=1}^{N} (\hat{y_i} - y_i)^2
\end{equation}
where $\hat{y_i}$ is the predicted number of hosts for a given sample
and $y_i$ is the true number of hosts.

\subsubsection{Joint Flow Correlation and Chain Length Prediction with \espresso}

The second approach is to directly integrate chain length prediction
into the \espresso\ flow correlation model, enabling it to
simultaneously perform flow correlation and chain length prediction in
a multi-task learning setup~\cite{caruana1997_oldmultitask}. This has
two potential benefits:
\begin{itemize}
    \item This can save on computation compared with training a separate
    model for each task. Furthermore, at inference time, only one model
    is needed.
    \item Additionally, it may boost overall performance, as the model
    needs to learn a more robust representation of the features that
    can work in all problems.
\end{itemize}
Researchers have demonstrated the effectiveness of this strategy in
domains such as computer
vision~\cite{kokkinos2017_ubernet,yang2018_multitaskships} and
NLP~\cite{tang2017_multitaskspeech,collobert2008_MultitaskNLP}.

Due to the transformer-based design of \espresso, extending the model
for multiple outcomes is relatively straightforward. We propose the
addition of a special token to the input sequence whose final output
will be used for chain length prediction. This is similar to the class
token used in the original ViT construction~\cite{dosovitskiy2021an}.
This token can be connected to a Multi-Layer Perceptron (MLP) that
performs the upstream and downstream host count prediction.

The chain length prediction is treated as a regression task with mean
squared error (MSE) as the loss function. The MSE loss for chain length
prediction is added to the triplet loss used for flow correlation,
allowing the model to optimize both objectives simultaneously. A weight
hyperparameter $\lambda$ controls the relative strength of the chain
length prediction error compared to the flow correlation objective:

\begin{equation}
    \mathcal{L} = \mathcal{L}_{\text{triplet}} + \lambda \cdot
    \mathcal{L}_{\text{MSE-chain-length}}
\end{equation}

\subsubsection{Experimental Results}

\begin{table}[t]
\centering
\begin{tabular}{l|ccc}
\toprule
\toprule
\textbf{Dataset} & \textbf{Up Acc.} & \textbf{Down Acc.} & \textbf{Avg. Acc.} \\
\midrule
SSH   & 0.950 & 0.890 & 0.920 \\
SOCAT & 0.800 & 0.710 & 0.755 \\
ICMP  & 0.925 & 0.963 & 0.944 \\
DNS   & 0.798 & 0.797 & 0.798 \\
Mixed & 0.814 & 0.770 & 0.792 \\
\bottomrule
\bottomrule
\end{tabular}
\caption{Performance of the standalone chain-only prediction model
across various datasets.}
\label{tab:chain_only_performance}
\end{table}

\begin{table*}[t]
\centering
\begin{tabular}{llcccccccc}
\toprule
\toprule
& & \multicolumn{3}{c}{\textbf{Chain-Length Prediction}} &
\multicolumn{5}{c}{\textbf{Correlation Performance}} \\
\cmidrule(lr){3-5} \cmidrule(lr){6-10}
& & & & & & \multicolumn{2}{c}{\textbf{FPR $\leq 10^{-5}$}} &
\multicolumn{2}{c}{\textbf{FPR $\leq 10^{-4}$}} \\
\cmidrule(lr){7-8} \cmidrule(lr){9-10}
\textbf{Protocol} & \textbf{$\lambda$} & \textbf{Up Acc.} &
\textbf{Down Acc.} & \textbf{Avg. Acc.} & \textbf{AUC} &
\textbf{pAUC} & \textbf{max TPR} & \textbf{pAUC} & \textbf{max TPR} \\
\midrule
\multirow{3}{*}{\textbf{SSH}}
& 0.1 & 0.808 & 0.790 & 0.799 & 0.998 & 0.702 & 0.898 & 0.929 & 0.972 \\
& 0.5 & 0.814 & 0.838 & 0.826 & 0.998 & 0.702 & 0.914 & 0.928 & 0.970 \\
& 1.0 & 0.812 & 0.832 & 0.822 & 0.998 & 0.554 & 0.914 & 0.930 & 0.976 \\
\midrule
\multirow{3}{*}{\textbf{SOCAT}}
& 0.1 & 0.810 & 0.806 & 0.808 & 0.999 & 0.888 & 1.000 & 0.888 & 1.000 \\
& 0.5 & \textbf{0.820} & 0.808 & 0.814 & 0.9999 & 0.6190 & 1.000 & 0.6190 & 1.000 \\
& 1.0 & 0.804 & 0.810 & 0.807 & 1.000 & \textbf{1.000} & 1.000 & 1.000 & 1.000 \\
\midrule
\multirow{3}{*}{\textbf{ICMP}}
& 0.1 & 0.748 & 0.790 & 0.769 & 0.999 & \textbf{0.826} & 0.980 & 0.969 & 0.988 \\
& 0.5 & 0.770 & 0.786 & 0.778 & 0.999 & 0.772 & 0.988 & 0.968 & 0.992 \\
& 1.0 & \textbf{0.800} & \textbf{0.810} & \textbf{0.805} & 0.999 & 0.802 & 0.988 & 0.977 & 0.998 \\
\midrule
\multirow{3}{*}{\textbf{DNS}}
& 0.1 & \textbf{0.674} & 0.654 & \textbf{0.664} & 0.960 & 0.002 & 0.004 & 0.012 & 0.020 \\
& 0.5 & 0.658 & 0.624 & 0.641 & 0.957 & 0.002 & 0.004 & 0.011 & 0.020 \\
& 1.0 & 0.656 & 0.628 & 0.642 & 0.958 & 0.002 & 0.004 & 0.012 & 0.020 \\
\midrule
\multirow{3}{*}{\textbf{Mixed}}
& 0.1 & 0.788 & 0.794 & 0.791 & 0.991 & \textbf{0.252} & \textbf{0.379} & 0.501 & 0.685 \\
& 0.5 & 0.790 & 0.816 & 0.803 & 0.991 & 0.231 & 0.354 & 0.499 & 0.680 \\
& 1.0 & 0.784 & 0.814 & 0.799 & 0.990 & 0.222 & 0.339 & 0.490 & 0.664 \\
\bottomrule
\bottomrule
\end{tabular}%
\caption{Performance of the jointly learned model for direct
stepping-stone chain-length prediction and flow correlation. The weight
$\lambda$ balances the contribution of the chain-length prediction loss
against the primary correlation loss.}
\label{tab:direct_chain_length_prediction}
\end{table*}

\paragraph{Chain-length Prediction}
The performance of the two chain-length prediction approaches is
detailed in Table~\ref{tab:chain_only_performance} and
Table~\ref{tab:direct_chain_length_prediction}. A key observation is
the superior performance of the standalone CNN model, which is trained
exclusively for length prediction using packet-level features. As shown
in Table~\ref{tab:chain_only_performance}, this model achieves
significantly higher average accuracy on the SSH (0.950), ICMP
(0.925), and DNS (0.798) datasets compared to the best results from
the joint-learning model. The performance is more comparable for the
SOCAT and mixed-protocol datasets. We speculate that the availability
of richer, packet-level features, particularly per-packet timestamps,
provides the standalone model with more granular timing information
that is crucial for accurately inferring chain length. Effectively, the
chain-only model operates at a higher resolution of timing features
than \espresso, possibly enabling improvements in prediction accuracy
against difficult samples.

In the joint learning approach
(Table~\ref{tab:direct_chain_length_prediction}), the results show a
complex trade-off between chain-length prediction and flow correlation.
No single value of $\lambda$ consistently emerges as the ideal choice
across all datasets. Furthermore, increasing the weight of the
chain-length prediction loss does not reliably lead to better
prediction accuracy either. Encouragingly, the integration of the
chain-length prediction task appears to have a minimal overall impact
on the primary task of flow correlation, with performance neither
significantly degrading nor improving across the different weights.
This suggests that multi-task learning is a viable approach, though the
standalone model demonstrates that a specialized architecture with more
granular features currently offers the best performance for this
specific task.

\paragraph{Inferring Chain Length via Full-Path Correlation.}
An alternative to predicting chain length directly is to infer it by
reconstructing the entire connection path. This approach leverages the
underlying pair-correlation model to determine if every link in a
stepping-stone chain can be successfully identified. Instead of
treating length prediction as a regression task, this method reframes
it as a test of complete recall across all nodes in the chain. A chain
is considered successfully traced only if every constituent
stepping-stone pair is correlated above a given confidence threshold,
and accuracy is calculated as the percentage of chains for which every
single stepping-stone pair was correctly identified.

\begin{table*}[t]
\centering
\begin{tabular}{l|cc|cc|cc}
\toprule
\toprule
\multirow{2}{*}{\textbf{Dataset}} &
\multicolumn{2}{c|}{\textbf{FPR} $\leq 10^{-3}$} &
\multicolumn{2}{c|}{\textbf{FPR} $\leq 10^{-4}$} &
\multicolumn{2}{c}{\textbf{FPR} $\leq 10^{-5}$} \\
\cmidrule(lr){2-3} \cmidrule(lr){4-5} \cmidrule(lr){6-7}
& \textbf{Avg. TPR} & \textbf{Chain Acc.} &
\textbf{Avg. TPR} & \textbf{Chain Acc.} &
\textbf{Avg. TPR} & \textbf{Chain Acc.} \\
\midrule
SSH   & 0.993 $\pm$ 0.080 & 0.992 & 0.983 $\pm$ 0.123 & 0.980 & 0.947 $\pm$ 0.214 & 0.938 \\
SOCAT & 1.000 $\pm$ 0.000 & 1.000 & 1.000 $\pm$ 0.000 & 1.000 & 1.000 $\pm$ 0.000 & 1.000 \\
ICMP  & 1.000 $\pm$ 0.000 & 1.000 & 0.996 $\pm$ 0.055 & 0.994 & 0.994 $\pm$ 0.070 & 0.992 \\
DNS   & 0.000 $\pm$ 0.000 & 0.000 & 0.000 $\pm$ 0.000 & 0.000 & 0.000 $\pm$ 0.000 & 0.000 \\
Mixed & 0.842 $\pm$ 0.307 & 0.758 & 0.310 $\pm$ 0.414 & 0.220 & 0.000 $\pm$ 0.000 & 0.000 \\
\bottomrule
\bottomrule
\end{tabular}
\caption{Performance of correlation-based (full-chain) prediction at
various FPR thresholds. The table shows the average True Positive Rate
(TPR) with standard deviation, and the full chain accuracy for each
dataset.}
\label{tab:full_chain_correlation_based}
\end{table*}

\begin{figure}[h]
    \centering
    \includegraphics[width=0.8\linewidth]{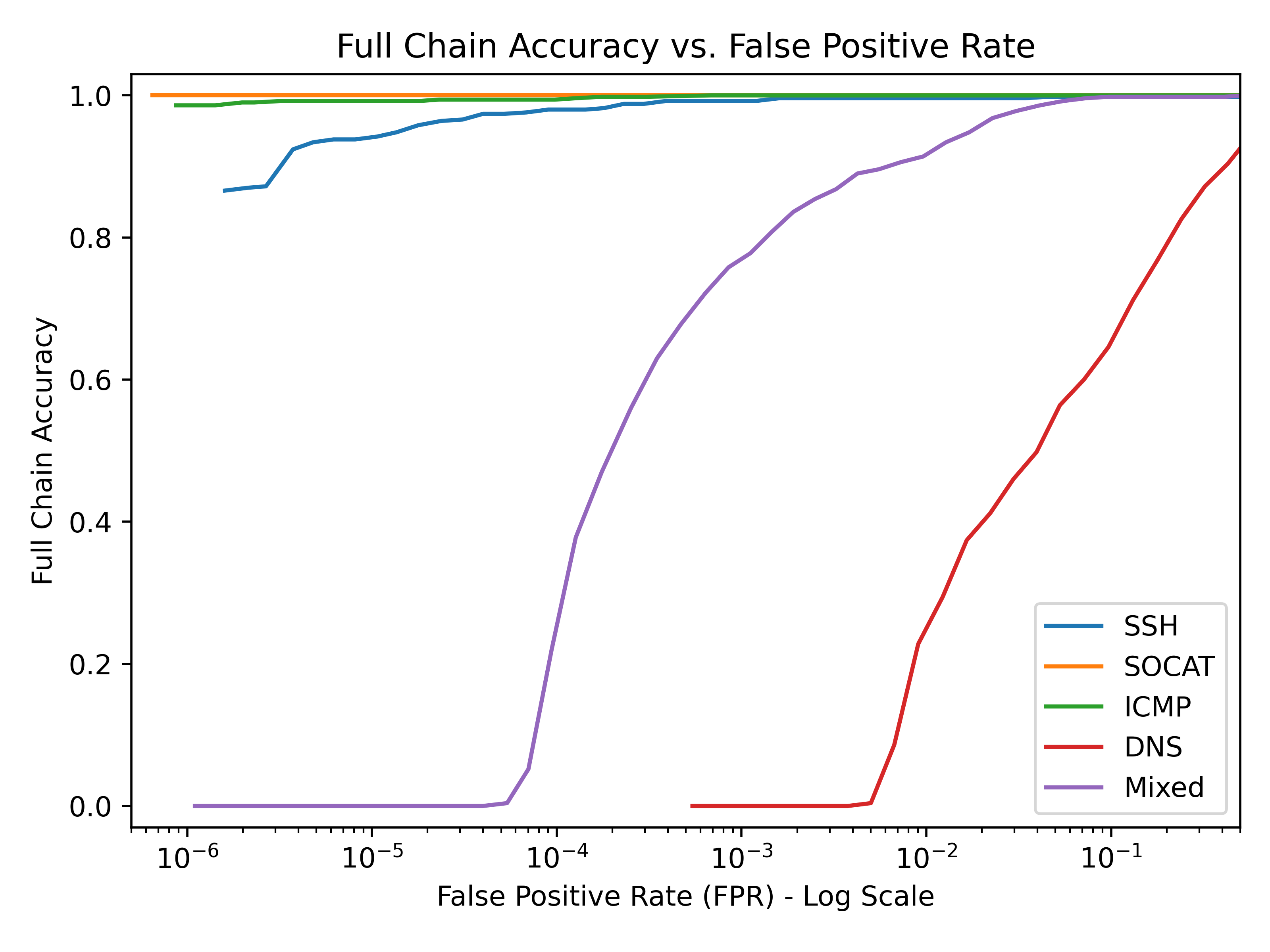}
    \caption{Full chain correlation accuracy as a function of FPR with
    \espresso.}
    \label{fig:ssh_chain_metrics}
\end{figure}

The results of this experiment are summarized in
Table~\ref{tab:full_chain_correlation_based} and in
Figure~\ref{fig:ssh_chain_metrics}. The performance patterns observed
unsurprisingly mirror those of the standard network-mode correlation
task. The model achieves near-perfect chain accuracy for the SOCAT and
ICMP datasets, and very high accuracy for SSH, particularly at an FPR
of $10^{-3}$. However, performance drops significantly for the more
challenging datasets. The DNS protocol results in zero correctly
identified full chains. Similarly, the Mixed protocol dataset shows a
steep decline in performance as the FPR threshold becomes more
restrictive, failing to identify any complete chains at an FPR of
$10^{-5}$. This indicates that while this method is highly effective
for protocols with strong correlation signals, its utility as a
reliable chain-length counter is limited in scenarios with more diverse
or ambiguous traffic patterns, where even a single missed link results
in failure.

Comparing the two strategies reveals a clear trade-off between
robustness and certainty. The direct prediction models, particularly
the standalone CNN, provide a reasonable estimate of chain length, even
in challenging scenarios such as DNS, where the full-chain correlation
method fails entirely. In contrast, the full-chain correlation method,
when successful, provides not just the length but the complete
composition of the chain with high confidence, but requires
extensive network-level monitoring. Therefore, the two approaches are
complementary: direct prediction serves as a robust tool for general
assessment, while full-chain correlation offers a high-fidelity
reconstruction method for traffic where correlation signals are strong.

\subsection{Impact of Padding Obfuscation and Delays}

To evaluate the robustness of our correlation model, we investigate its
performance against standard traffic obfuscation techniques that an
adversary might employ to evade detection. This section details our
experiments applying two fundamental noising strategies (packet padding
and timing perturbations) to a standard protocol where such
manipulations would be feasible, such as SSH. By systematically
increasing the intensity of these defenses, we can measure the
resilience of both \espresso\ and the DCF baseline and identify their
breaking points.

\subsubsection{Obfuscation Strategies}
We implemented two distinct obfuscation methods designed to disrupt the
statistical features that correlation systems rely on.

\paragraphX{Varying-Rate Padding}
Adding dummy (or chaff) packets is a foundational technique to defend
against traffic analysis by masking the true volume and burst patterns
of a connection~\cite{juarez2016toward}. Simple, constant-rate padding
can be trivial to identify and filter. To create a more sophisticated
and less predictable defense, we developed a varying-rate injection
strategy that introduces dummy packets in bursts.

A correlation model must evaluate a vast number of potential
incoming-outgoing flow pairings. A sophisticated padding defense should
therefore generate chaff that not only masks the original flow's
patterns but also creates new, plausible-looking patterns. A
varying-rate strategy creates dummy streams that can exhibit a higher
degree of coincidental correlation with uncorrelated flows, forcing the
detection model to disambiguate the true correlated pair from a
multitude of compelling, but false, alternatives.

The padding algorithm begins by determining the number of dummy packets
to inject, $N'_d$, by sampling from a normal distribution centered on
a target overhead percentage. The total duration of the traffic, $T$,
is then divided into $K$ temporal segments, where $K$ is uniformly
sampled from $\{K_{\min}=5, K_{\max}=15\}$. To create a bursty
traffic pattern, the $N'_d$ packets are distributed unevenly across
these segments using a Dirichlet distribution:
\begin{equation}
(w_1, \dots, w_K) \sim \text{Dir}(\alpha) \quad \text{where } \alpha = (1, \dots, 1)
\end{equation}
\begin{equation}
n_k = \text{round}(w_k \cdot N'_d) \quad \text{such that} \quad \sum_{k=1}^{K} n_k = N'_d
\end{equation}
Within each segment $k$ of duration $T_k = T/K$, the inter-packet
arrival times for its $n_k$ allocated dummy packets are modeled as a
Poisson process by sampling from an exponential distribution with rate
$\lambda_k = n_k / T_k$. Finally, to make the dummy traffic
statistically similar to the real traffic, the sizes of the injected
packets are sampled from the empirical distributions of the original
flow's incoming and outgoing packet sizes.

\paragraphX{Timing Perturbation}
The second strategy directly targets the fine-grained timing features
that are often the most powerful signal for traffic
correlation~\cite{kiyavash2008multi,dyer2012peek}. By introducing
random, artificial latency (jitter), this method aims to desynchronize
the temporal relationship between an incoming and outgoing flow.

For each packet $p_i$ in a flow with an original timestamp $t_i$, a
random delay $d_i$ is introduced with a set probability,
$P_{\text{delay}}$. The delay is drawn from a uniform distribution up
to a maximum value, $D_{\max}$. The new timestamp, $t'_i$, is
calculated as:

\begin{equation}
t'_i = t_i + d_i, \quad \text{where } d_i =
\begin{cases}
\mathcal{U}(0, D_{\max}) & \text{with } P_{\text{delay}} \\
0 & \text{with } 1 - P_{\text{delay}}
\end{cases}
\end{equation}

After applying these delays to all packets in the flow, the entire
sequence is re-sorted by the new timestamps to maintain chronological
order. We tested this strategy at several intensities by varying the
parameters $(P_{\text{delay}}, D_{\max})$, corresponding to ``light''
(two variants) and ``heavy'' obfuscation profiles. The first light
variant delays packets with a 25\% probability up to a maximum of 1.0
second. The second light variant increases the likelihood to 50\% but
reduces the maximum delay to 0.5 seconds. The ``heavy'' profile uses a
75\% probability with a 1.0-second maximum delay.

\subsubsection{Experimental Results}

\begin{figure}[h!]
    \centering
    \begin{subfigure}[b]{0.9\linewidth}
        \centering
        \includegraphics[width=\linewidth]{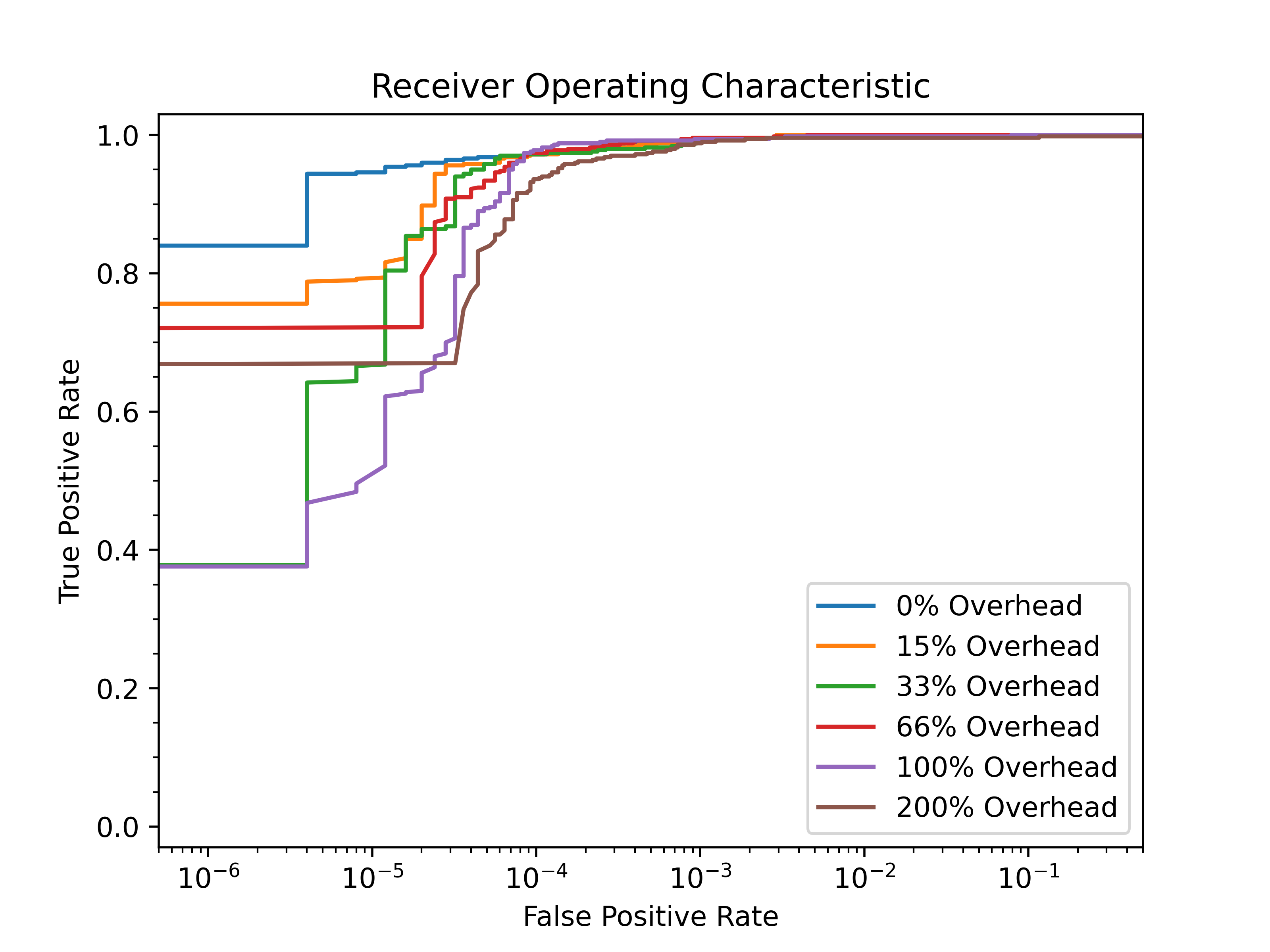}
        \caption{\espresso\ against padding (no delays)}
        \label{fig:ssid-obfs1-espresso}
    \end{subfigure}
    \hfill
    \begin{subfigure}[b]{0.9\linewidth}
        \centering
        \includegraphics[width=\linewidth]{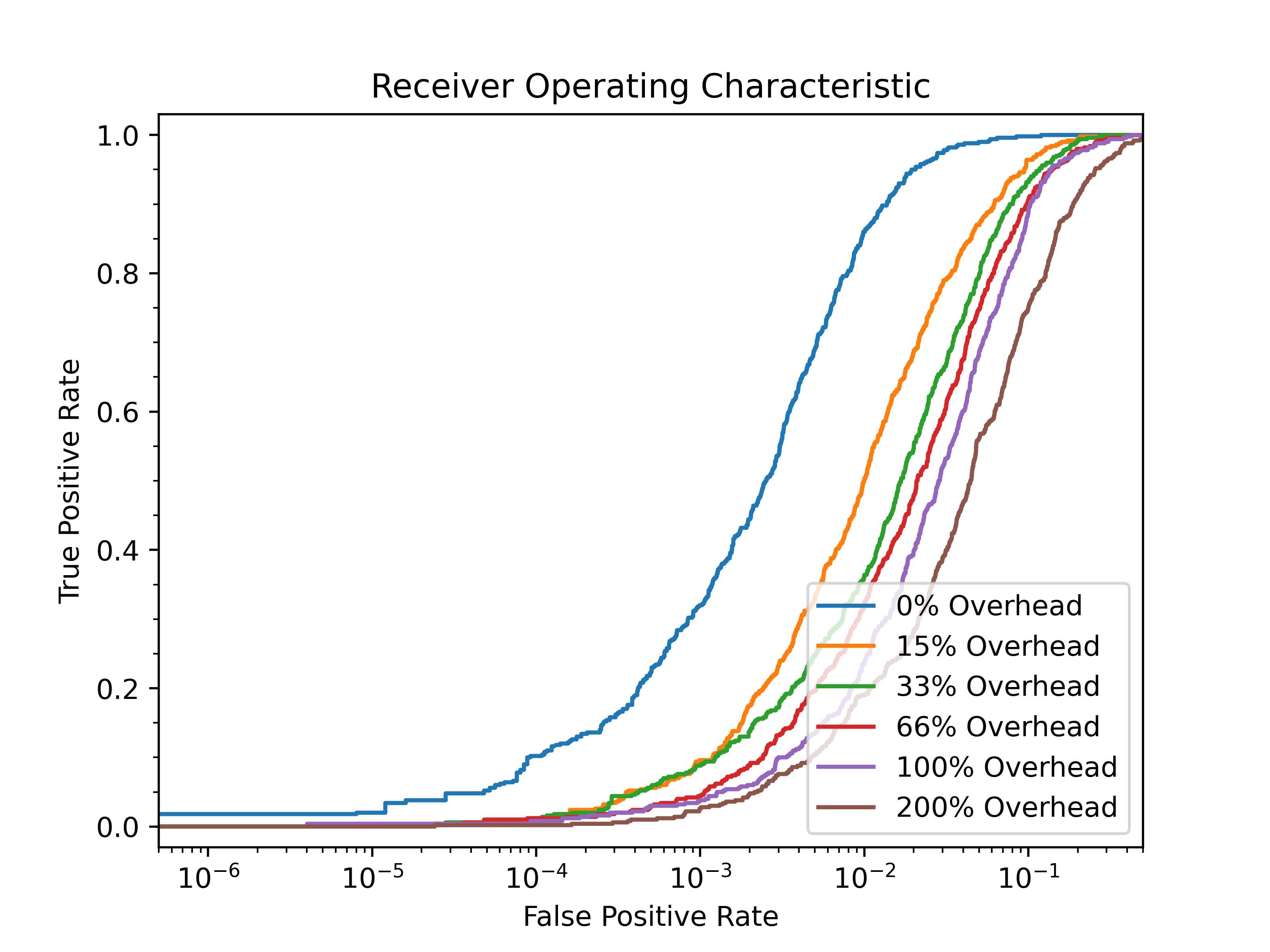}
        \caption{DCF against padding (no delays)}
        \label{fig:ssid-obfs1-dcf}
    \end{subfigure}
    \caption{ROC curves for (a) \espresso\ and (b) DCF on the SSH-only
    dataset under traffic padding obfuscation in the
    \textit{network-mode} detection scenario.}
    \label{fig:ssid-obfs1}
\end{figure}

\begin{figure}[h!]
    \centering
    \begin{subfigure}[b]{0.8\linewidth}
        \centering
        \includegraphics[width=\linewidth]{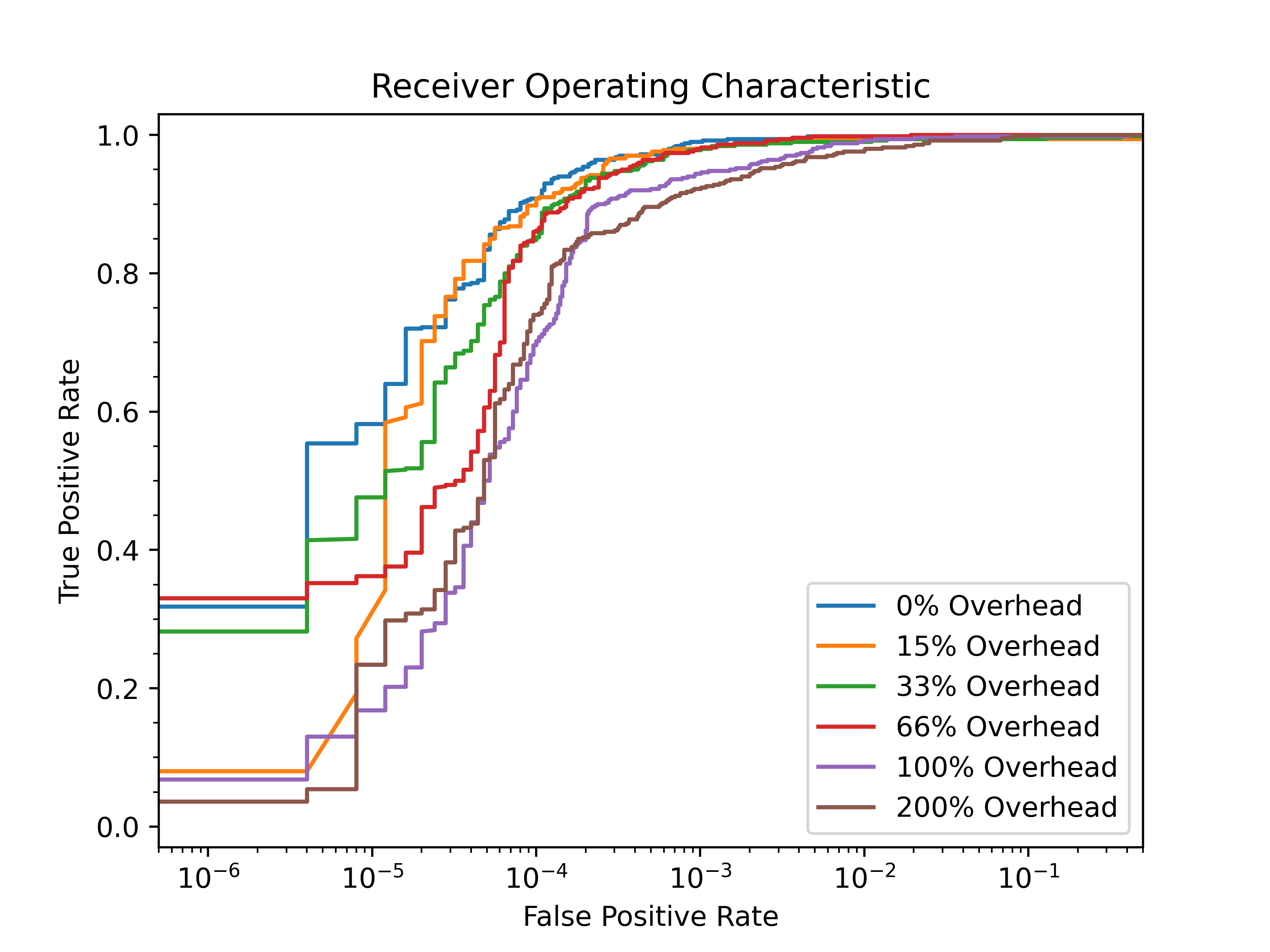}
        \caption{\espresso\ against light delays (v1)}
        \label{fig:ssid-obfs2-espresso}
    \end{subfigure}
    \hfill
    \begin{subfigure}[b]{0.8\linewidth}
        \centering
        \includegraphics[width=\linewidth]{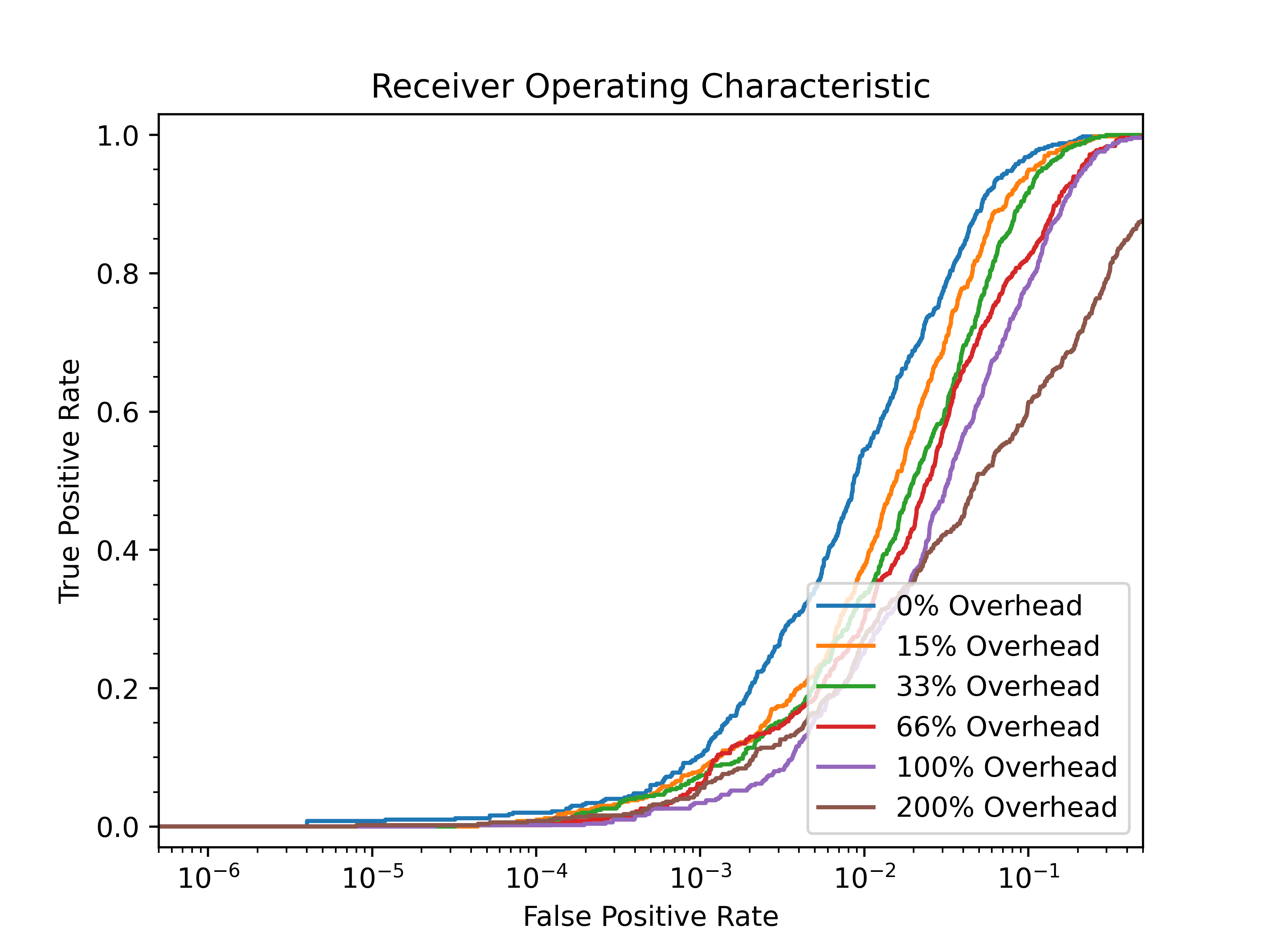}
        \caption{DCF against light delays (v1)}
        \label{fig:ssid-obfs2-dcf}
    \end{subfigure}
    \caption{ROC curves for (a) \espresso\ and (b) DCF on the SSH-only
    dataset under padding with light delays (v1) in the
    \textit{network-mode} detection scenario.}
    \label{fig:ssid-obfs2}
\end{figure}

\begin{figure}[h!]
    \centering
    \begin{subfigure}[b]{0.8\linewidth}
        \centering
        \includegraphics[width=\linewidth]{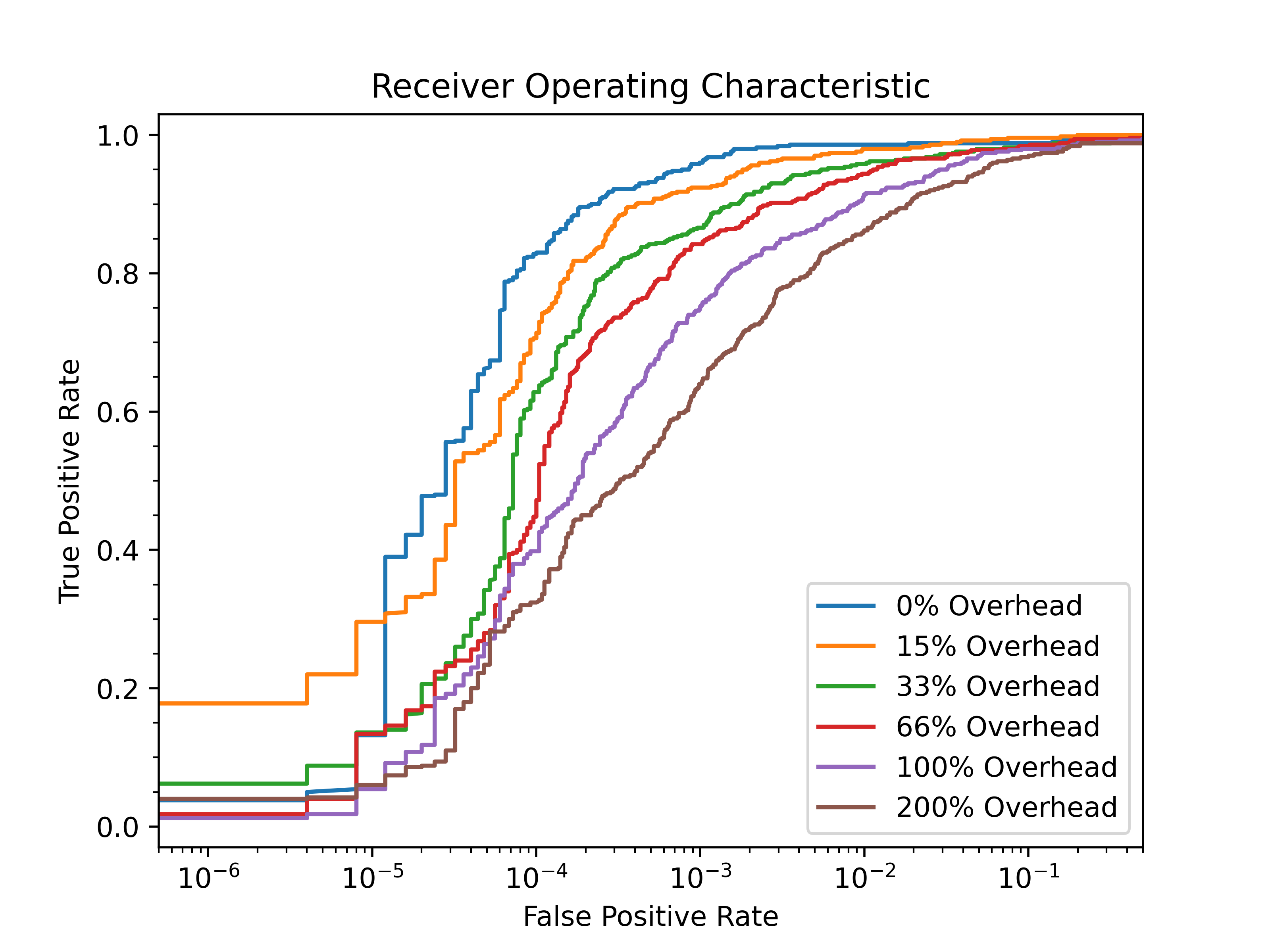}
        \caption{\espresso\ against light delays (v2)}
        \label{fig:ssid-obfs3-espresso}
    \end{subfigure}
    \hfill
    \begin{subfigure}[b]{0.8\linewidth}
        \centering
        \includegraphics[width=\linewidth]{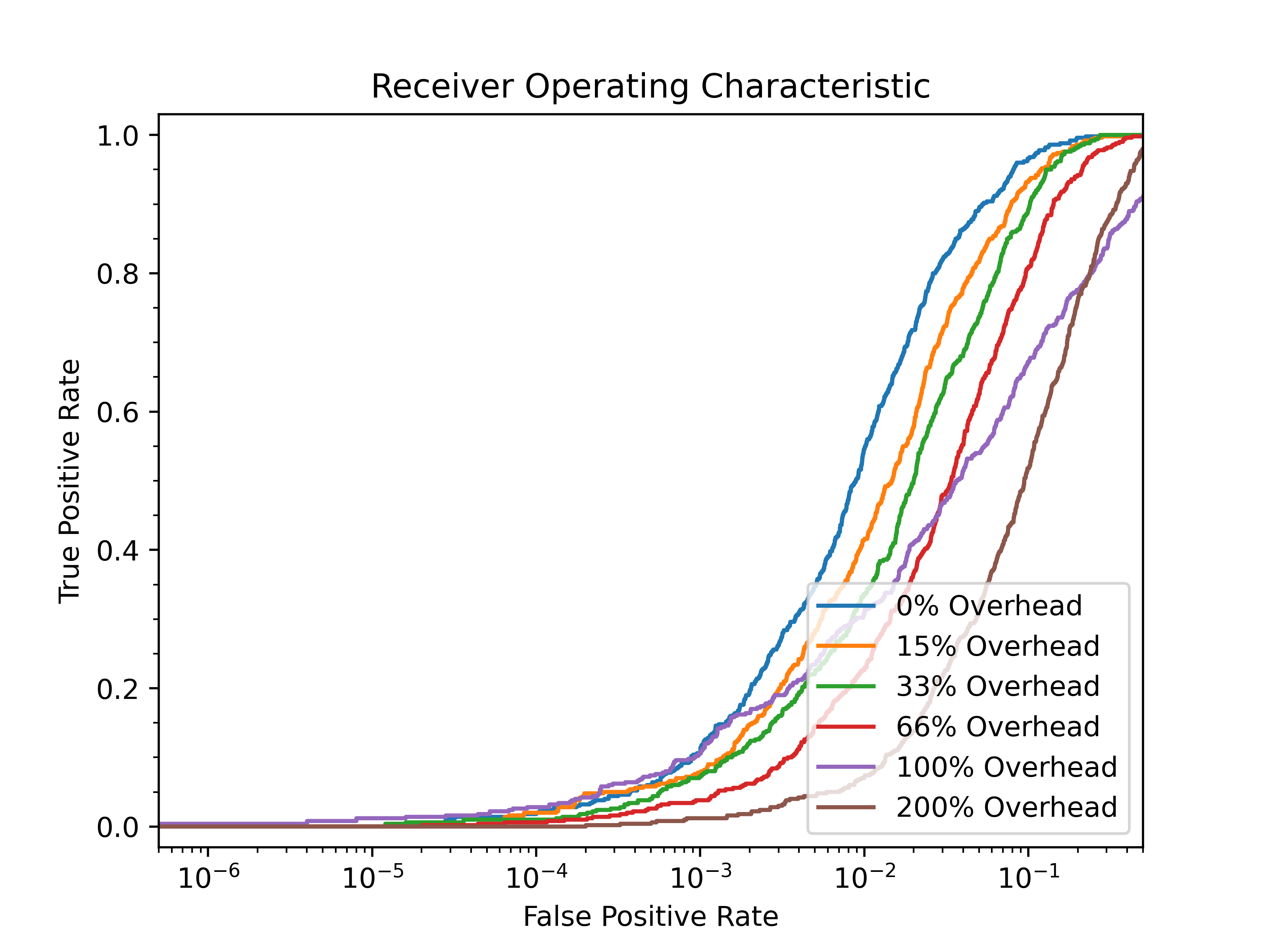}
        \caption{DCF against light delays (v2)}
        \label{fig:ssid-obfs3-dcf}
    \end{subfigure}
    \caption{ROC curves for (a) \espresso\ and (b) DCF on the SSH-only
    dataset under padding with light delays (v2) in the
    \textit{network-mode} detection scenario.}
    \label{fig:ssid-obfs3}
\end{figure}

\begin{figure}[h!]
    \centering
    \begin{subfigure}[b]{0.8\linewidth}
        \centering
        \includegraphics[width=\linewidth]{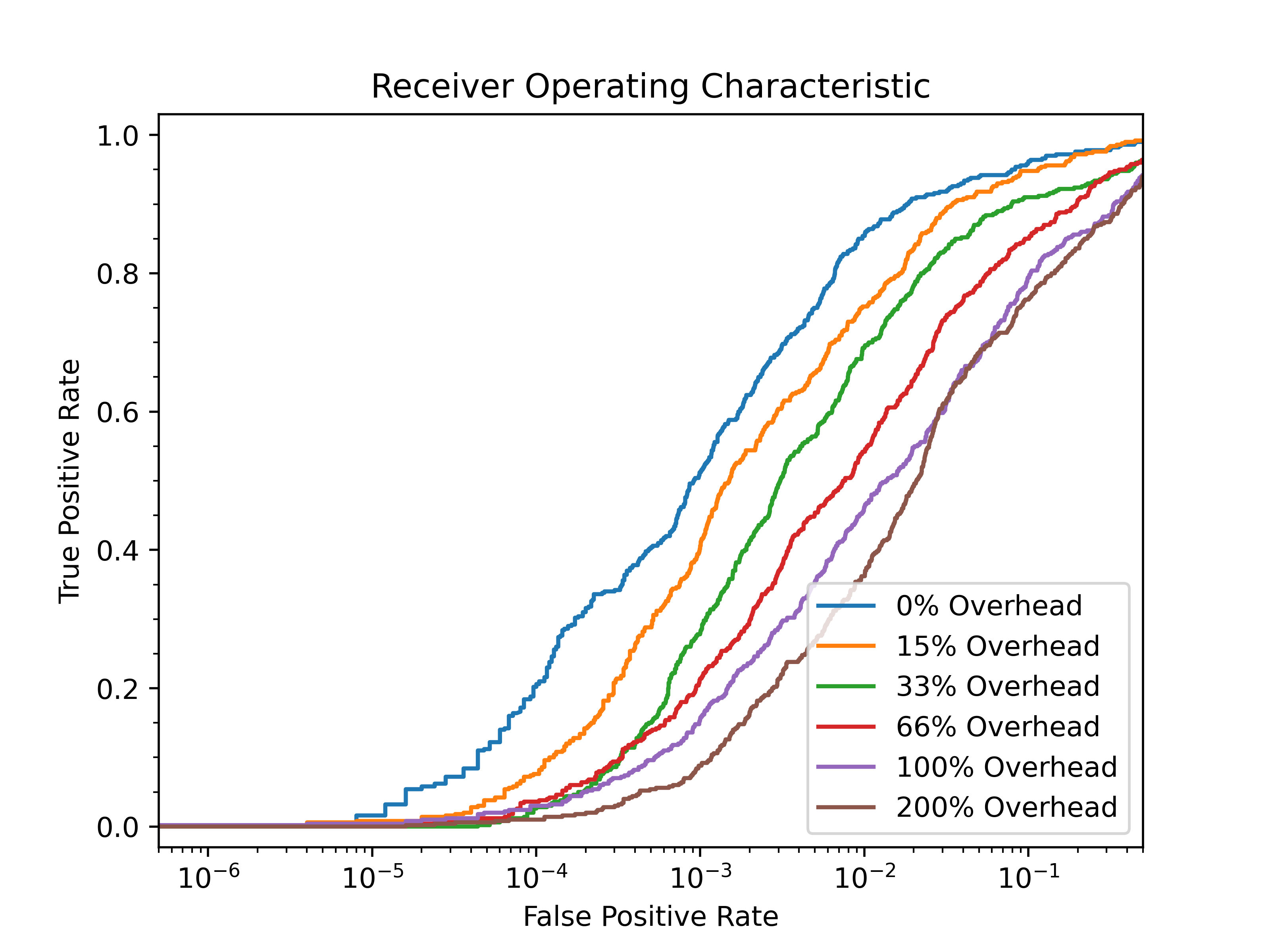}
        \caption{\espresso\ against heavy delays}
        \label{fig:ssid-obfs4-espresso}
    \end{subfigure}
    \hfill
    \begin{subfigure}[b]{0.8\linewidth}
        \centering
        \includegraphics[width=\linewidth]{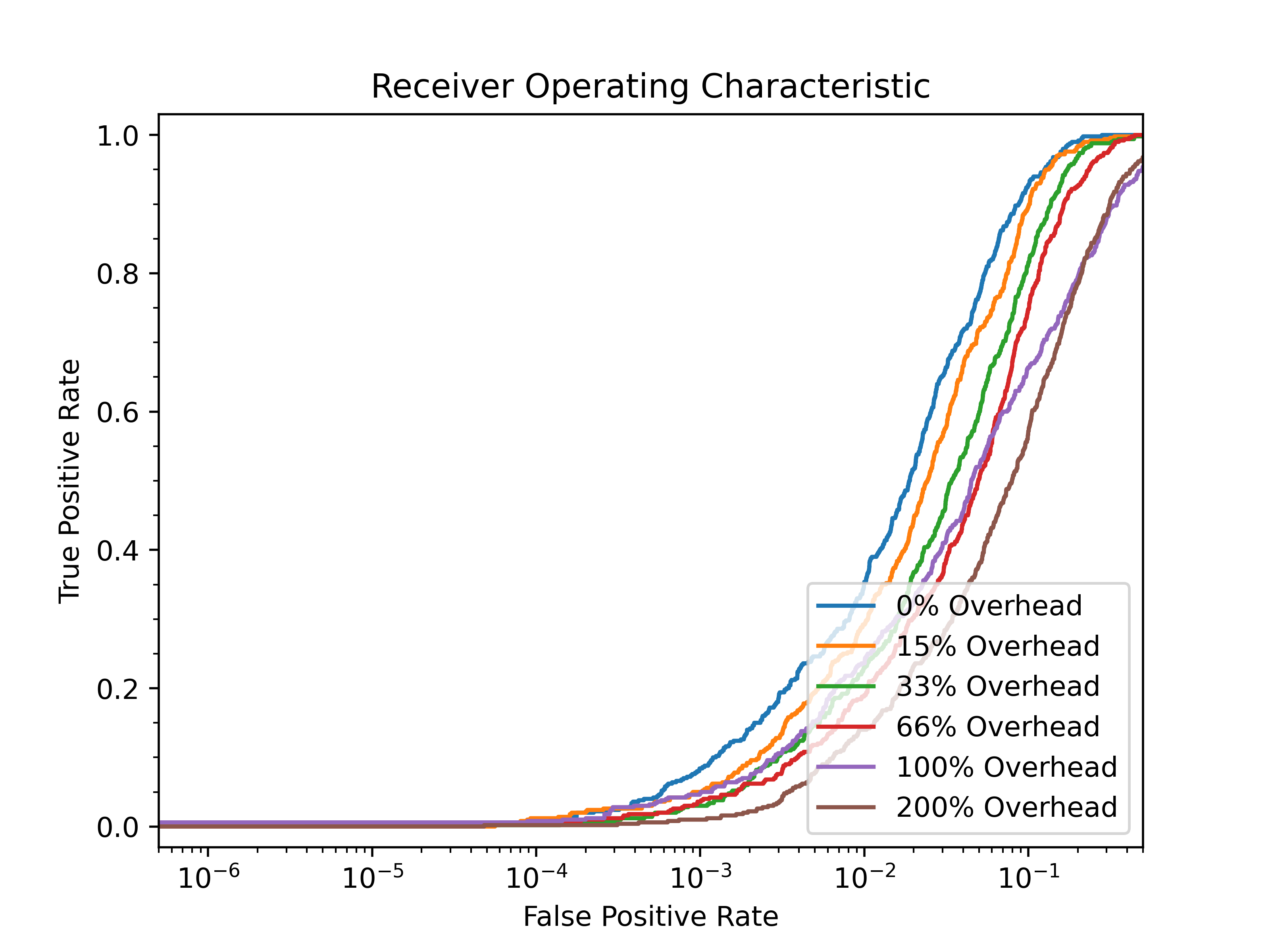}
        \caption{DCF against heavy delays}
        \label{fig:ssid-obfs4-dcf}
    \end{subfigure}
    \caption{ROC curves for (a) \espresso\ and (b) DCF on the SSH-only
    dataset under padding with heavy delays in the \textit{network-mode}
    detection scenario.}
    \label{fig:ssid-obfs4}
\end{figure}

The following analysis evaluates the impact of these obfuscation
techniques on the SSH dataset in the network-mode detection scenario.
The performance degradation for \espresso\ and DCF is shown in
Figures~\ref{fig:ssid-obfs1} through~\ref{fig:ssid-obfs4}.

Our results demonstrate that \espresso\ is highly resilient to
padding-only defenses. As shown in Figure~\ref{fig:ssid-obfs1}, even
with a 200\% padding overhead (tripling the number of packets),
correlation performance sees only a modest decrease. For instance, at
100\% overhead, \espresso's max TPR at an FPR of $10^{-5}$ is 0.496, a
noticeable but not catastrophic drop from the baseline of 0.710. DCF,
with its lower initial performance, is more significantly impacted by
padding alone.

The introduction of timing perturbations, however, proves to be a far
more effective countermeasure. The ``light delays (v1)'' profile (25\%
probability) combined with padding already causes significant
performance degradation for both models, as seen in
Figure~\ref{fig:ssid-obfs2}. At 100\% padding overhead, \espresso's
max TPR at FPR $10^{-5}$ falls to 0.168. The ``light delays (v2)''
profile (50\% probability) degrades performance even further
(Figure~\ref{fig:ssid-obfs3}), reducing \espresso's max TPR to 0.054
under the same conditions.

The ``heavy delays'' profile (75\% probability) is devastatingly
effective, as illustrated in Figure~\ref{fig:ssid-obfs4}. When
combined with 100\% padding overhead, this strategy almost completely
nullifies \espresso's correlation capabilities, reducing its max TPR at
an FPR of $10^{-5}$ to just 0.004. This dramatic performance collapse
under severe timing jitter provides insight into why the DNS-tunneled
traffic is so difficult to correlate. The heavy delays artificially
recreate the disruptive effect of the DNS tunnel's polling-based
communication, confirming that radical timing alterations are the most
effective method for breaking traffic correlation systems like
\espresso.

\section{Discussion}
\label{sec:discussion}

\paragraphX{Key Findings.}
Our experiments establish several clear conclusions about the
application of deep learning flow correlation to stepping-stone
intrusion detection.
First, \espresso\ substantially and consistently outperforms the DCF
baseline across all five tunneling protocols and in both detection
modes, demonstrating that the combination of full-sequence
context and time-aligned interval features is the primary driver
of correlation performance---rather than any specific architectural
choice.
Second, the nature of the tunneling protocol has an outsized impact on
detectability: standard bursty protocols (SSH, SOCAT, ICMP) are highly
correlatable, while the polling-based behavior of DNS tunneling
fundamentally disrupts the temporal patterns that correlation methods
rely on.
Third, chain length prediction is feasible and operationally useful,
with the standalone CNN model offering robust estimates even in the
DNS case where flow correlation fails entirely.
Fourth, and most critically, timing-based perturbations are far more
effective at defeating correlation than volume-based padding, pointing
to a specific and actionable vulnerability that defenders and future
researchers must address.

\paragraphX{The Dominance of Timing Features.}
Our obfuscation experiments revealed a critical insight: \espresso's
performance is far more sensitive to timing perturbations than to
volume-based padding. Even a massive 200\% increase in packet count
had a less detrimental effect than modest, probabilistic delays.
This finding, combined with the poor performance on DNS-tunneled
traffic, strongly suggests that \espresso\ relies heavily on timing
patterns. An adversary aware of this could focus their efforts on
introducing jitter to evade detection effectively. Future work should
explore architectures that are inherently more robust to timing noise
or that can explicitly model and filter out such perturbations.

\paragraphX{Practical FPR Implications.}
The introduction motivated the stringent FPR requirements for
stepping-stone detection with a concrete scenario: in a network
observing one million connection pairs per hour with only ten true
stepping-stone chains, a 1\% FPR would produce approximately 10,000
false alarms.
It is worth closing the loop by examining what the FPR levels achieved
in our experiments mean in this context.
At the FPR threshold of $10^{-5}$---the tightest threshold in our
tables---\espresso\ achieves a TPR of 0.710 on SSH traffic and 0.790
on the mixed-protocol dataset in network-mode detection.
At this FPR, the same one-million-pair gateway would generate only
approximately 10 false positives per hour while catching the majority
of true intrusions.
For well-structured bursty protocols, the situation is even more
favorable: the SOCAT and ICMP datasets yield perfect TPR at
$10^{-5}$ FPR, and the DNS-excluded mixed-protocol setting
achieves TPR $>$ 0.95.
These results suggest that for networks where attackers use standard
tunneling protocols, the system is operationally viable.
The DNS case remains a meaningful gap: at any FPR threshold below
$10^{-3}$ in network-mode, detection collapses.
A realistic deployment would therefore benefit from combining
\espresso\ with protocol-aware pre-filtering that routes DNS traffic to
specialized detection logic.

\paragraphX{Limitations of the Synthetic Dataset.}
The most significant limitation of this work is its exclusive reliance
on synthetic data. While our collection tool was carefully
parameterized using empirical distributions derived from real-world
traffic, several important gaps remain.
Most critically, all negative examples in our training and evaluation
sets are drawn from other stepping-stone chains generated by the same
simulator. In a real network, the vast majority of traffic is benign
background activity---web browsing, file transfers, video streams,
database queries---and the model has never been trained to distinguish
correlated attack traffic from this diverse population.
It is plausible that benign flows exhibiting bursty, interactive
patterns (e.g., SSH administrative sessions, database connections)
could produce spuriously high correlation scores, inflating the
effective FPR in deployment beyond what our controlled experiments
suggest. A rigorous evaluation in a realistic environment would require
capturing true benign negative traffic alongside synthetic attack
chains, which we leave as an important direction for future work.
A secondary concern is the fidelity of the traffic generation model
itself. Our simulator parameterizes burst sizes and inter-burst
intervals from a capture-the-flag competition dataset, which may not
fully represent the range of behaviors exhibited by sophisticated,
long-running intrusion campaigns (e.g., slow-and-low exfiltration,
automated lateral movement, or C2 beaconing). The Docker-based
simulation also abstracts away network effects present in real
deployments, such as asymmetric routing, variable MTU, and
real-stack TCP retransmissions.

\paragraphX{Comparison Scope.}
Our primary deep learning baseline is DCF~\cite{oh2022deepcoffea}, and
our ablation introduces a Modified DCF variant. We did not benchmark
against traditional passive SSID methods such as Poisson-based
approaches~\cite{blum2004_PoissonRWSSID,he2007_PoissonSSID} or
wavelet-based
methods~\cite{donoho2002_multiscaleSSID}. The rationale is twofold.
First, these methods were not designed for the low-FPR operating regime
that is the focus of this work; they typically report TPR and FPR at a
single threshold rather than across a continuous ROC curve, making a
direct comparison of their performance in our evaluation setting
difficult to construct fairly.
Second, the prior literature has already established that deep
learning methods substantially outperform classical statistical
approaches on flow correlation
tasks~\cite{Nasr18DeepCorr,oh2022deepcoffea}, and DCF is the
strongest published representative of the deep learning family.
Benchmarking against classical methods would be informative for
completeness but is unlikely to change the qualitative conclusions.
We acknowledge this as a limitation and encourage future work to
establish a broader comparison, particularly in settings where dataset
or computational constraints may favor simpler baselines.

\paragraphX{Scalability and Deployment.}
Deploying a neural flow correlation system on a live network raises
practical scalability concerns that our experiments do not fully
address. The core challenge is that the number of candidate flow pairs
grows quadratically with the number of concurrent connections observed
at a network monitor. \espresso's embedding-based architecture partially
mitigates this: flows are encoded into fixed-size vectors in a single
forward pass, and correlation is then reduced to a distance computation
in embedding space, which is much cheaper than running the FEN on every
pair. Nevertheless, the FEN itself is a 9-layer transformer that
processes sequences of 1,200 time bins; embedding a single flow
requires a non-trivial amount of computation. In our training setup,
an optimization step over a batch of 64 samples takes approximately
253\,ms on an NVIDIA RTX 2080 Ti. A careful analysis of throughput,
latency, and memory requirements in a realistic deployment
environment---and the potential use of quantization or knowledge
distillation to reduce model
size~\cite{shen2021efficient,wang2020linformer}---is an important
direction that we leave to future work.

\paragraphX{Limitations of Loss Enhancements.}
While the exploration of temporal alignment and feature decorrelation
losses was an interesting exercise, the results were mixed.
Temporal alignment provided a small benefit, but none of the loss
augmentation strategies fundamentally solved the challenges posed by
complex protocols like DNS. This suggests that simply encouraging
different feature properties through auxiliary loss terms may not be
sufficient. A more promising direction could be to design network
architectures that are structurally biased towards the properties of
each protocol, rather than relying on the loss function to guide the
learning process. Further development and refinement of the
multi-model strategy is another direction of exploration; however,
this strategy will still be fundamentally limited by the poor
single-model performance against DNS-tunneled traffic.

\paragraphX{Beyond Per-Connection Correlation.}
Our findings suggest that some protocols and sophisticated obfuscation
techniques may render per-connection correlation practically impossible
in isolation. This reality calls for a strategic shift from a narrow
focus on individual correlated pairs to a broader, more holistic
detection framework. Such a system would treat correlation scores as
just one of many signals in a larger analytical model. Future systems
should aim to integrate data streams across longer timescales and
diverse protocols, looking for patterns of attacker behavior rather
than just traffic similarities. For example, correlating a suspicious
DNS C2 channel with a subsequent SSH login from an unexpected IP could
provide a much stronger signal of intrusion than either event would
alone. This approach moves toward modeling attacker tactics, techniques,
and procedures (TTPs) directly, where flow correlation serves as a
crucial but not solitary piece of evidence.

\paragraphX{Modeling Advanced Attacker Behavior.}
A key limitation of this project lies in our benchmark dataset. Our
dataset contains stepping-stone intrusion chains with relatively
simplistic simulations of attacker behavior. In the current
experiments, stepping stone chains are created statically. However,
sophisticated attackers are likely to engage in more dynamic behaviors,
such as incrementally extending or shrinking the chain over time by
adding or removing intermediate hosts. Additionally, the activities
within a tunnel may vary significantly based on the phase of the attack
(e.g., reconnaissance, data exfiltration). Future work must focus on
simulating these more complex and dynamic attacker behaviors to create
more realistic and challenging benchmarks.

\paragraphX{Robustness Against Adaptive Adversaries.}
A critical challenge in deploying learning-based correlation systems is
the threat of an adaptive adversary who possesses a copy of the
detection model (white-box access) or a reasonable proxy of it
(black-box access). In this scenario, the adversary can utilize the
model as an oracle, iteratively modifying their traffic shaping
strategy (e.g., altering packet timing and sizing) to maximize the
embedding distance between the attacker and target flows. Our
experimental results in Section~\ref{sec:ssid} clearly indicate the
strategy an adversary would likely adopt. We observed that \espresso\
struggles to correlate traffic from DNS tunnels (specifically
\texttt{dnscat2}) compared to bursty protocols like SSH or SOCAT. We
hypothesized that this performance gap is driven by the ``polling''
nature of the DNS tunneling protocol, which transmits data in highly
periodic, non-bursty intervals. An adaptive adversary, therefore, need
not invent entirely new obfuscation techniques; they could simply
reshape their traffic to mimic these high-regularity, periodic patterns
to evade detection. While the defender could attempt to build
adversarial training strategies to gain some robustness, such an
adversary is likely to be exceptionally difficult to detect using
traffic correlation methods, as evidenced by our DNS results. Instead,
these advanced obfuscation techniques are perhaps better addressed by
anomaly detection methods. The distribution of traffic produced by
standard usage of protocols like SSH and DNS is likely to appear
distinctively different than what is produced when a layer of
regularizing traffic obfuscation is applied. This suggests that a
holistic defense strategy must pair flow correlation with statistical
anomaly detection to effectively cover the blind spots created by
highly regularized traffic shaping.

\section{Conclusion}
\label{sec:conclusion}

Stepping-stone intrusions remain one of the most effective and
underappreciated evasion techniques available to network attackers.
By routing malicious sessions through chains of compromised
intermediary hosts, adversaries can make forensic attribution
extremely difficult---yet the problem has received comparatively
little attention from the deep learning community. A key reason is
the absence of suitable large-scale datasets and the demanding
operational requirement that any practical detector must achieve
extremely low false positive rates to avoid burying true alerts in
noise.

In this paper, we addressed both obstacles.
We developed a flexible synthetic data collection tool that simulates
stepping-stone attack chains across five tunneling protocols---SSH,
SOCAT, ICMP, DNS, and mixed multi-protocol chains---using
traffic distributions derived from real-world captures, producing
the largest labeled dataset for this problem to date.
We then applied \espresso, a transformer-based deep learning flow
correlation model that combines time-aligned interval features,
full-sequence context, and online triplet learning, to the
stepping-stone detection task.
Our baseline experiments demonstrated that \espresso\ substantially
outperforms the state-of-the-art DeepCoFFEA (DCF) method across all
five protocols and in both host-mode and network-mode detection
settings, establishing a new performance benchmark for this domain.
In the most challenging network-mode setting, \espresso\ achieves a
TPR exceeding 0.99 at an FPR of $10^{-3}$ for SSH, SOCAT, and
ICMP traffic, and remains competitive on the more difficult
mixed-protocol dataset.

Beyond flow correlation, we demonstrated that chain length prediction
is both feasible and operationally meaningful, providing defenders
with a tool for distinguishing benign administrative pivoting from
multi-hop attack chains. A standalone CNN trained on packet-level
features proved most effective for this task, while a full-chain
reconstruction approach offered high-confidence path tracing for
protocols where correlation signals are strong.
Our investigation into adversarial robustness revealed that \espresso\
is highly resilient to packet padding but is significantly degraded by
timing-based perturbations---a vulnerability that mirrors its
difficulty with DNS-tunneled traffic and confirms that fine-grained
temporal features are both the primary strength and the primary attack
surface of the model.
Temporal alignment loss augmentation provided modest but consistent
improvements in the most challenging settings at negligible
computational cost.

The central challenge limiting further progress in this domain is the
absence of real-world traffic to validate synthetic benchmarks and
guide robust model development.
Our evaluations operate entirely in a closed world of synthetic attack
traffic; in realistic deployments, the model must distinguish
stepping-stone flows from a rich background of benign interactive
sessions that have never been seen during training.
Closing this gap---through longitudinal captures in monitored research
networks, collaboration with operational security teams, or advances
in privacy-preserving data sharing---is the most impactful step the
community can take.
We hope that the dataset generation tools, model, and evaluation
framework introduced here serve as a foundation for that work.

\bibliographystyle{IEEEtran}
\bibliography{references}

\appendices


\section{ESPRESSO on Tor Traffic Correlation}
\label{app:tor}

\espresso\ was originally developed and evaluated as a flow correlation
attack against the Tor anonymity network, where the goal is to link a
user's ingress connection (entering a Tor guard node) with the
corresponding egress connection (exiting toward the destination
server) in order to deanonymize the user.
This appendix documents that original evaluation, which established
\espresso's performance characteristics and directly motivated its
application to stepping-stone detection in the main body of this paper.
The model architecture---including the time-interval feature
representation, transformer-based FEN, post-FEN windowing, online
triplet mining, and MLP classification head---is described in full in
Section~\ref{sec:espresso}.

\subsection{Evaluation Methodology}

\subsubsection{Metrics}
Our primary metric for evaluating correlation performance is the True
Positive Rate (TPR) vs.\ False Positive Rate (FPR) curve, also known
as the Receiver Operating Characteristic (ROC) curve. To emphasize
performance at extremely low FPR values, which are critical for
practical flow correlation applications, we plot ROC curves using a
logarithmic scale on the FPR axis. This visualization makes it easier
to discern model performance in the low-FPR regime.

In addition to the ROC curve, we use two specialized metrics:
\begin{itemize}
    \item \textbf{Maximum TPR at a Maximum FPR Threshold:} The maximum
    true positive rate the model achieves when the false positive rate
    is constrained to a predefined threshold. This metric is
    particularly useful for comparing models where minimizing FPR is
    of paramount importance.
    \item \textbf{Partial Area Under the Curve (pAUC):} The area under
    the ROC curve for FPR values below a specified maximum threshold.
    By focusing on the low-FPR sections of the ROC, this metric allows
    more precise comparisons between models in scenarios where false
    positive rates must be kept extremely low.
\end{itemize}

\subsubsection{Dataset}
We evaluate on the same dataset collected and used in the DCF
paper~\cite{oh2022deepcoffea}. This dataset consists of network
traffic flows collected using a Tor client and an SSH proxy to capture
both ingress and egress flows. From this dataset, we use the data
collected in June 2022, using 8,662 flow pairs for training, 764 pairs
for validation, and 811 pairs entirely unseen for testing.

In addition to the original DCF dataset, we simulate the TrafficSliver
defense with two splits applied to the ingress flows. This defense
introduces a more challenging evaluation scenario by splitting the
ingress traffic across two independent network paths, making it more
difficult for the correlation model to accurately associate incoming
and outgoing flows. By including this defense mechanism, we ensure that
\espresso\ is tested against both standard and adversarial network
conditions.

\subsection{Results \& Analysis}

We present the results of evaluating \espresso\ compared to DCF and
the architectural ablations GreenTea and HotWater. The results are
demonstrated through ROC curves with a particular emphasis on
performance at very low FPR rates, and summarized in
Table~\ref{tab:max-tpr-pauc-2}.

\subsubsection{ROC Curve Analysis}

\begin{figure}[t]
  \centering
  \includegraphics[width=0.8\linewidth]{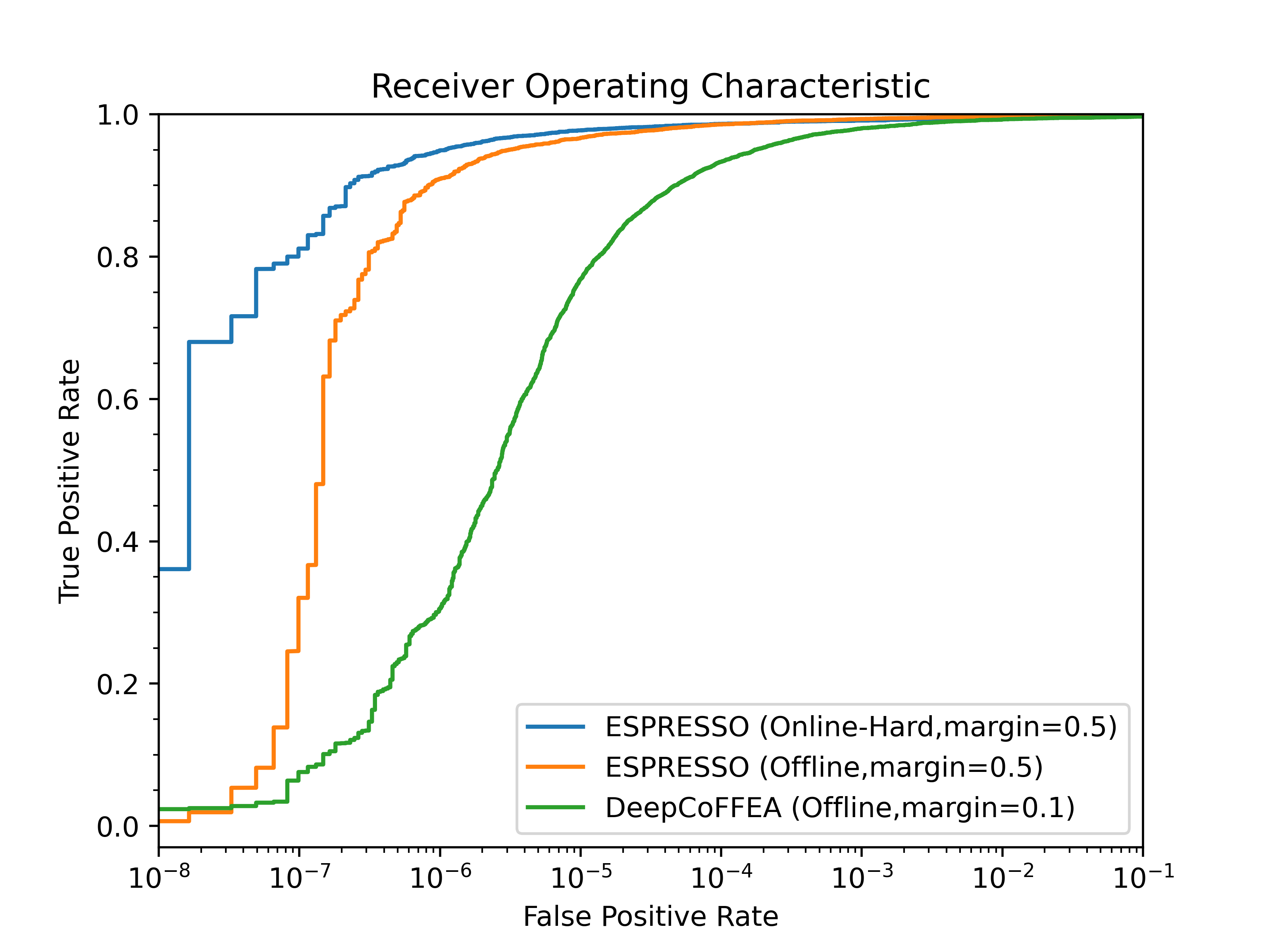}
  \caption{ROC curves comparing \espresso\ (online mining), \espresso\
  (offline mining), and DeepCoFFEA.}
  \label{fig:multi-roc}
\end{figure}

\paragraphX{Comparison of ESPRESSO and DeepCoFFEA.}
Figure~\ref{fig:multi-roc} shows the ROC curves comparing \espresso\
(trained with online and offline mining) against DeepCoFFEA.
The results demonstrate that \espresso\ significantly outperforms DCF,
particularly at very low FPRs (below $10^{-6}$). The online training
strategy combined with hard triplet mining gives \espresso\ an
advantage in both efficiency and correlation accuracy. Notably,
\espresso\ achieves a TPR of over 0.9 at FPR values below $10^{-5}$,
while DCF lags behind in both the offline and online mining setups.

\begin{figure}[t]
  \centering
  \includegraphics[width=0.8\linewidth]{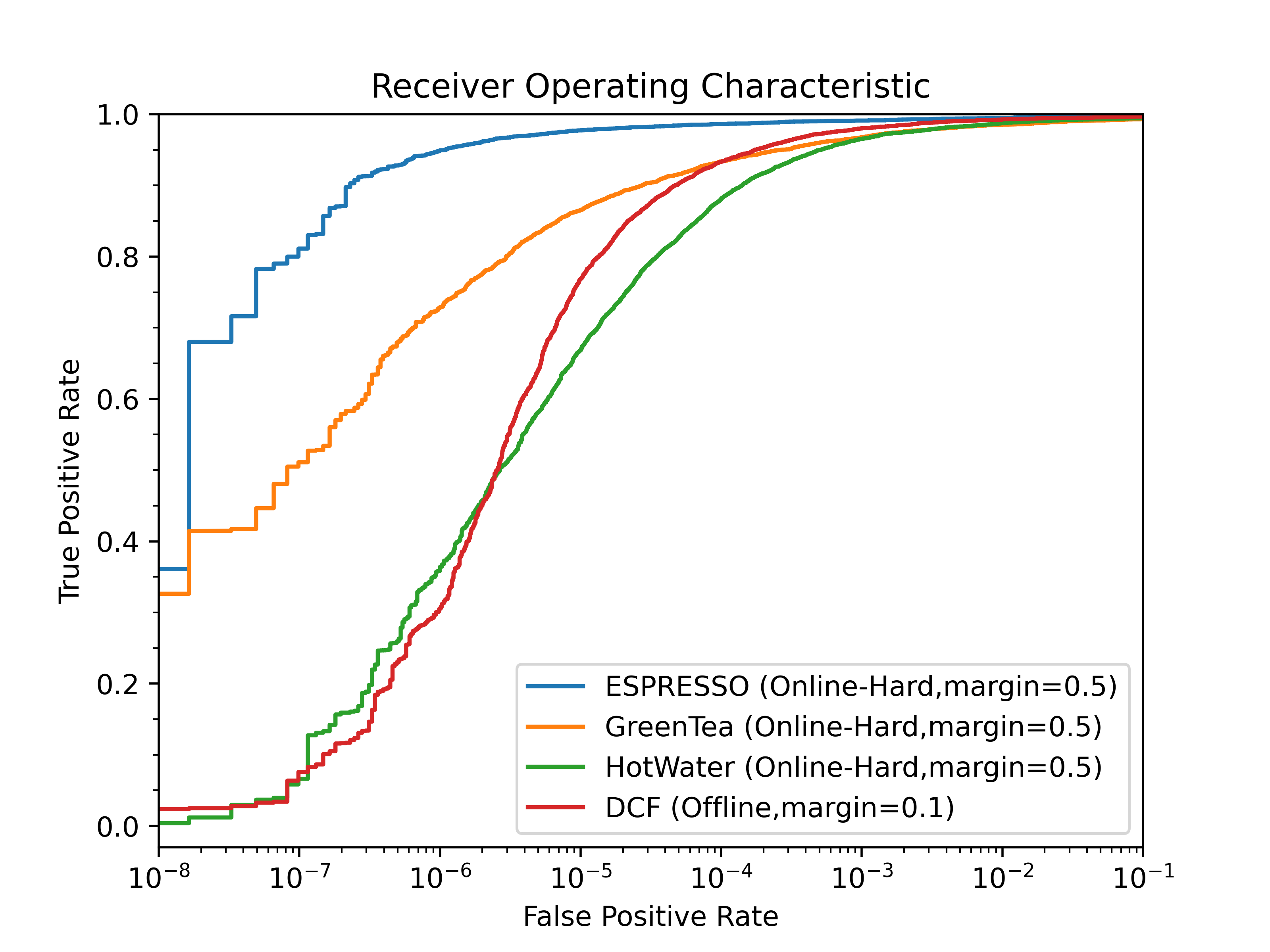}
  \caption{ROC curves comparing \espresso, GreenTea, HotWater, and
  DeepCoFFEA models.}
  \label{fig:arch-roc}
\end{figure}

\paragraphX{Impact of Architecture.}
Figure~\ref{fig:arch-roc} compares the architectures of \espresso\
(transformer backbone), GreenTea (local convolutional mixing), HotWater
(identity mixing), and DCF (CNN backbone, isolated windowing strategy).
\espresso\ outperforms all other architectures. Interestingly, GreenTea
performs competitively at very low FPRs, achieving performance close to
\espresso\ below an FPR of $10^{-8}$. HotWater and DCF perform
similarly, with both models struggling to achieve a TPR of over 0.5 at
very low FPR values. It is notable that the HotWater variant does as
well as it does, given the complete lack of spatial feature processing
inside the transformer blocks.

\begin{figure}[t]
  \centering
  \includegraphics[width=0.8\linewidth]{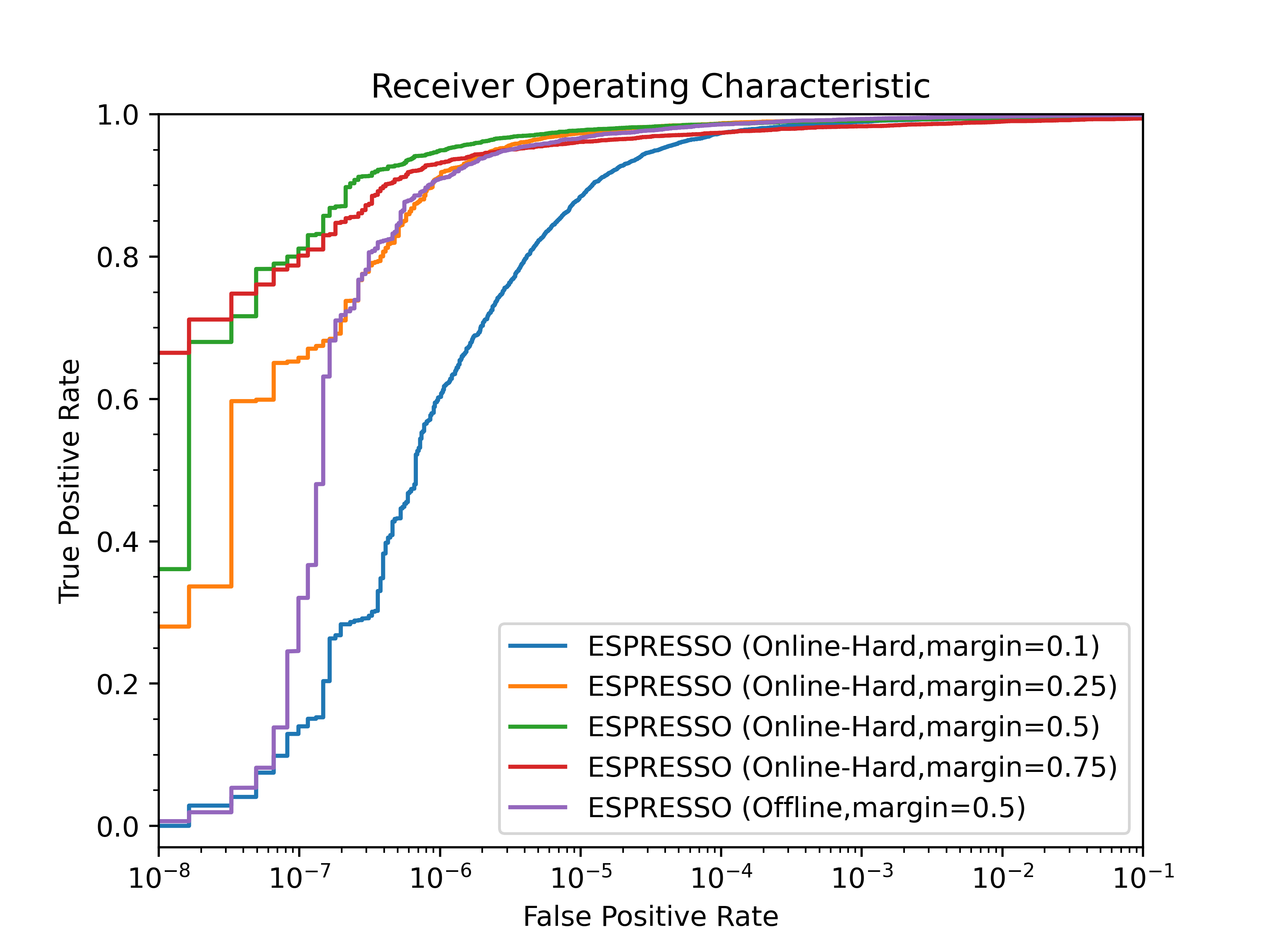}
  \caption{ROC curves comparing \espresso\ trained with different loss
  margins in the online mining scenario.}
  \label{fig:espresso-roc}
\end{figure}

\paragraphX{Effect of Loss Margins on ESPRESSO.}
Figure~\ref{fig:espresso-roc} shows the performance of \espresso\ when
trained using online mining with varying loss margins ($0.1$, $0.25$,
$0.5$, and $0.75$) compared to \espresso\ trained in offline mode with
a margin of $0.5$. A margin of $0.5$ yields the best overall
performance. The smallest margin ($0.1$) performs significantly worse,
highlighting the importance of selecting an appropriate margin for
correlation tasks. \espresso\ with a margin of $0.75$ also shows strong
results, suggesting that larger margins are generally more effective
for this task.

\begin{figure}[t]
  \centering
  \includegraphics[width=0.8\linewidth]{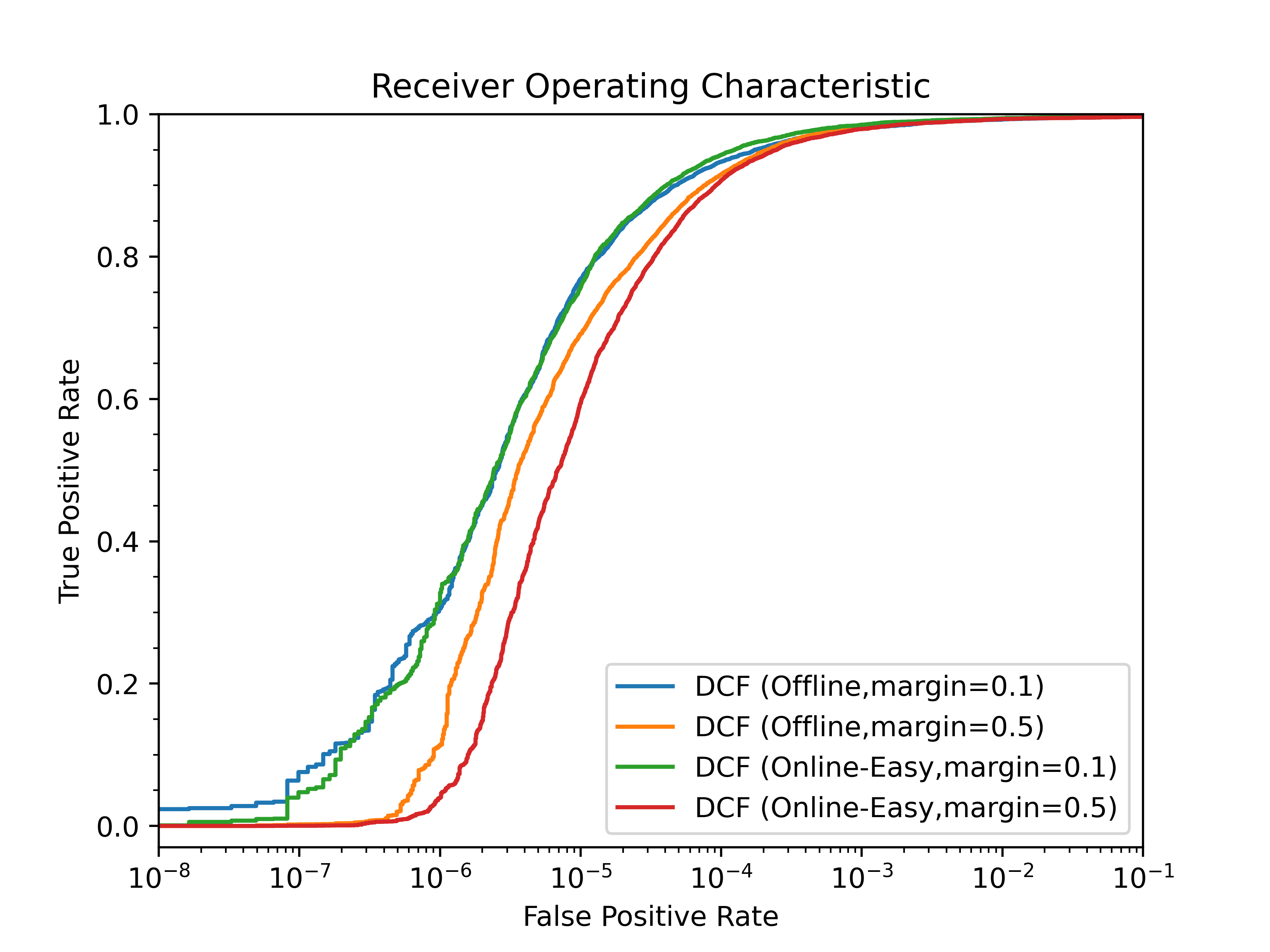}
  \caption{ROC curves comparing DeepCoFFEA trained with various
  margins and mining strategies.}
  \label{fig:dcf-roc}
\end{figure}

\paragraphX{DCF Training Analysis.}
Given the significant impact that loss margin has on \espresso, we
re-evaluate DCF using variable loss margins in both the online and
offline mining strategies (Figure~\ref{fig:dcf-roc}). Online
batch-hard mining was completely ineffective when training DCF, with
the model unable to learn at all---likely because the selected triplets
are too difficult at the per-window level. We therefore use the easier
batch-all strategy for DCF.  Both online and offline training perform
similarly, though the smaller margin of $0.1$ offers a slight advantage
over $0.5$, particularly at low FPR. This shows that DCF, which lacks
the global context-awareness of \espresso, benefits more from a tighter
loss margin. We hypothesize that with the limited information available
within any given window, DCF fails to effectively use the greater
distance ranges possible with a larger embedding margin.

\begin{figure}[t]
  \centering
  \includegraphics[width=0.8\linewidth]{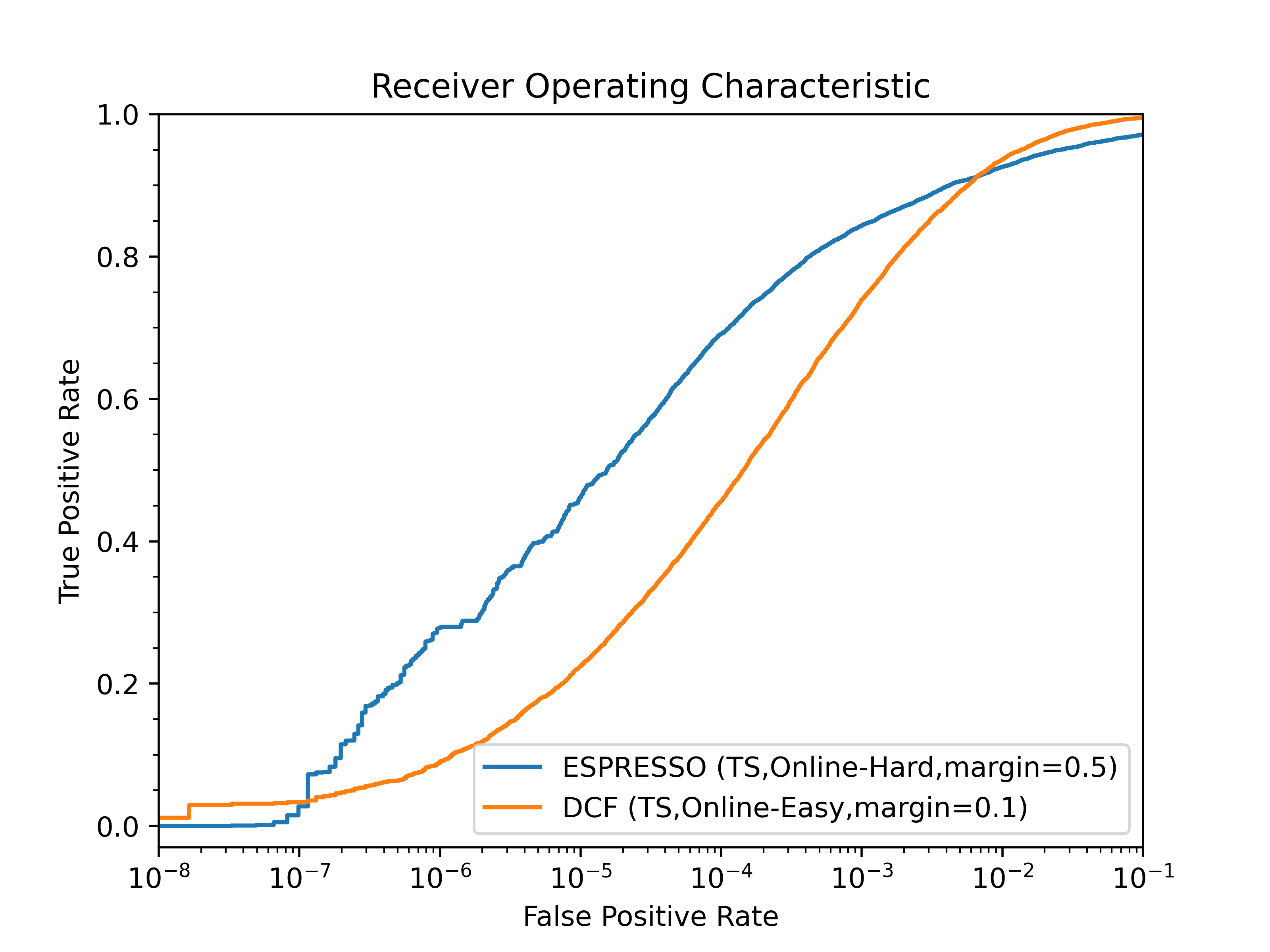}
  \caption{ROC curves comparing \espresso\ and DeepCoFFEA when
  benchmarked on the TrafficSliver dataset.}
  \label{fig:ts-roc}
\end{figure}

\paragraphX{TrafficSliver Defense Results.}
Figure~\ref{fig:ts-roc} compares the ROC curves of \espresso\ and DCF
against the TrafficSliver dataset. The TrafficSliver defense
significantly degrades correlation performance for both models.
However, \espresso\ demonstrates resilience, achieving a TPR of over
0.2 at an FPR of $10^{-6}$, indicating that non-trivial vulnerability
remains even when up to half of the network traffic is absent in the
ingress flow. This result underscores the strength of \espresso's
correlation model even under adversarial conditions.

\subsubsection{Summary Table}

\begin{table*}[]
    \centering
    \begin{tabular}{ccc|cc|cc|cc}
    \toprule
    \toprule
    \multirow{2}{*}{\textbf{Model}} & \textbf{Training} & \textbf{Loss} & \multicolumn{2}{c|}{FPR $\leq 10^{-5}$} & \multicolumn{2}{c|}{FPR $\leq 10^{-6}$} & \multicolumn{2}{c}{FPR $\leq 10^{-7}$} \\
    & \textbf{Mode} & \textbf{Margin} & \textbf{TPR} & \textbf{pAUC} & \textbf{TPR} & \textbf{pAUC} & \textbf{TPR} & \textbf{pAUC} \\
    \midrule
    DCF & Offline       & 0.1  & 0.768 & 0.584 & 0.304 & 0.197 & 0.076 & 0.034 \\
    DCF & Online        & 0.1  & 0.756 & 0.581 & 0.313 & 0.177 & 0.047 & 0.012 \\
    DCF & Offline       & 0.5  & 0.691 & 0.491 & 0.112 & 0.038 & 0.002 & 0.001 \\
    DCF & Online        & 0.5  & 0.594 & 0.357 & 0.039 & 0.011 & 0.000 & 0.000 \\ \midrule
    ESPRESSO & Online   & 0.1  & 0.884 & 0.762 & 0.602 & 0.384 & 0.139 & 0.061 \\
    ESPRESSO & Online   & 0.25 & \textbf{0.973} & \textit{0.900} & \textit{0.912} & \textit{0.783} & 0.657 & 0.511 \\
    ESPRESSO & Online   & 0.5  & \textbf{0.977} & \textbf{0.922} & \textbf{0.949} & \textbf{0.885} & \textbf{0.811} & \textit{0.678} \\
    ESPRESSO & Online   & 0.75 & \textbf{0.961} & \textbf{0.924} & \textbf{0.931} & \textbf{0.854} & \textbf{0.801} & \textbf{0.731} \\
    ESPRESSO & Offline  & 0.5  & \textbf{0.967} & \textbf{0.931} & \textit{0.909} & \textit{0.733} & 0.320 & 0.090 \\ 
    \midrule
    GreenTea & Online   & 0.5  & 0.865 & 0.805 & 0.728 & 0.630 & 0.511 & 0.425 \\
    HotWater & Online   & 0.5  & 0.668 & 0.538 & 0.359 & 0.238 & 0.066 & 0.030 \\
    \bottomrule
    \bottomrule
    \end{tabular}
    \caption{Maximum TPR and pAUC for FPR thresholds of $10^{-5}$, $10^{-6}$, and $10^{-7}$.}
    \label{tab:max-tpr-pauc-2}
\end{table*}

\paragraphX{Analysis of Max TPR and pAUC.}
From Table~\ref{tab:max-tpr-pauc-2}, \espresso\ consistently
outperforms DCF and other variants, particularly at very low FPR
thresholds. \espresso\ trained with a margin of 0.5 and online
batch-hard mining achieves the highest Max TPR values at all FPR
levels. At an FPR threshold of $10^{-7}$, \espresso\ achieves a Max
TPR of 0.811 and a pAUC of 0.678, while DCF (offline) only achieves a
Max TPR of 0.076 and a pAUC of 0.034.

The choice of margin significantly impacts performance. \espresso\ with
larger margins (0.5 and 0.75) consistently yields better results than
with smaller margins, whereas DCF struggles to maintain performance at
low FPR thresholds with a larger margin of 0.5. At an FPR threshold of
$10^{-7}$, a loss margin of 0.75 has a small advantage over 0.5 in the
pAUC score, indicating that even higher loss margins may be useful when
targeting the lowest possible FPR.

The GreenTea variant performs competitively at lower FPR thresholds,
achieving a Max TPR of 0.511 at $10^{-7}$, but still lags behind
\espresso. HotWater exhibits much lower performance across all metrics,
confirming the benefit of global self-attention for capturing
long-range dependencies in traffic sequences. These findings
highlight the importance of both margin selection and mining strategy
in optimizing the model's performance in the low-FPR regime.

\subsection{Discussion}

\paragraphX{Generalization Across Different Network Settings.}
While \espresso\ has demonstrated strong performance in controlled
experimental settings, its generalization to more diverse network
conditions remains an open question. In real-world applications,
network traffic can be affected by various factors such as mobile
networks with higher latency, packet loss, or network jitter. These
conditions introduce additional variability that may impact correlation
efficacy. Collecting a more realistic dataset with samples from various
network conditions and locations would be required to fully assess
this.

\paragraphX{Impact of Larger Dataset Sizes.}
As with many machine learning models, the size and diversity of the
training dataset has a significant effect on the model's ability to
generalize to new data. While the dataset used for this evaluation is
substantial, real-world networks will generate far larger and more
diverse data. It remains an open question how \espresso\ will scale
with even larger datasets with greater traffic variety. One possibility
is that the model could plateau or encounter diminishing returns
depending on the variety of traffic patterns encountered.

\paragraphX{Improving Model Efficiency.}
An important direction for further development of \espresso\ is
exploring lightweight versions of the architecture to improve
computational efficiency. Linear attention
mechanisms~\cite{shen2021efficient,wang2020linformer} may offer a
better balance between computational cost and efficacy, making
\espresso\ more suitable for large-scale deployment and real-time
systems.

\section{Supplemental Experiments on the DCF Architecture}
\label{app:dcf_ablation}

In Section~\ref{sec:ssid}, we presented the performance of the Deep
Correlation Framework (DCF) using a fixed input window dimension
optimized for our primary dataset. However, a critical question
regarding the deployment of such models is their sensitivity to window
length---specifically, whether larger input contexts improve
correlation accuracy or simply introduce latency and computational
overhead. Furthermore, while \espresso\ leverages a transformer-based
Feature Extraction Network to process full-length traffic sequences,
the original DCF relied on processing isolated windows using a standard
Convolutional Neural Network inherited from the DeepFingerprinting
method.

This appendix explores these questions by expanding our experimental
evaluation in two directions. First, we assess the impact of scaling
the DCF input window size by factors of $2\times$ and $3\times$.
Second, we explore a structural variant of the DCF framework where we
replace its window-based processing with a ``whole-sequence'' approach.
In this experiment, we modify the DCF method and model---originally
designed for fixed-length individual-window correlation---to serve as a
sequential Feature Extraction Network capable of processing full
traffic streams. This experiment allows us to isolate the impact of the
backbone architecture (CNN vs.\ Transformer) from the input processing
strategy (windowed vs.\ sequential).

\subsection{DeepCoFFEA Window and Framework Modifications}
\label{sec:appendix_dcf_windows}

\subsubsection{Experimental Setup}

To rigorously evaluate the impact of windowing and backbone
architecture, we devised two distinct experimental configurations.

\paragraph{Input Window Scaling}
In our baseline experiments, the DCF model operates on a fixed window
size $W$, which we set to 8s by default. To test sensitivity to this
hyperparameter, we retrained and evaluated the model on window sizes of
$2W$ and $3W$ while keeping the remaining architecture (feature
dimension, depth) constant. The input sequence length was increased by
the same factor. The objective was to observe if a larger receptive
field allows the model to capture long-range dependencies missed by the
shorter baseline window, or if the local burst structures captured by
$W$ are sufficient for robust correlation.

\paragraph{Adapting Deep Fingerprinting as a Feature Extraction Network}
The standard DCF approach processes traffic by slicing flows into
independent windows before feature extraction. In this experiment, we
modified the DCF pipeline to operate more like \espresso: ingesting the
entire traffic sequence at once and generating a corresponding sequence
of feature vectors. To achieve this, we adapted the Deep Fingerprinting
architecture~\cite{Sirinam2018DeepFingerprinting} to serve as the
backbone FEN.

\paragraphX{Architecture Modification.}
The standard DF architecture consists of a stack of convolutional
blocks followed by a flattening layer and a dense classification head.
To adapt this for flow correlation,
we modified it as follows:

\begin{enumerate}
    \item \textbf{Removal of Prediction Head:} We truncated the network
    by removing the final flattening operation and the subsequent dense
    classification layers.
    \item \textbf{Sequence Output:} The model output is taken from the
    final convolutional layer, resulting in a sequence of feature maps
    (e.g., a shape of $L \times 256$, where $L$ is the sequence length
    determined by the input size and pooling ratios).
    \item \textbf{Dimensionality Reduction:} We appended a
    time-distributed linear layer to project the 256-dimensional
    convolutional features down to our target embedding dimension of
    64, allowing the modified DF backbone to output a sequence of
    64-dimensional vectors and treating the convolutional stride as the
    windowing mechanism.
\end{enumerate}

\paragraphX{Input Modalities.}
We evaluated this ``Modified DCF'' using two different input
representations to test the importance of temporal alignment:

\begin{enumerate}
    \item \textbf{Per-Packet Features:} The model was fed raw sequences
    of packet metadata with the standard DCF feature representation
    (e.g., sizes $\times$ direction and inter-arrival times). In this
    modality, the convolutional filters operate on indices of packets.
    Consequently, the output ``windows'' correspond to fixed counts of
    packets rather than fixed durations of time. This causes a lack of
    time alignment between windows when comparing different traces,
    which has severe performance repercussions.
    \item \textbf{Time-Interval Features:} The model was fed traffic
    aggregated into fixed time intervals. In this modality, the
    convolutional filters operate on time bins just as in the standard
    \espresso\ variants, ensuring temporal alignment.
\end{enumerate}

Using these configurations, we run benchmarks on the stepping-stone
datasets within the more difficult \emph{network-mode} detection
setting for each protocol grouping, without any of the loss
augmentation strategies discussed in Section~\ref{sec:ssid}.

\subsubsection{Experimental Results}

\begin{figure*}[]
    \centering
    \begin{subfigure}[b]{0.48\linewidth}
        \centering
        \includegraphics[width=\linewidth]{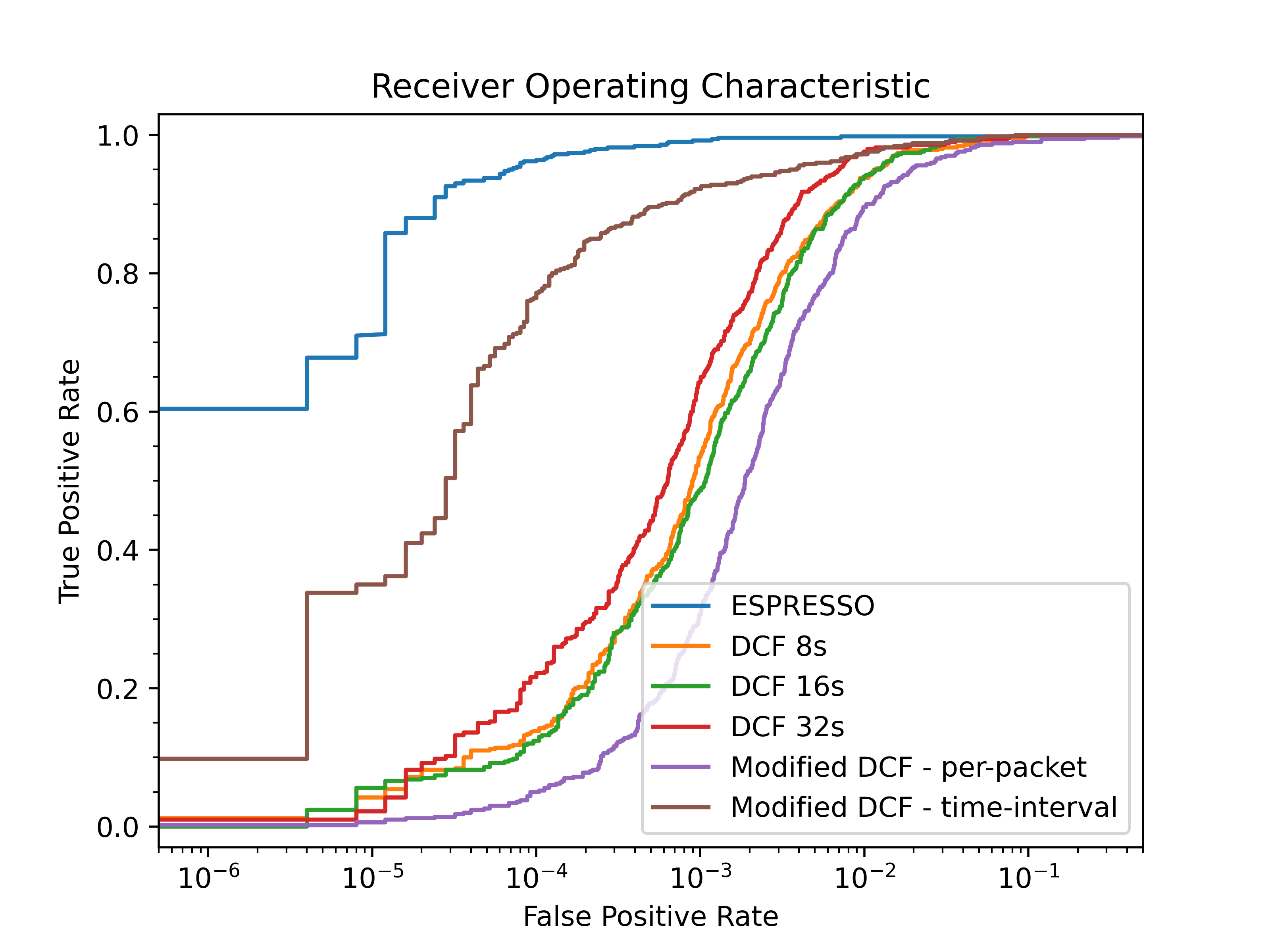}
        \caption{SSH-only ROC}
        \label{fig:ssid-roc-1}
    \end{subfigure}
    \hfill
    \begin{subfigure}[b]{0.48\linewidth}
        \centering
        \includegraphics[width=\linewidth]{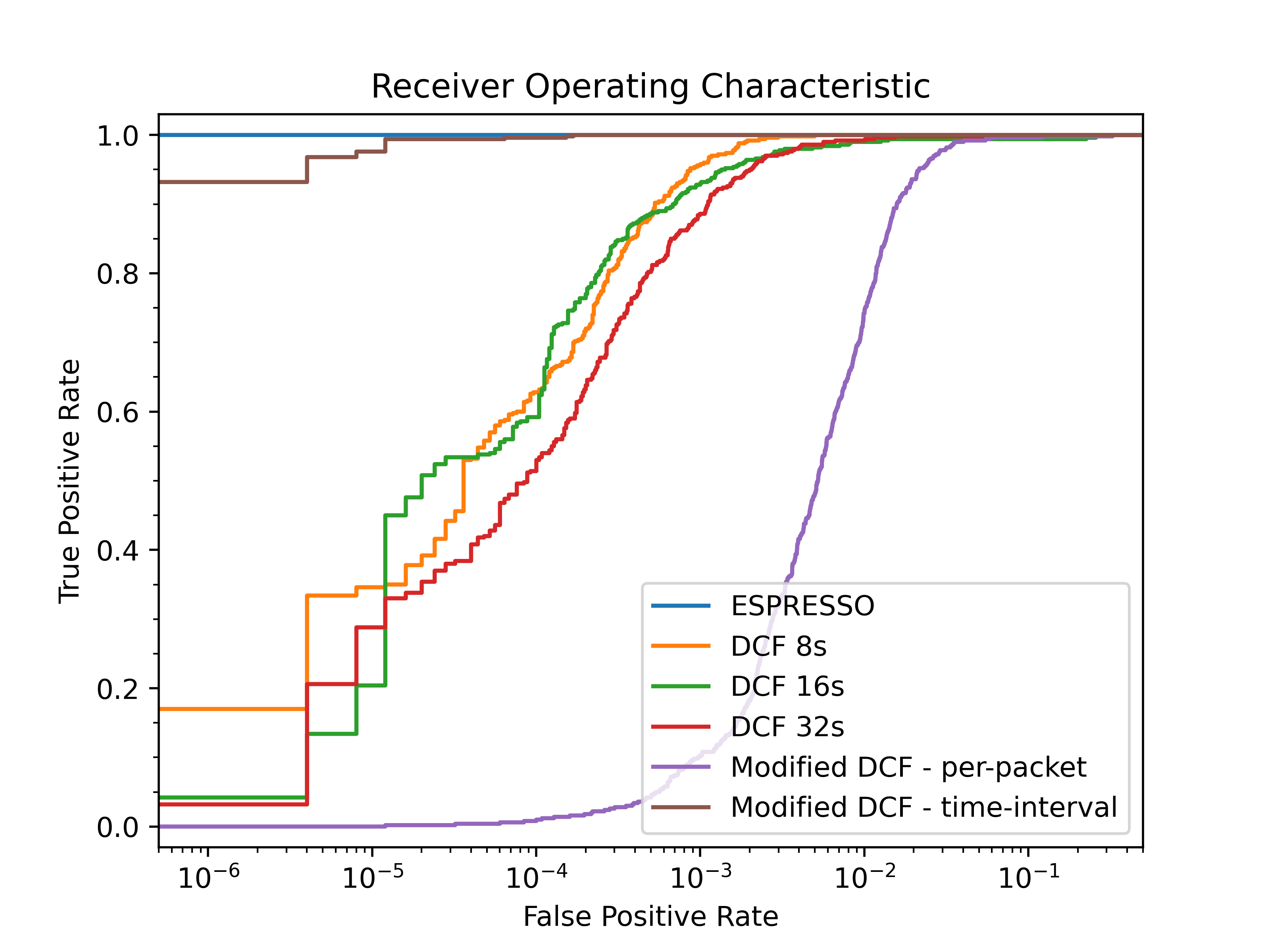}
        \caption{SOCAT-only ROC}
        \label{fig:ssid-roc-2}
    \end{subfigure}


    \begin{subfigure}[b]{0.48\linewidth}
        \centering
        \includegraphics[width=\linewidth]{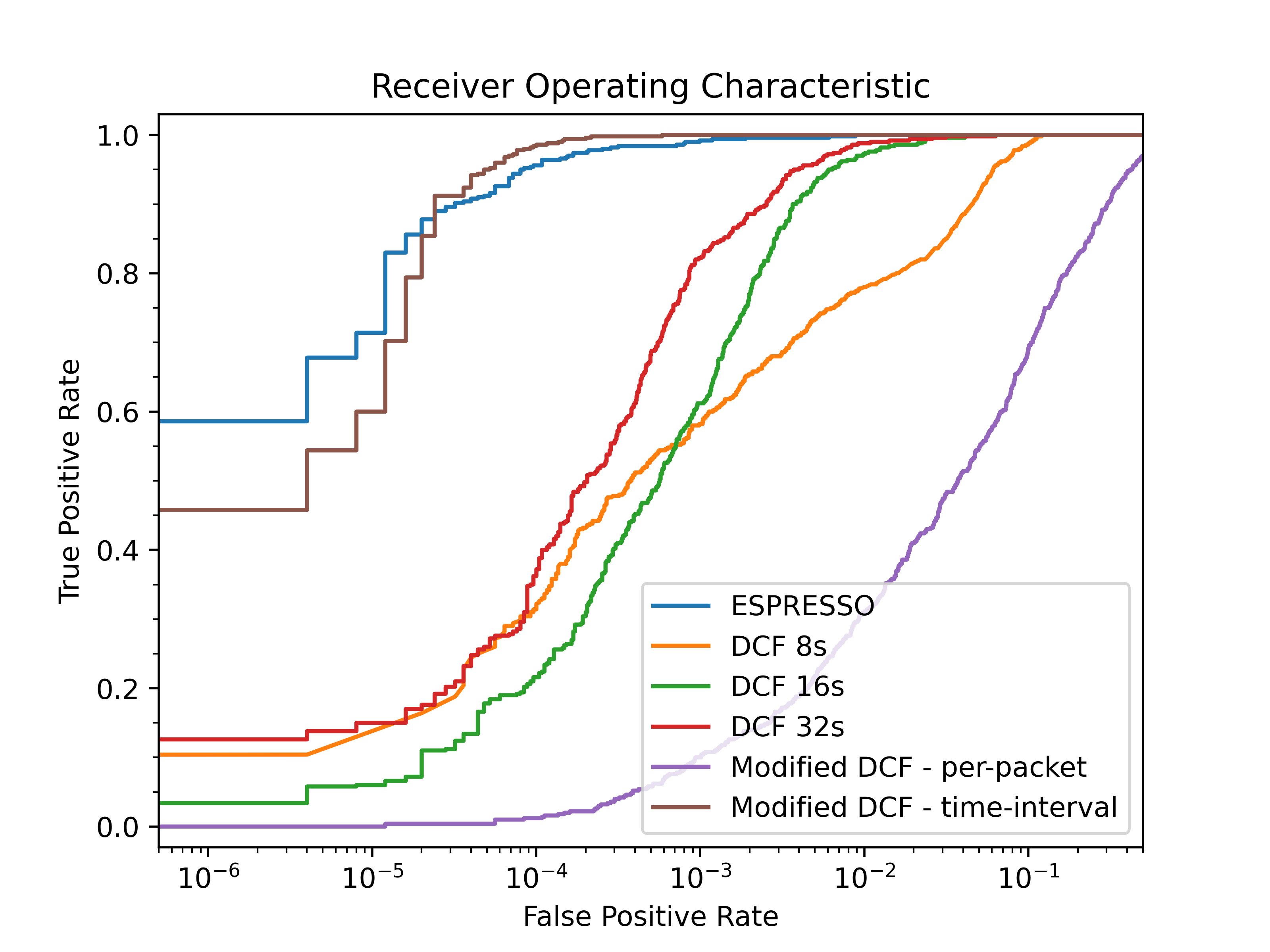}
        \caption{ICMP-only ROC}
        \label{fig:ssid-roc-3}
    \end{subfigure}
    \hfill
    \begin{subfigure}[b]{0.48\linewidth}
        \centering
        \includegraphics[width=\linewidth]{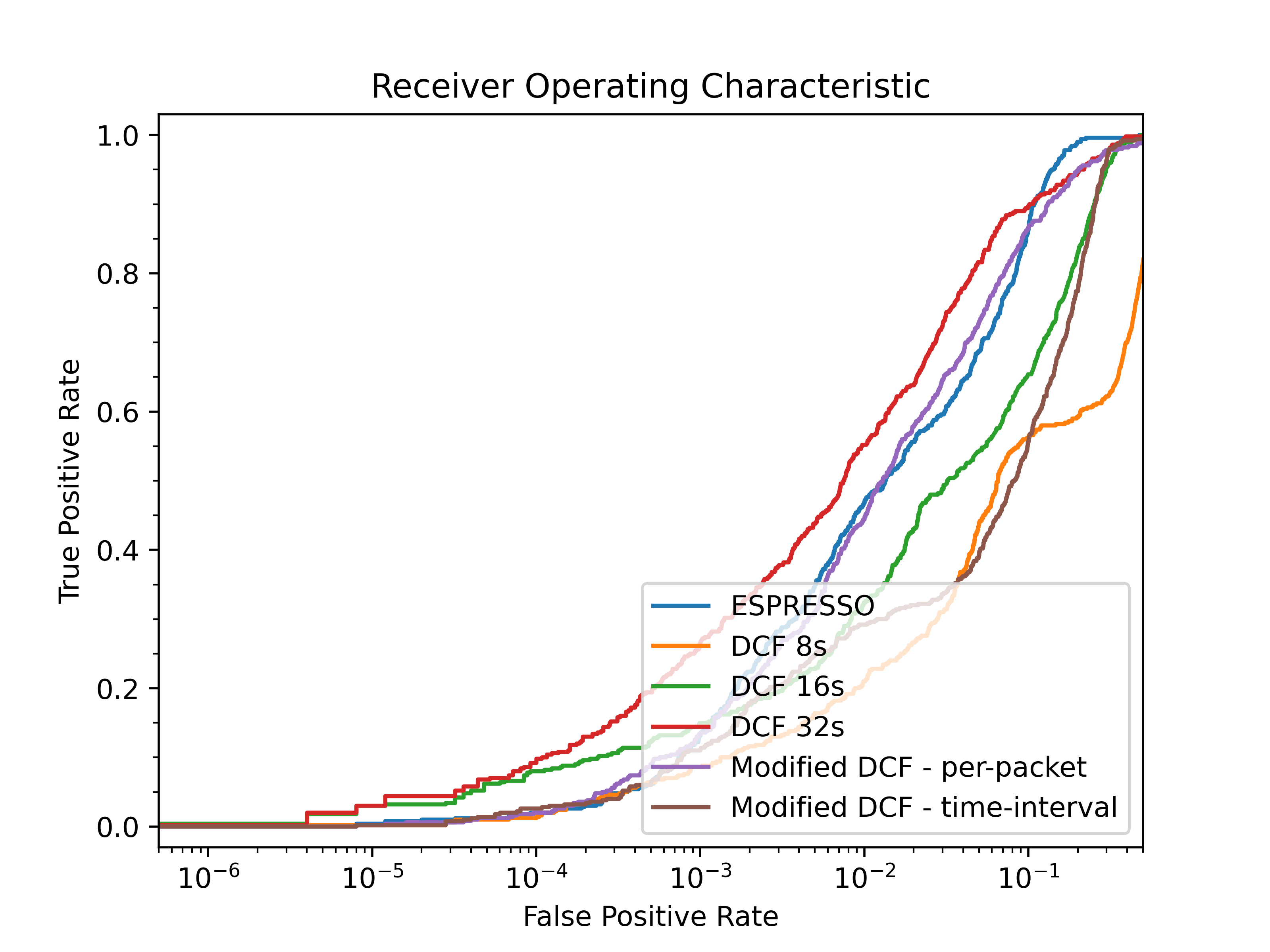}
        \caption{DNS-only ROC}
        \label{fig:ssid-roc-4}
    \end{subfigure}


    \begin{subfigure}[b]{0.48\linewidth}
        \centering
        \includegraphics[width=\linewidth]{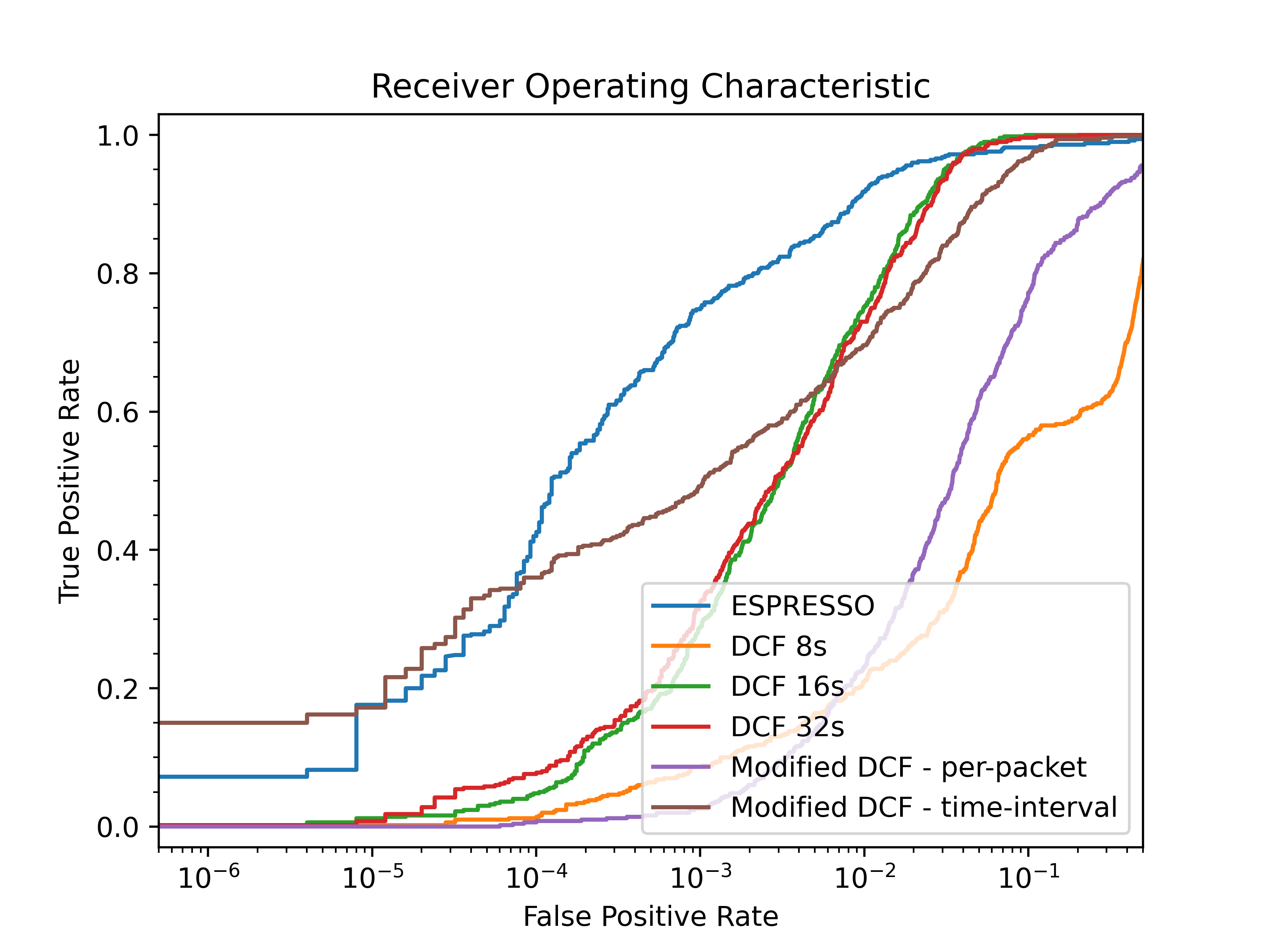}
        \caption{Mixed protocol ROC}
        \label{fig:ssid-roc-5}
    \end{subfigure}

    \caption{ROC curves presenting correlation efficacy across five
    datasets in the \textit{network-mode} detection scenario for DCF
    window scaling and framework modification experiments. \espresso\
    is included as a reference.}
    \label{fig:ssid-net-roc-combined-appendix}
\end{figure*}

Figure~\ref{fig:ssid-net-roc-combined-appendix} presents the ROC
curves for each dataset under each experimental setting, with the
standard \espresso\ configuration included as a reference.

The first observation is that the impact of window size is highly
dataset-dependent. Increasing the window size of DCF yields significant
performance improvements against DNS and ICMP tunneled traffic, a
marginal improvement against SSH, and no improvement against SOCAT.
For the mixed-protocol dataset, longer windows show improvement at
higher TPR targets but are outperformed by the small default window at
lower FPR targets. The most notable outcome is that the 32-second
window size outperforms \espresso\ against DNS-only traffic by a
notable margin; however, the overall performance remains poor and is
likely unsuitable for practical applications.

The Modified DCF configuration with per-packet input features performs
comparably to standard DCF on DNS-only traffic but is substantially
degraded on SOCAT, ICMP, and mixed datasets. This is likely due to
misaligned time windows when using per-packet features, preventing the
model from learning robust representations. When we instead use
\emph{time-interval} features, performance is significantly improved
and is competitive with \espresso\ across most protocols---at the low
end of FPR, this configuration meets or slightly exceeds \espresso's
performance. This indicates that the key drivers of \espresso's
performance are the combination of full-sequence input and
time-interval feature representation, rather than improvements in the
backbone neural network architecture. These results also again
highlight the distinctiveness of DNS tunneled traffic, in which
large-window per-packet feature representations appear key to
maximizing correlation efficacy.

\subsection{Evaluation of Loss Augmentation Strategies on the DCF Architecture}
\label{sec:appendix_dcf_loss}

In Section~\ref{sec:ssid}, we introduced several auxiliary loss
functions---specifically Temporal Alignment, Orthogonality, and
Covariance Decorrelation loss---designed to regularize the feature
space of \espresso. While the efficacy of these losses was demonstrated
on the transformer-based architecture of \espresso, their impact on the
CNN-based DCF architecture remains unclear. In this section, we provide
an abbreviated benchmarking of the loss augmentation strategies on the
Modified DCF architecture using the best-performing loss weights
identified for \espresso.

\subsubsection{Results}

\begin{figure}[]
    \centering
    \begin{subfigure}[b]{0.8\linewidth}
        \centering
        \includegraphics[width=\linewidth]{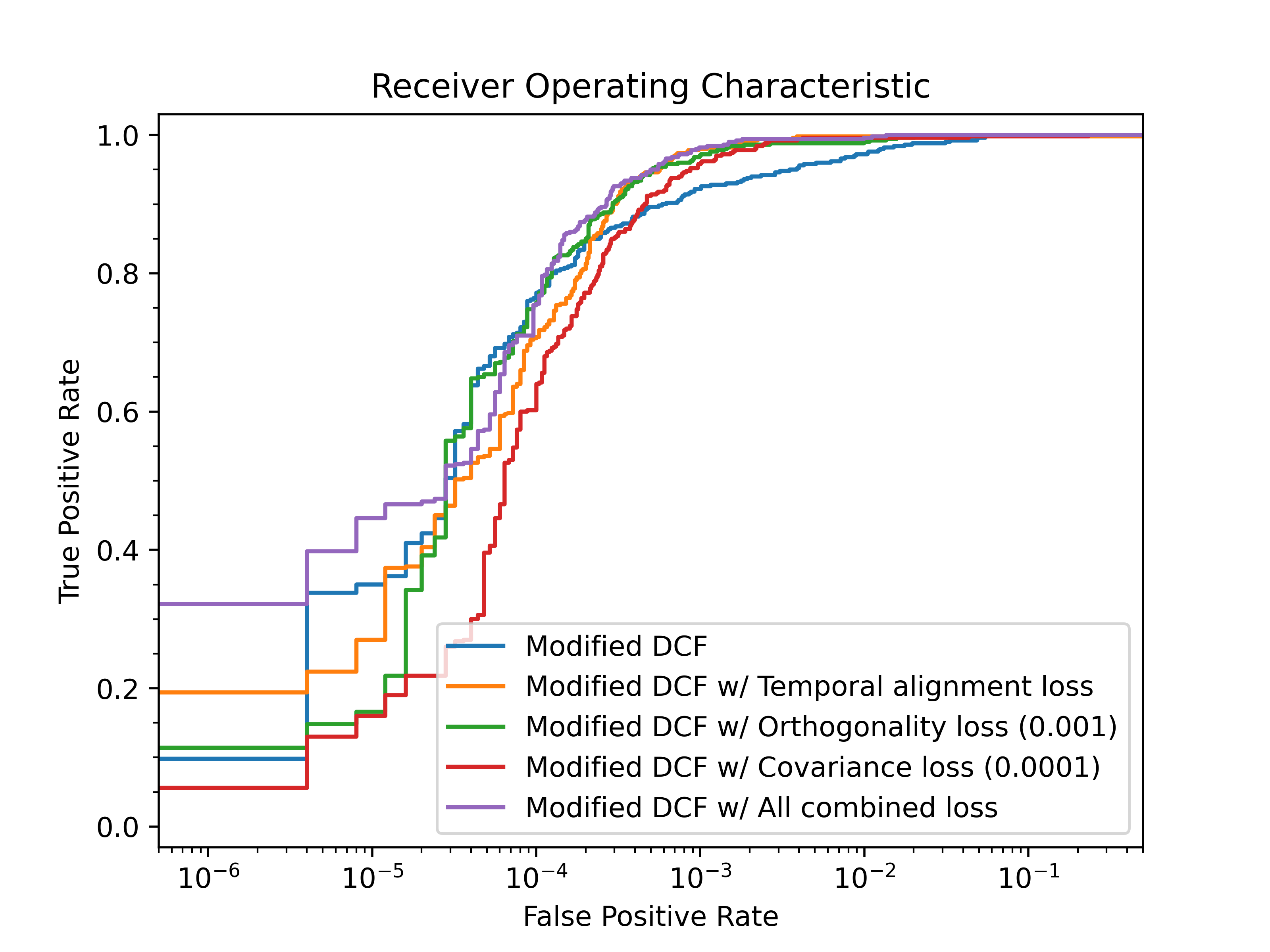}
        \caption{SSH-only ROC}
        \label{fig:roc-ssh-dcf}
    \end{subfigure}
    \hfill
    \begin{subfigure}[b]{0.8\linewidth}
        \centering
        \includegraphics[width=\linewidth]{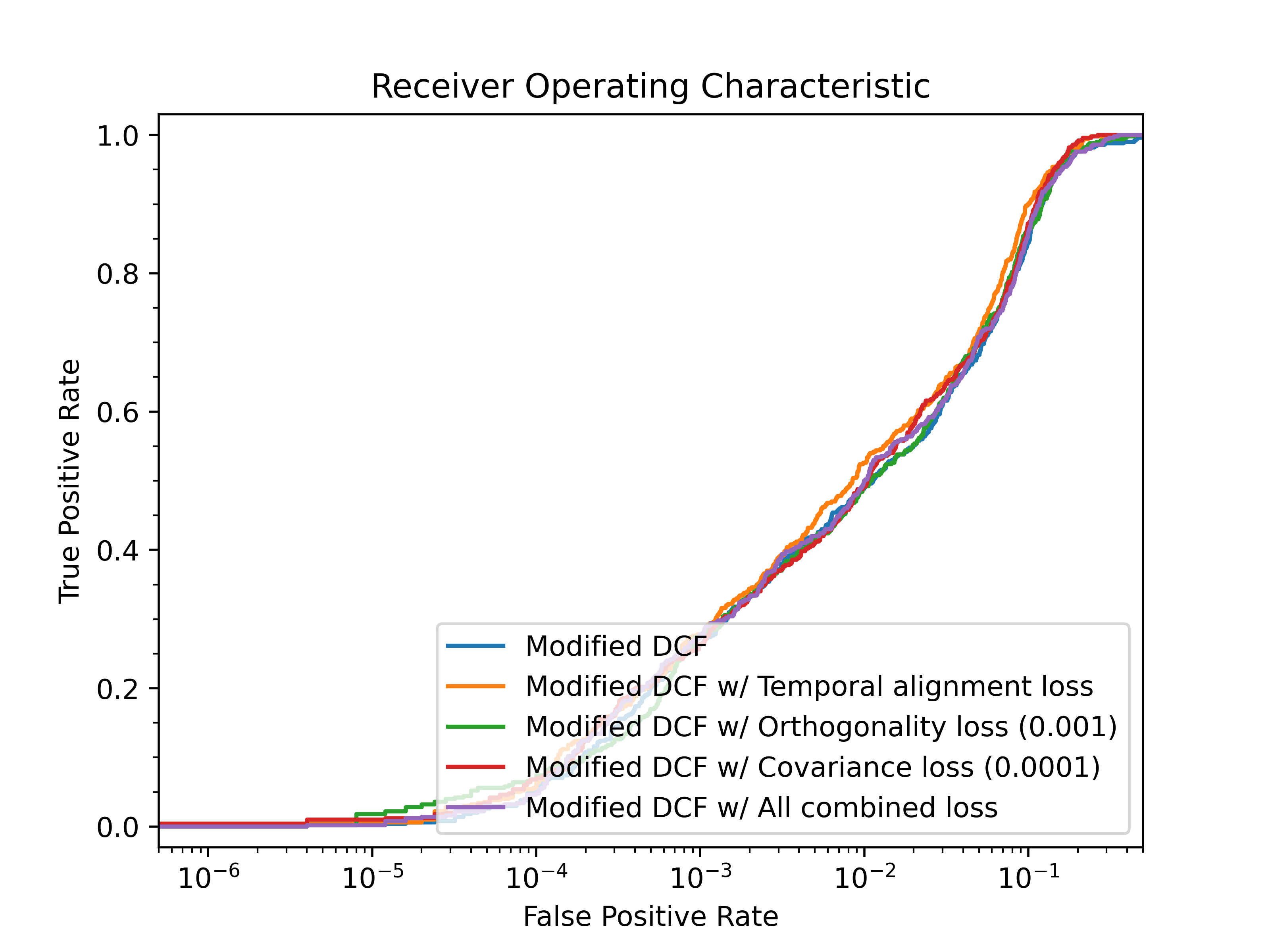}
        \caption{DNS-only ROC}
        \label{fig:roc-dns-dcf}
    \end{subfigure}


    \begin{subfigure}[b]{0.8\linewidth}
        \centering
        \includegraphics[width=\linewidth]{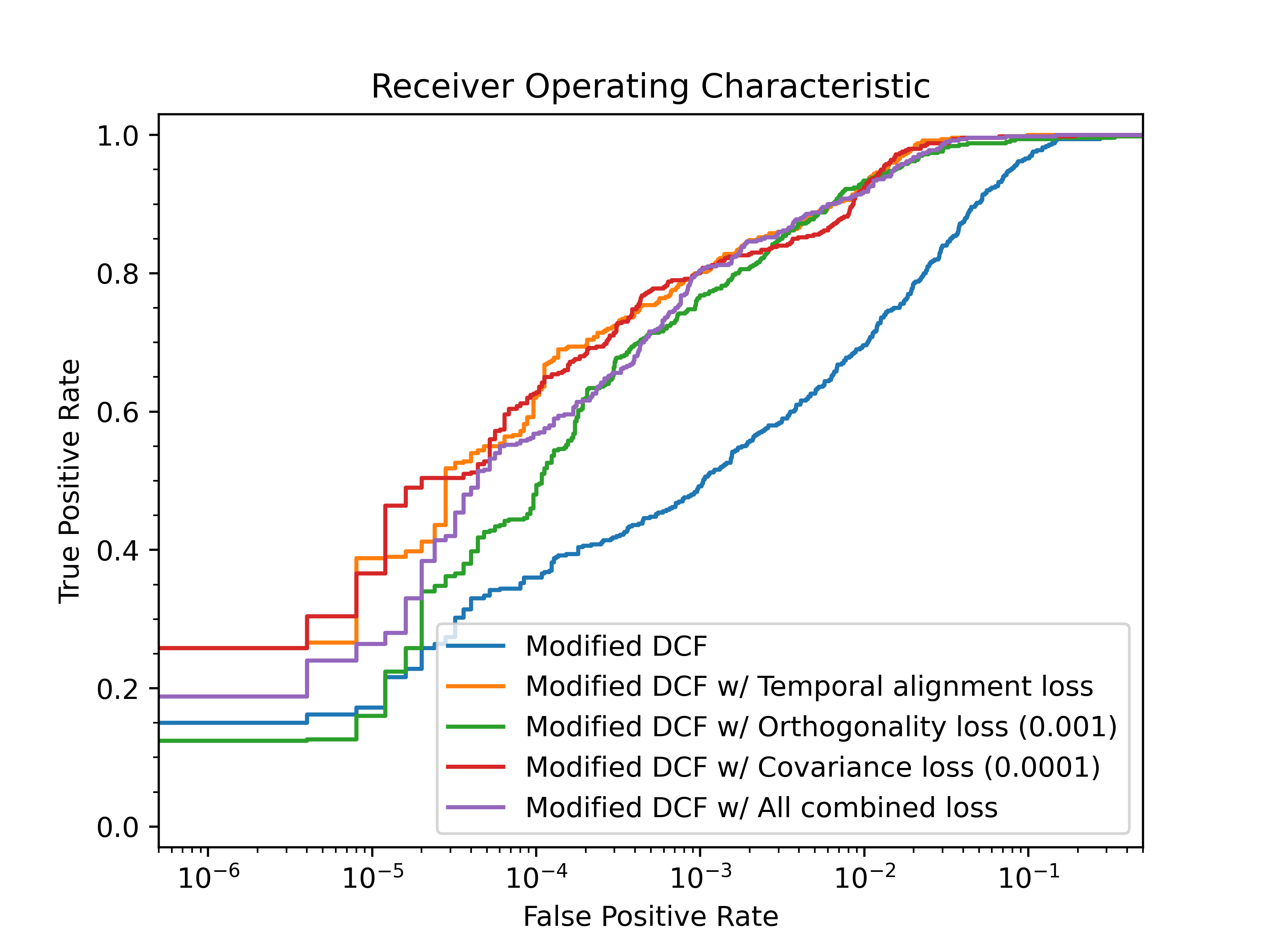}
        \caption{Mixed Protocol ROC}
        \label{fig:roc-mixed-dcf}
    \end{subfigure}

    \caption{ROC curves demonstrating correlation performance of
    Modified DCF with loss augmentation strategies on (a) SSH-only,
    (b) DNS-only, and (c) Mixed protocol datasets.}
    \label{fig:roc-protocols-combined}
\end{figure}

Figure~\ref{fig:roc-protocols-combined} shows the ROC curves for
Modified DCF applied with various loss configurations on the SSH-only,
DNS-only, and Mixed-protocol datasets.

Similar to the results for \espresso, the impact of the loss on
correlation is highly variable. On SSH-only data, the combined loss
augmentation outperforms other models at $10^{-5}$ FPR, achieving a
TPR of 0.446 compared to the 0.350 TPR of the baseline model. It
outperforms other loss augmentations at this FPR threshold but drops
off steeply. Under the DNS-only traffic setting, all training losses
perform nearly identically to one another, demonstrating minimal
performance separation. Against the Mixed-protocol dataset, the
baseline performed relatively poorly, leading most loss augmentation
strategies to outperform it across a wide range of FPR targets. In
these trials, the temporal alignment and covariance losses came out
ahead, scoring 0.620 and 0.626 TPR at $10^{-4}$ FPR, compared to
0.360 TPR achieved by the baseline. Overall, performance deviations
among the different loss training strategies appear more spurious than
indicative of any clear trend.

\end{document}